\documentclass{aa}

\usepackage{amsmath, amsthm, amssymb, amsfonts}
\usepackage[T1]{fontenc}
\usepackage{graphicx}
\usepackage{natbib}
\usepackage{subfig}
\usepackage{siunitx}
\usepackage{cleveref}

\crefname{section}{§}{§§}
\Crefname{section}{§}{§§}
\crefformat{section}{§#2#1#3}
\bibpunct{(}{)}{;}{a}{}{,} 
\title{}
\author{zofi}

\begin{document}
        
           \title{Structure of the outer Galactic disc with \textit{Gaia} DR2}
        
        \subtitle{}
        
        \author{\v{Z}. Chrob\'{a}kov\'{a}\inst{1,2}, R. Nagy\inst{3}, M. L\'{o}pez-Corredoira\inst{1,2}}
        
        \institute{Instituto de Astrofísica de Canarias, E-38205 La Laguna, Tenerife, Spain
                \and
                Departamento de Astrofísica, Universidad de La Laguna, E-38206 La Laguna, Tenerife, Spain
                \and
                Faculty of Mathematics, Physics, and Informatics, Comenius University, Mlynsk\'{a} dolina, 842 48 Bratislava, Slovakia
        }
        
        \date{Received xxxx; accepted xxxx}
        
        
        \abstract
        {The structure of outer disc of our Galaxy is still not well described, and many features need to be better understood. The second Gaia data release (DR2) provides data in unprecedented quality that can be analysed to shed some light on the outermost parts of the Milky Way.}
        {We calculate the stellar density using star counts obtained from Gaia DR2 up to a Galactocentric distance R=20 kpc with a deconvolution technique for the parallax errors. Then we analyse the density in order to study the structure of the outer Galactic disc, mainly the warp.}
        {In order to carry out the deconvolution, we used the Lucy inversion technique for recovering the corrected star counts. We also used the Gaia luminosity function of stars with $M_G<10$ to extract the stellar density from the star counts.}
        {The stellar density maps can be fitted by an exponential disc in the radial direction $h_r=2.07\pm0.07$ kpc, with a weak dependence on the azimuth, extended up to 20 kpc without any cut-off. The flare and warp are clearly visible. The best fit of a symmetrical S-shaped warp gives $z_w\approx z_\odot+(37\pm4.2(stat.)-0.91(syst.)) pc\cdot\left(R/R_\odot\right)^{2.42\pm 0.76(stat.) + 0.129 (syst.)}sin(\phi+\ang{9.3}\pm\ang{7.37} (stat.) +\ang{4.48} (syst.))$ for the whole population. When we analyse the northern and southern warps separately, we obtain an asymmetry of an $\sim25\%$ larger amplitude in the north. This result may be influenced by extinction because the Gaia G band is quite prone to extinction biases. However, we tested the accuracy of the extinction map we used, which shows that the extinction is determined very well in the outer disc. Nevertheless, we recall that we do not know the full extinction error, and neither do we know the systematic error of the map, which may influence the final result. \\
        The analysis was also carried out for very luminous stars alone ($M_G<-2$), which on average represents a younger population. We obtain similar scale-length values, while the maximum amplitude of the warp is $20-30\%$ larger than with the whole population. The north-south asymmetry is maintained.    
        }
        {}
        
        \maketitle

\section{Introduction}\label{intro}
Studying the Galactic structure is crucial for our understanding of the Milky Way. Star counts are widely used for this purpose \citep{paul}, and the importance of this tool has increased in the past decades with the appearance of wide-area surveys \citep{bahcall, majewski}, which made it possible to obtain reliable measurements of the Galactic thin- and thick-disc and halo \citep{chen_counts,juric,bovy,robin_bulge}. It is common to simplify the Galactic disc as an exponential or hyperbolic secant form, but there are many asymmetries such as the flare and warp that need to be taken into account. These structures can be seen from 3D distribution of stars, as shown by \cite{liu}, who mapped the Milky Way using the LAMOST (The Large Sky Area Multi-Object Fibre Spectroscopic Telescope) RGB (red-giant branch) stars; \cite{skowron}, who constructed a map of the Milky Way from classical Cepheids; or \cite{anders}, who used the second Gaia data release (DR2).

The warp was first detected in the Galactic gaseous disc in 21 cm HI observations \citep{kerr,oort}. Since then, the warp has also been discovered in the stellar disc \citep{carney,martin_warp,reyle,amores,chen_warp}, and the kinematics of the warp has been studied as well \citep{dehnen,drimmel,martin_warp_kin,schonrich}. \\

Vertical kinematics in particular can reveal much about the mechanism behind the formation of warp. \cite{poggio} found a gradient of $5-6$ km/s in the vertical velocities of upper main-sequence stars and giants located from 8 to 14 kpc in Galactic radius using Gaia DR2 data, revealing the kinematic signature of the warp. Their findings suggest that the warp is principally a gravitational phenomenon. \cite{skowron_kin_warp} also found a strong gradient in vertical velocities using classical Cepheids supplemented by the OGLE (Optical Gravitational Lensing Experiment) survey. \cite{nas_clanok} investigated the dynamical effects produced by different mechanisms that can explain the radial and vertical components of extended kinematic maps of \cite{martin}, who used Lucy's deconvolution method (see Sect. \ref{ch6}) to produce kinematical maps up to a Galactocentric radius of 20 kpc. \cite{nas_clanok} found that vertical motions might be dominated by
external perturbations or mergers, although with a minor component due to a warp whose amplitude is evolving with time. However, the kinematic signature of the warp is not enough to explain the observed velocities. \\

To date, the shape of the warp has been constrained only roughly, and the kinematical information is not satisfying enough to reach consensus about the mechanism causing the warp. Theories include accretion of intergalactic matter onto the disc \citep{martin_accretion}, interaction with other satellites \citep{kim}, the intergalactic magnetic field \citep{battaner}, a misaligned rotating halo \citep{debattista}, and others.

We now have a new opportunity to improve our knowledge about the Milky Way significantly through the Gaia mission of the European Space Agency \citep{gaia2}. Gaia data provide unprecedented positional and radial velocity measurements and an accurate distance determination, although the error of the parallax measurement increases with distance from us. It brings us the most accurate data about the Galaxy so far, ideal to advance in all branches of Galactic astrophysics and study our Galaxy in greater detail than ever before. Gaia DR2 has been used by \cite{anders}, who provided photo-astrometric distances, extinctions, and astrophysical parameters up to magnitude G=18, making use of the Bayesian
parameter estimation code {\tt StarHorse}. After introducing the observational data and a number of priors, their code finds the Bayesian stellar parameters, distances, and extinctions. The authors also present density maps, which we compare with our results in Section \ref{ch8}. Gaia data have also been used to study the structure of outer Galactic disc, especially the warp and the flare. The first Gaia data release brought some evidence of the warp \citep{schonrich}, but the more extensive second data release provides a better opportunity to study the warp attributes. \cite{poggio} combined Gaia DR2 astrometry with 2MASS (Two Micron All-Sky Survey) photometry and revealed the kinematic signature of the warp up to 7 kpc from the Sun. \cite{li} found the flare and the warp in the Milky Way, using only OB stars of the Gaia DR2. In this work, we make use of Gaia DR2 data as described in Section \ref{ch2} and use star counts to obtain the stellar density by applying Lucy's inversion technique. Then we analyse the density maps to determine the warp. 

The paper is structured as follows: in Section \ref{ch2} we describe the Gaia data and extinction maps that we used, in Section \ref{ch4} we present the luminosity function used in our calculations, in Section \ref{ch5} we explain the methods for obtaining our density maps, and in Section \ref{results} we discuss the results. In Section \ref{chdensity} we present the exponential fits of the density, in Section \ref{ch11} we study the warp, and in Section \ref{ch12} we repeat the previous analysis of the young population.

\section{Data selection}\label{ch2}

We used data of the second Gaia data release \citep{gaia} here, which were collected during first 22 months of observation. We are interested in stars with known five-parameter astrometric solution: more than 1.3 billion sources. G magnitudes, collected by astrometric instrument in the white-light G-band of Gaia (330–1050 nm) are known for all sources, with precisions varying from around 1 millimag at the bright (G<13) end to around 20 millimag at G=20. For the details on the astrometric data processing and validation of these results, see \cite{lindegren}. We chose stars with apparent magnitude up to G=19, where the catalogue is complete up to 90\% \citep{arenou}. We chose data with a parallax in the interval [0,2] mas.

In our analysis, we did not consider any zero-point bias in the parallaxes of Gaia DR2, as found by some authors \citep{lindegren,arenou,stassun,zinn}, except in Sect. 4.3, where we repeat our main calculation including a non-zero value of the zero-point to prove that this effect is negligible in our results.

\subsection*{Extinction maps}\label{ch3}
We used two different extinction maps. For the luminosity function (Sect. \ref{ch4}), we used the extinction map of \cite{green_extinkcia} through its Python package \textit{dustmaps}, choosing the \textit{Bayestar17} version. This map covers 75\% of the sky (declinations of $\delta\gtrsim \ang{-30}$) and provides reddening in similar units as \citet[SFD]{sfd}. \\ 
To calculate the density (Sect. \ref{ch5}), we need to cover the whole sky, therefore we used the three-dimensional less accurate but full-sky extinction map of \cite{bovy_extinkcia} through its Python package \textit{mwdust}. This map combines the results of \cite{marshall}, \cite{green}, and \cite{drimmel} and provides reddening as defined in \cite{sfd}. \\
In order to convert the interstellar reddening of these maps into $E(B-V)$, we used coefficients \citep{hendy_ext_koef,rybizki_ext_koef}

\begin{eqnarray}\label{1}
\begin{split}
A_G/A_v&=0.859~, \\
R_V&=A_v/E(B-V)=3.1~.
\end{split}
\end{eqnarray}

\section{Luminosity function}\label{ch4}
To construct the luminosity function, we chose all stars with heliocentric distance $d<0.5$ kpc (distances determined as $1/\pi$, where $\pi$ is the parallax). We did not find many bright stars ($M_G<-5$) in this area, therefore we also chose a specific region with Galactic height $\lvert z \rvert <1$ kpc and Galactocentric distance $R<5$ kpc, in which we only selected stars with absolute magnitude $M_G<-5$. We normalised the counts of stars with high magnitude and then joined these two parts to create the luminosity function. \\
In the range of distance that we used for the luminosity function, the star counts are complete for the absolute magnitude that we are calculating, except perhaps for the possible loss of the brightest stars through saturation at $M_G<-5$. Moreover, the error in the parallax for these stars is negligible, so that the calculation of the absolute magnitude from the apparent magnitude is quite accurate. We did not take into account the variations of the luminosity function throughout the Galactic disc. We assumed that it does not change.

\begin{figure}
        \includegraphics[width=0.5\textwidth]{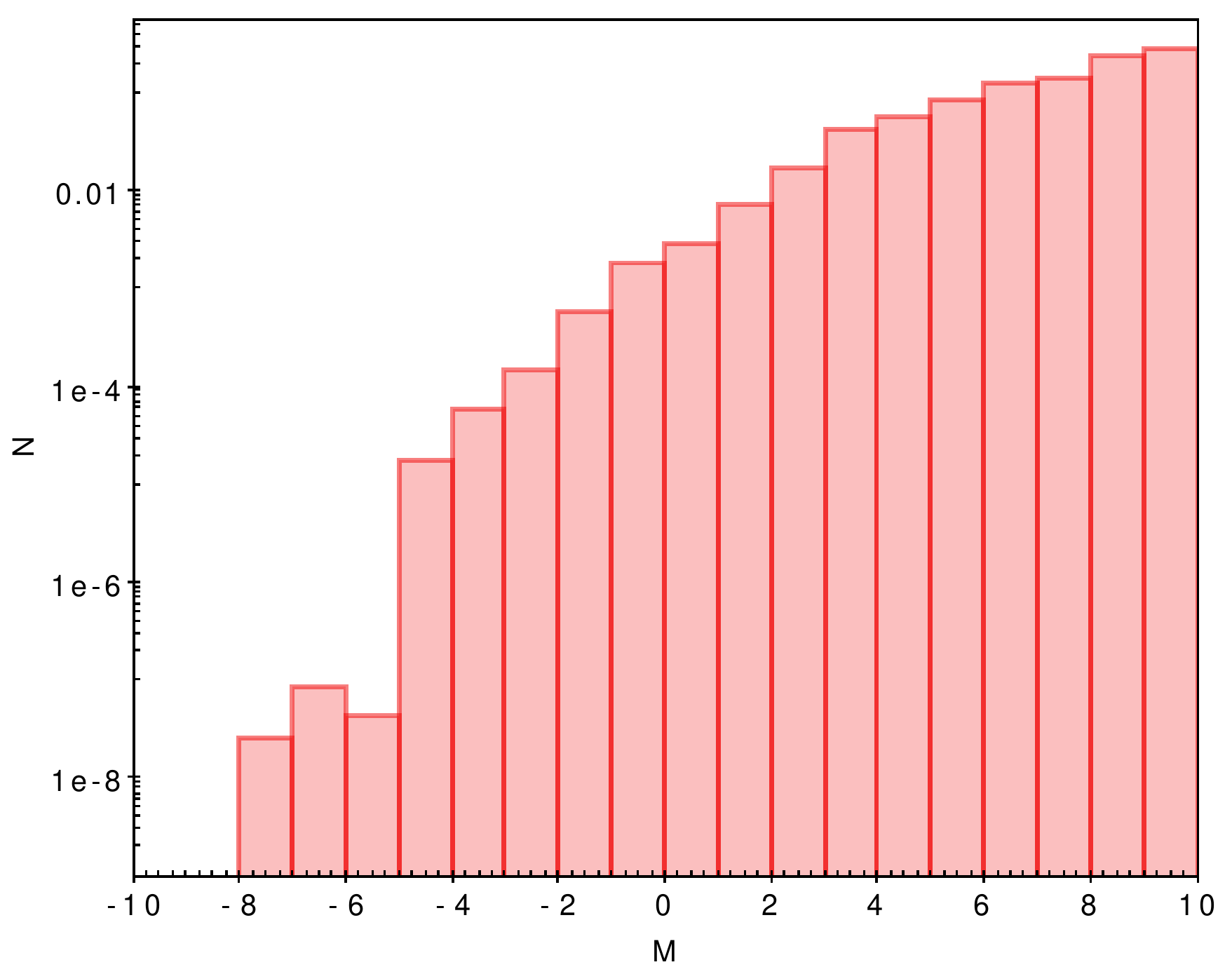}
        \caption{Luminosity function.}\label{o1}
\end{figure}

\begin{figure}
        \includegraphics[width=0.5\textwidth]{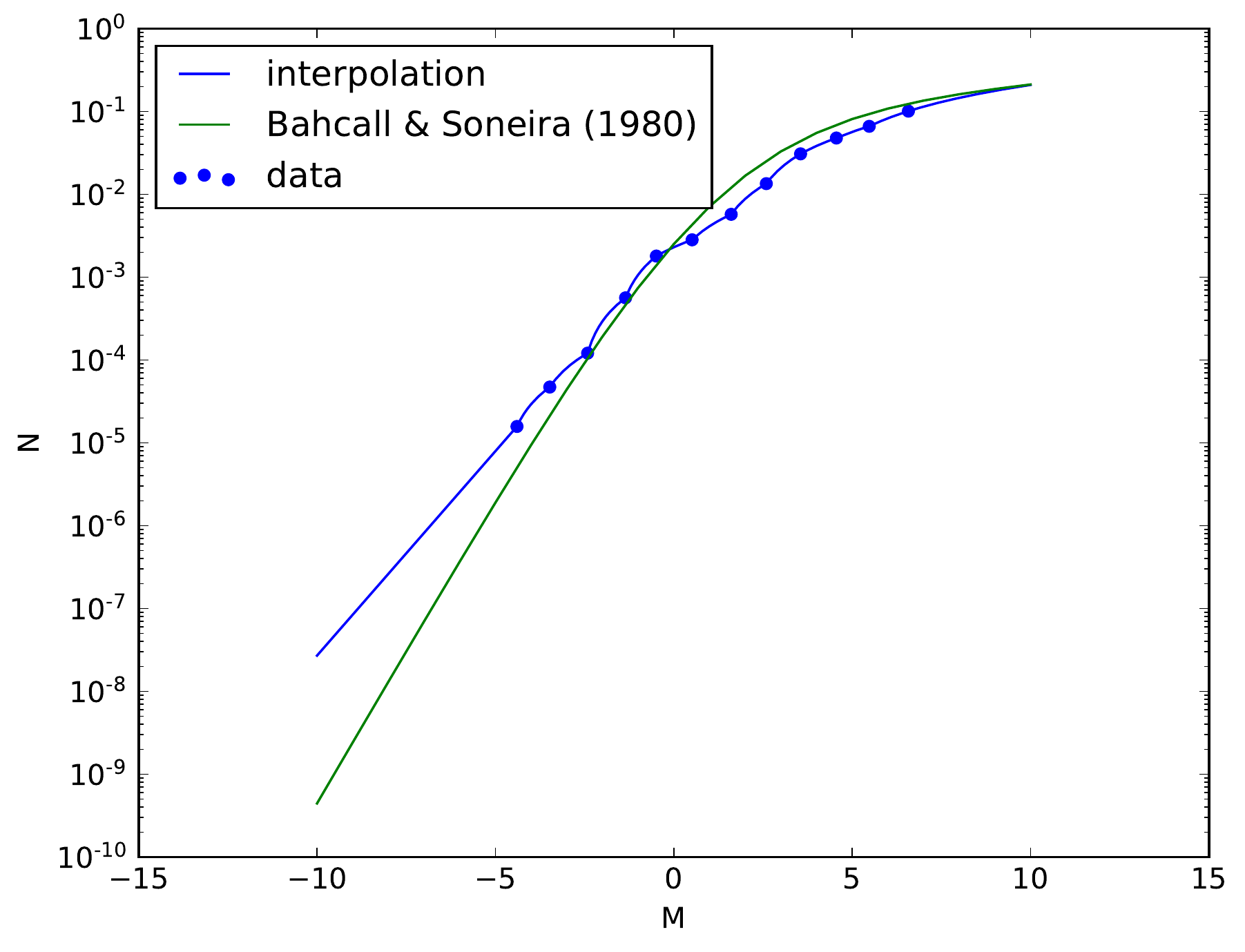}
        \caption{Interpolation of the luminosity function with a spline compared with the luminosity function of \cite{bahcall_lum}. These two functions are not directly comparable because \cite{bahcall_lum} used a slightly different filter in the visible, but it shows that our luminosity function is reasonable.}\label{o2}
\end{figure}

The luminosity function we obtained is shown in Fig.\ref{o1}. We interpolated the luminosity function with a spline $N=spl(M)$ of the first degree. The result is shown in Fig.\ref{o2}. For the interpolation, we used values between magnitudes $M=[-5,10]$ because the values outside this interval are unreliable, and we used the extrapolation of the spline function to lower magnitudes. The values of the luminosity function are listed in Table \ref{lumf}.

\begin{table}[h!]
        \centering
        \begin{tabular}{|c|c|c|c|c|} \hline
                $M_G$ & N  \\   \hline
                 -10 & $2.704\cdot 10^{-8}$   \\ \hline
                -9 & $8.424\cdot 10^{-8}$          \\ \hline
                -8 & $2.625\cdot 10^{-7}$       \\ \hline
                -7 & $8.177\cdot 10^{-7}$     \\ \hline
                -6 & $2.547\cdot 10^{-6}$    \\ \hline
                -5 & $7.936\cdot 10^{-6}$       \\ \hline
                -4 & $2.927\cdot 10^{-5}$     \\ \hline
                -3 & $8.028\cdot 10^{-5}$     \\ \hline
                -2 & $2.936 \cdot 10^{-4}$ \\ \hline
                -1 & $1.066 \cdot 10^{-3}$     \\ \hline
                0 & $2.299 \cdot 10^{-3}$       \\ \hline
                1 & $4.117 \cdot 10^{-3}$      \\ \hline
                2 & $8.805 \cdot 10^{-3}$      \\ \hline
                3 & $2.081\cdot 10^{-2}$      \\ \hline
                4 & $3.838\cdot 10^{-2}$  \\ \hline
                5 & $5.667\cdot 10^{-2}$  \\ \hline
                6 & $8.273\cdot 10^{-2}$      \\ \hline
                7 & 0.122    \\ \hline
                8 & 0.171     \\ \hline
                9 & 0.221     \\ \hline
                10 &  0.27      \\ \hline
        \end{tabular}
        \caption{Values of the luminosity function.}
        \label{lumf}
\end{table}

\section{Density maps}\label{ch5}

\subsection{Deconvolution of star counts}\label{ch6}
To calculate the stellar density, we need to measure star counts as a function of distance. However, the error of parallax increases with distance from us, which means that our analysis would be correct only within roughly $5$ kpc from the Sun. To be able to reach higher distances, we corrected for this effect using the method developed by \cite{martin}, who used Lucy's deconvolution method (Lucy 1974; see Appendix A) to obtain an accurate distance measurement up to $R=20$ kpc.
They expressed the observed number of stars per parallax $\overline{N}(\pi)$ as a convolution of the real number $N(\pi)$ of stars with a Gaussian function 

\begin{eqnarray}\label{2}
\overline{N}(\pi)=\int_{0}^{\infty} \mathrm{d}\pi^\prime N(\pi^\prime)G_{\pi^\prime}(\pi-\pi^\prime)~,
\end{eqnarray}
where

\begin{eqnarray}\label{3}
G_{\pi}(x)=\frac{1}{\sqrt{2\pi}\sigma_\pi}e^{-\frac{x^2}{2\sigma_\pi^2}}~.
\end{eqnarray}
For the error $\sigma_\pi$ we averaged errors of every bin, which we calculated from values given by Gaia DR2. \\
We only used the parallax between [0,2] mas. For the upper limit the relative error of parallax is very small and does not produce any bias. For the lower limit, the truncation avoiding the negative parallaxes affects the distribution of parallaxes and statistical properties (average, median, etc.) \citep[Section 3.3]{gaia_par}. However, in our method we do not calculate the average distance from the average parallax. We used Lucy's method, which iterates the counts of the stars with positive parallaxes until we obtained the final solution. This does not mean that we truncated the star counts with negative parallaxes. We used only the stars with positive parallaxes as is required by our method, explained in the Appendix A. $N(\pi)$ for negative values of $\pi$ can also be calculated and fitted, but they are not used in our calculation. In other words, we did not assume that the number of the stars with negative parallaxes is zero, we simply did not use this information because it is not necessary. The fact that this method does not produce any bias is tested in section 4.2.

\subsection{Monte Carlo simulation to test the Lucy inversion method}\label{ch7}
\begin{figure*}
        \centering
        \subfloat[Gamma distribution]{
                \includegraphics[width=0.5\textwidth]{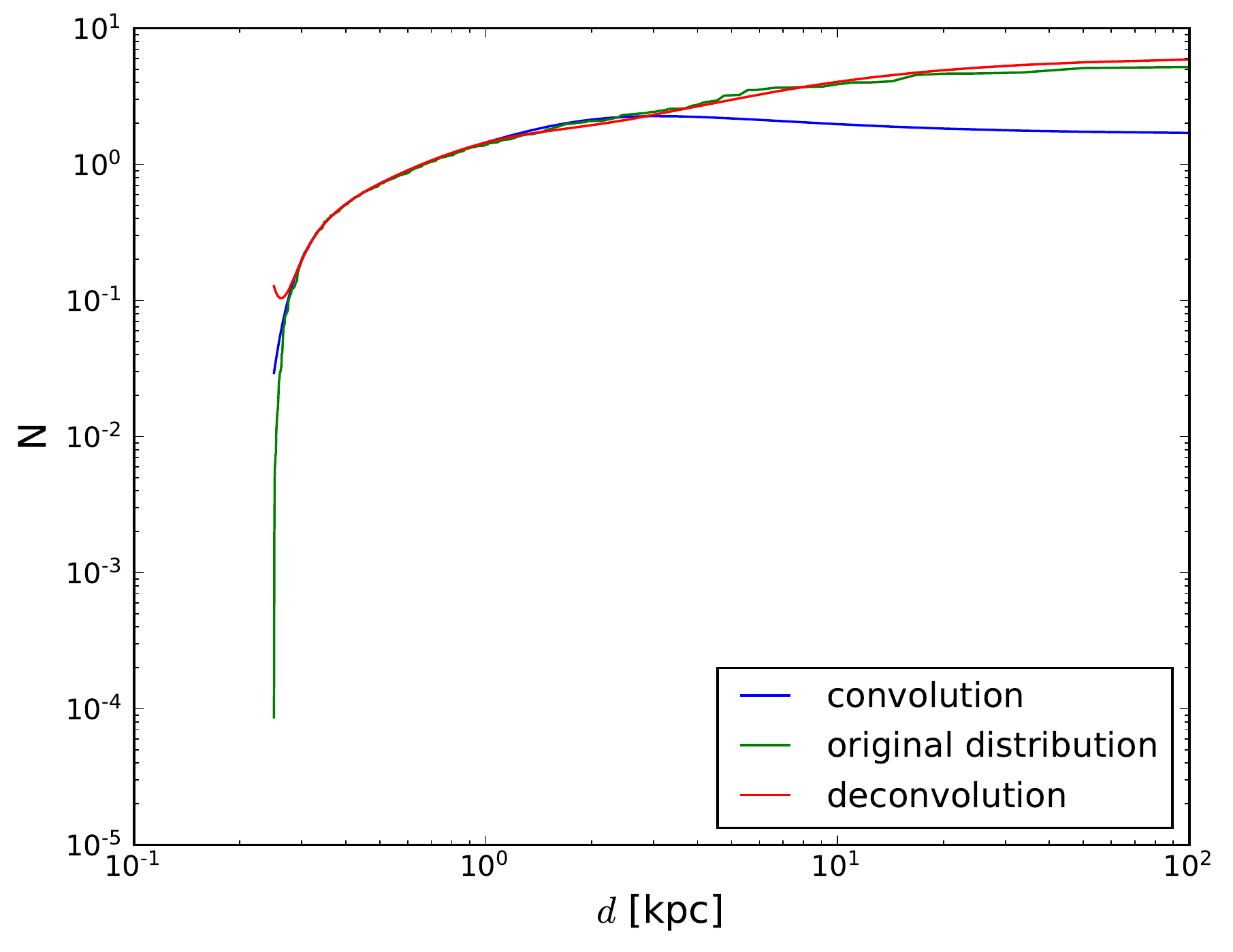}
        }
        \subfloat[Exponential distribution]{
                \includegraphics[width=0.5\textwidth]{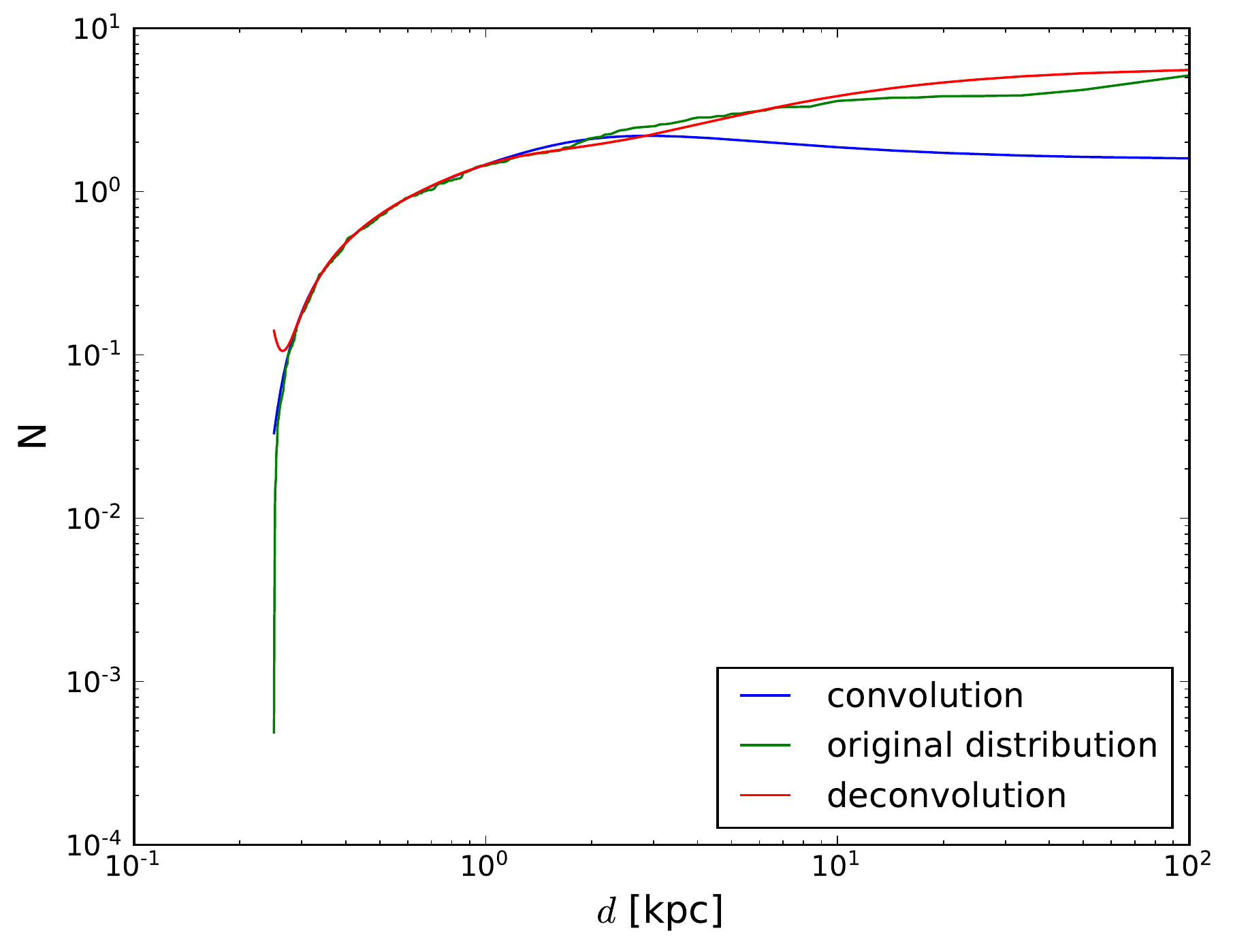}
        }
        \hspace{0mm}
        \subfloat[Geometric distribution]{
                \includegraphics[width=0.5\textwidth]{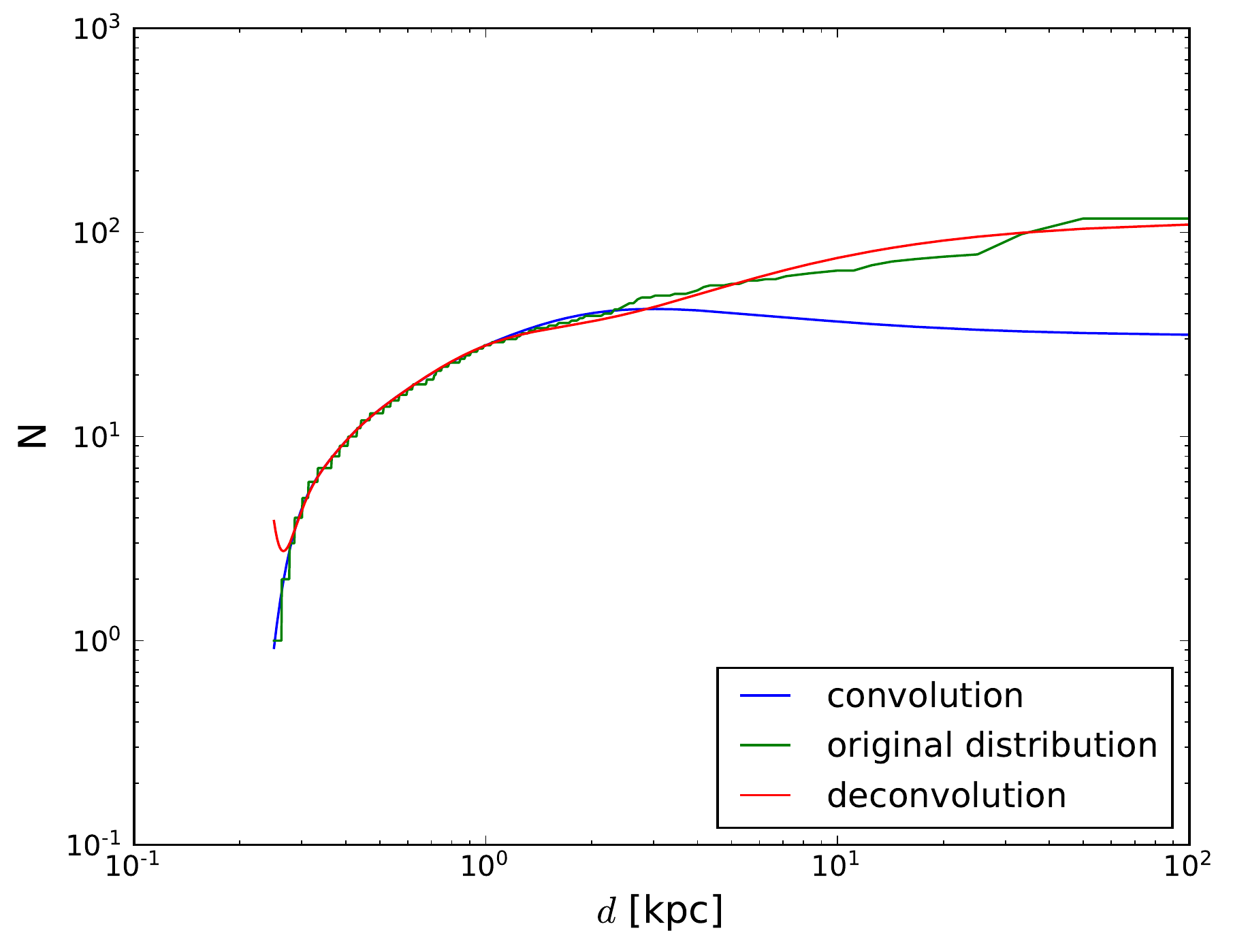}
        }
        \subfloat[Logarithmic distribution]{
                \includegraphics[width=0.5\textwidth]{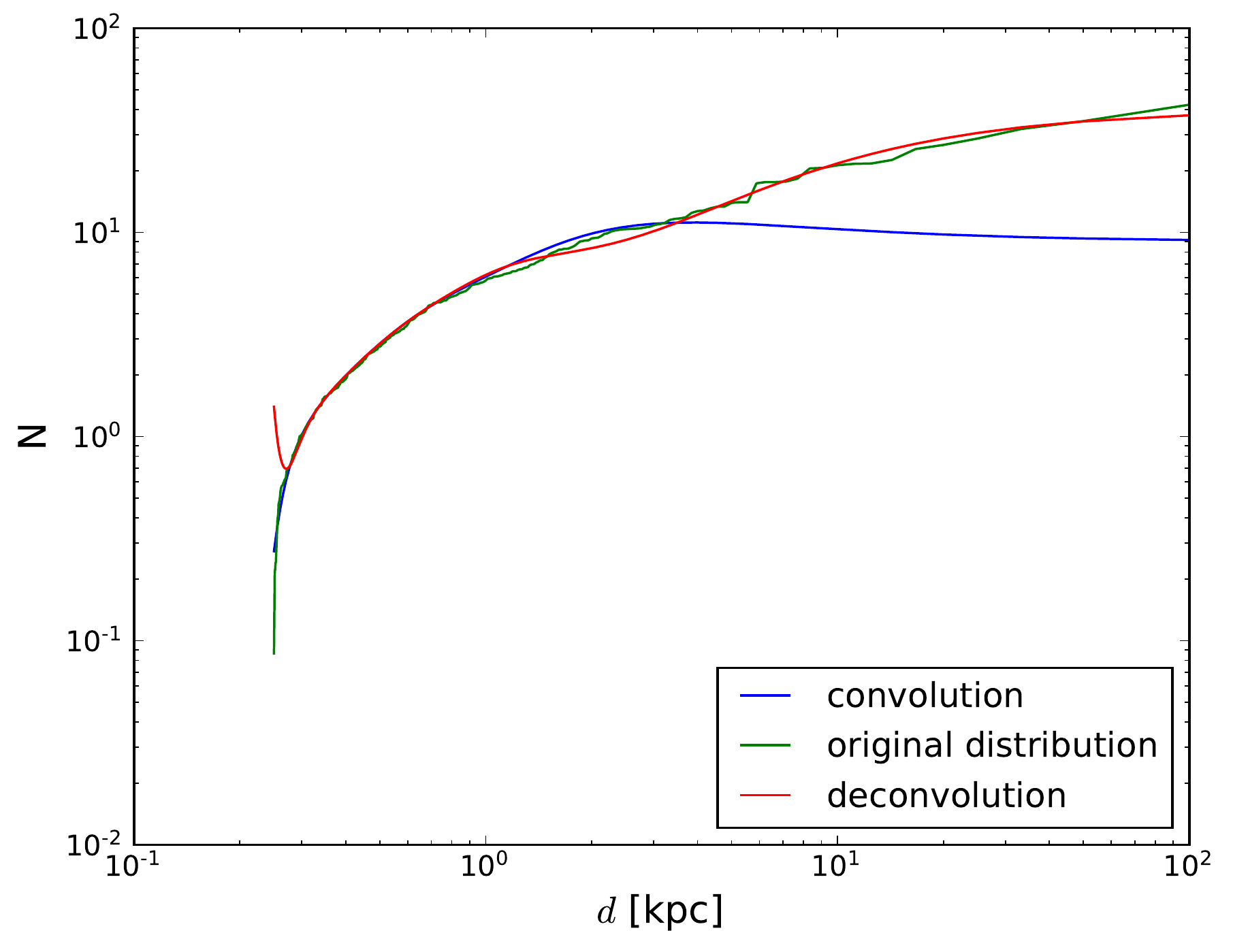}
        }
        \caption{Monte Carlo simulation of deconvolution. We recover random distributions, convolved with a Gaussian.}\label{o3}
\end{figure*}

\begin{figure*}
        \centering
        \subfloat[$\sigma_\pi=0.05$ mas]{
                \includegraphics[width=0.5\textwidth]{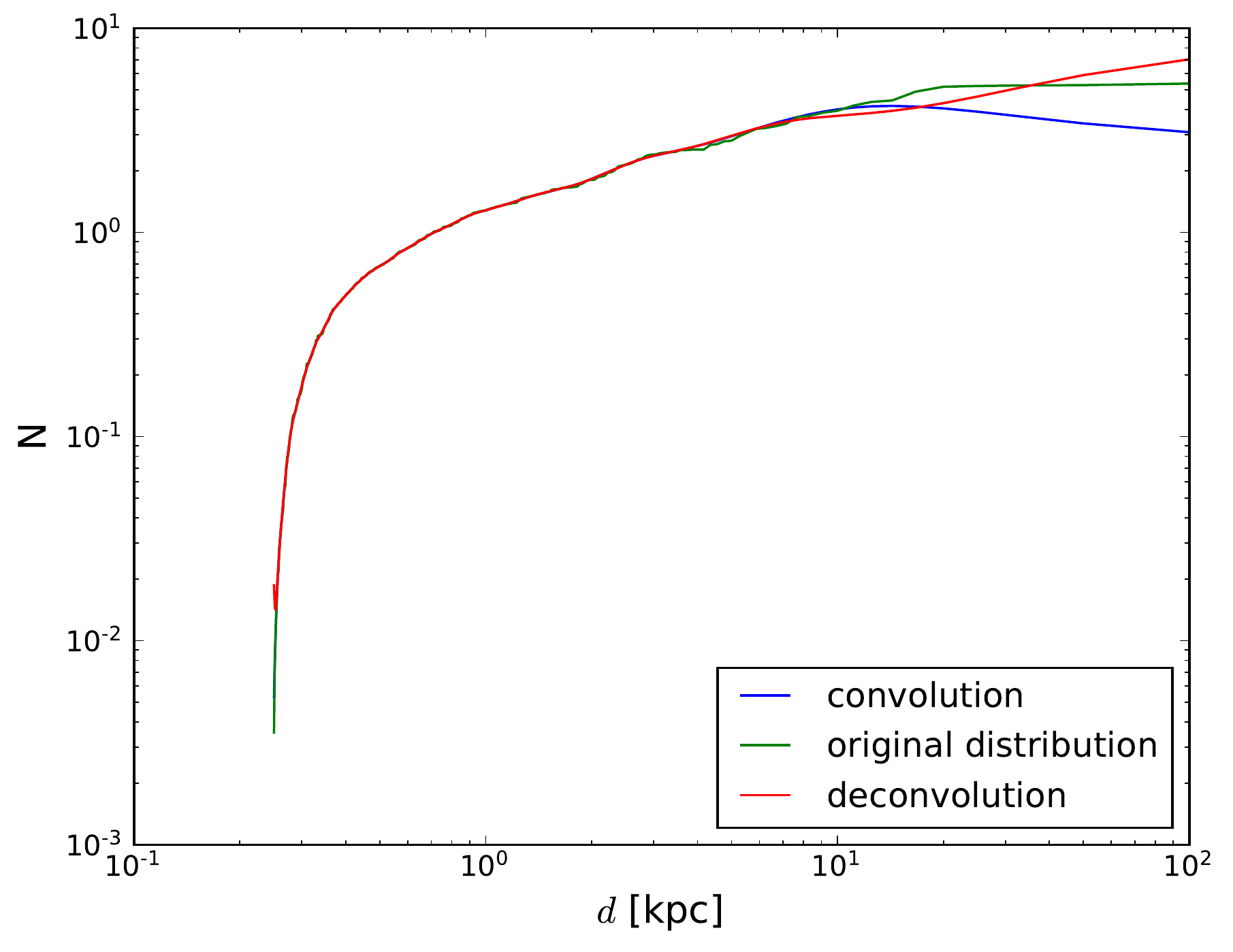}
        }
        \subfloat[$\sigma_\pi=0.1$ mas]{
                \includegraphics[width=0.5\textwidth]{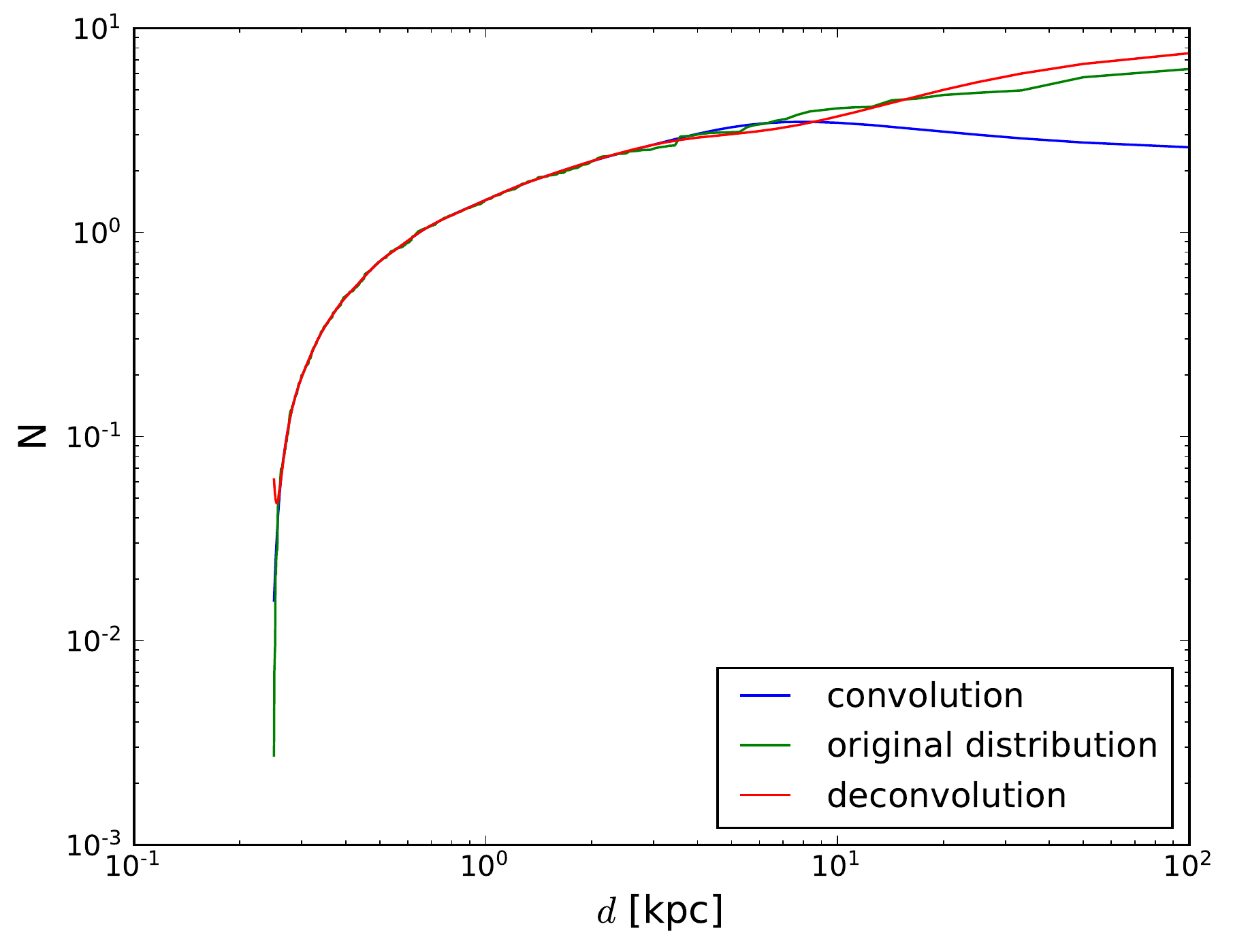}
        }
        \hspace{0mm}
        \subfloat[$\sigma_\pi=0.25$ mas]{
                \includegraphics[width=0.5\textwidth]{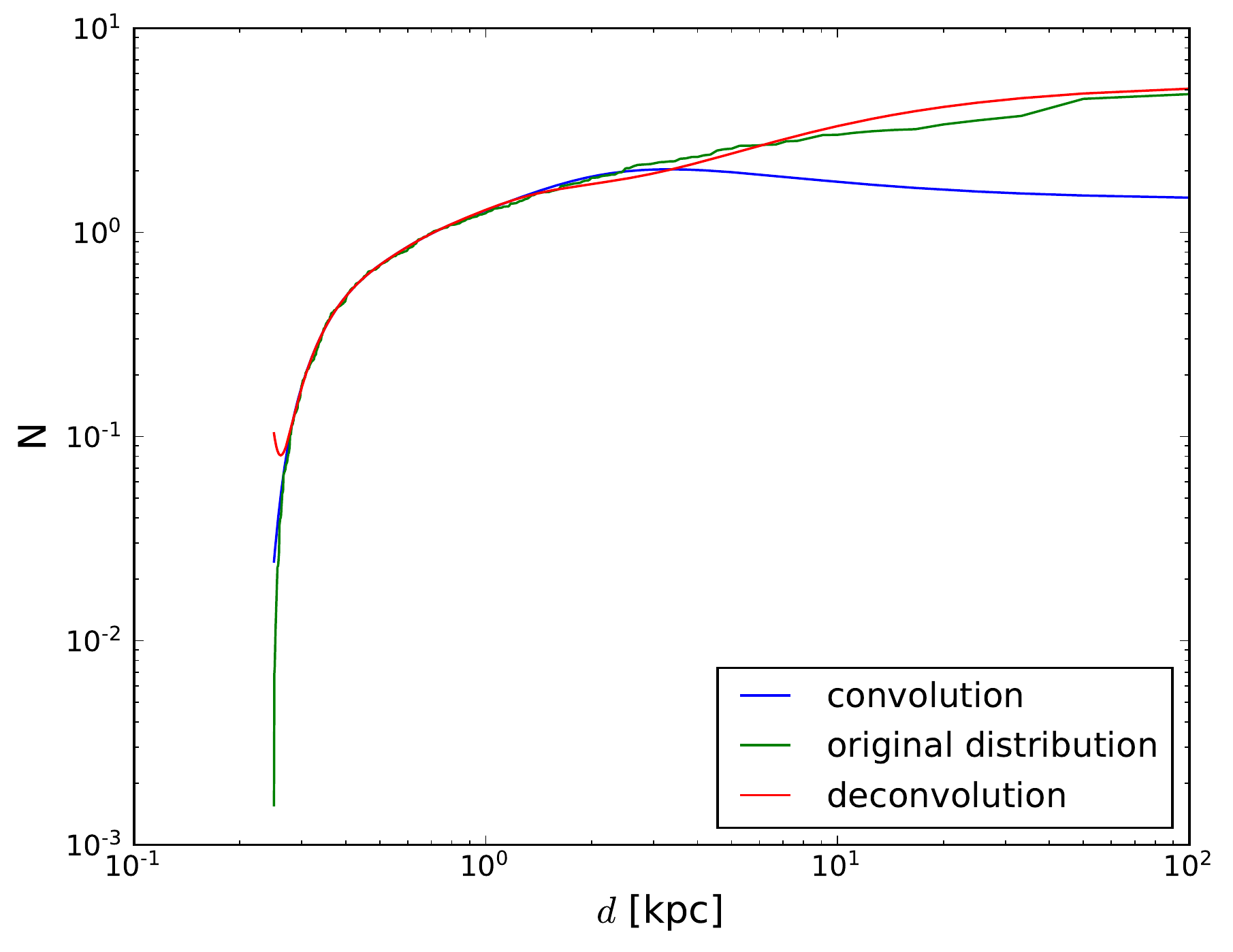}
        }
        \subfloat[$\sigma_\pi=0.4$ mas]{
                \includegraphics[width=0.5\textwidth]{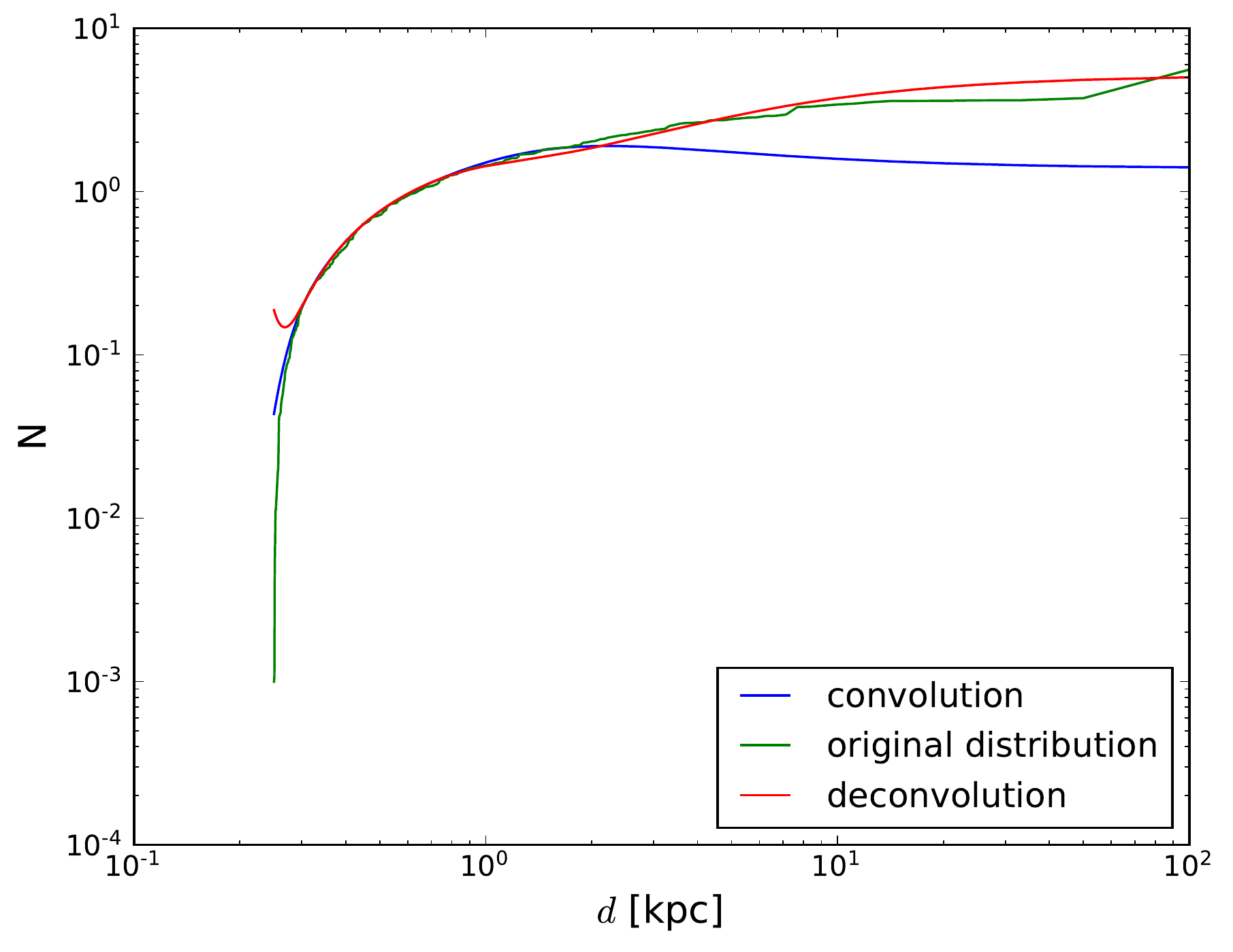}
        }
        \caption{Monte Carlo simulation of deconvolution. In all cases we recover a gamma distribution convolved with a Gaussian, varying the average error of parallax.}\label{chyba_par}
\end{figure*}

In order to test the reliability of the inversion method, we performed Monte Carlo simulations to determine whether we can recover the original function after deconvolution. We created datasets with randomly distributed particles. Then we convolved this distribution with a Gaussian. We applied Lucy's deconvolution method to the dataset to determine whether we can recreate the original distribution. The results are shown in Fig. \ref{o3}. We conclude that regardless of the original distribution, we can accurately recover the original data up to $50$ kpc or more, which is satisfying to study the Milky Way.
We also studied the dependence of the method on the parallax error. We used various values of the average parallax error in Eq. \ref{3} from the interval [0.05,0.4] mas, which are the most common values for the average parallax error in our data. In Fig. \ref{chyba_par} we plot the result, which shows that even though the precision of the method depends on the parallax error, we obtain a satisfying result up to 20 kpc even in the worst case with the highest parallax error.  

\subsection{Application to full-sky Gaia-DR2 data}\label{ch8}
We divided the data into bins of Galactic longitude $\ell$, Galactic latitude $b,$ and apparent magnitude $m$. For the values of $b$ we made bins of length $\ang{2}$ and corresponding $\ell$ in bins of $\ang{5}/cos(b)$. We divided each of the lines of sight in magnitude, binned with size $\Delta m=1.0$ between G=12 and G=19. We obtained 29 206 different areas in which we calculated the density independently. 

We made use of the fundamental equation of stellar statistics, where the number of stars $N(m)$ of apparent magnitude $m$ is expressed per unit solid angle and per unit magnitude interval \citep{chandrasekhar},

\begin{eqnarray}\label{5}
N(m)=\int_{0}^{\infty} \rho(r)\Phi(M)r^2\mathrm{d}r~,
\end{eqnarray}
where we substitute

\begin{eqnarray}
r(m)=(1/\pi)=10^{(m-M+5-A_G(1/\pi))/5}~,
\end{eqnarray}
which yields for the density

\begin{eqnarray}\label{4}
\rho(1/\pi)&=&\frac{N(\pi)\pi^4 }{\Delta\pi\omega  \int_{M_{G,low~lim}}^{M_{G,low~lim}+1} \mathrm{d}M_G\Phi(M_G)}~,
\end{eqnarray}

\begin{eqnarray}
M_{G,low~lim}&=&m_{G,low~lim}-5log_{10}(1/\pi)-10 \nonumber \\
&-&A_{G}(1/\pi)~,
\end{eqnarray}

where $\omega$ is the covered angular surface ($10 ~\mathrm{degrees}^2$ in our case), $\Delta\pi$ is the parallax interval (0.01 mas in our case), which must be added in the equation because we did not use the unit parallax, $\Phi(M_G)$ is the luminosity function in the G filter, $m_{G,low~lim}$ is the limiting maximum apparent magnitude, and $A_{G}(r)$ is the extinction, as a function of distance.

\begin{figure*}
        \centering
        \subfloat[$\ang{0}<\phi<\ang{30}$]{
                \includegraphics[width=0.5\textwidth]{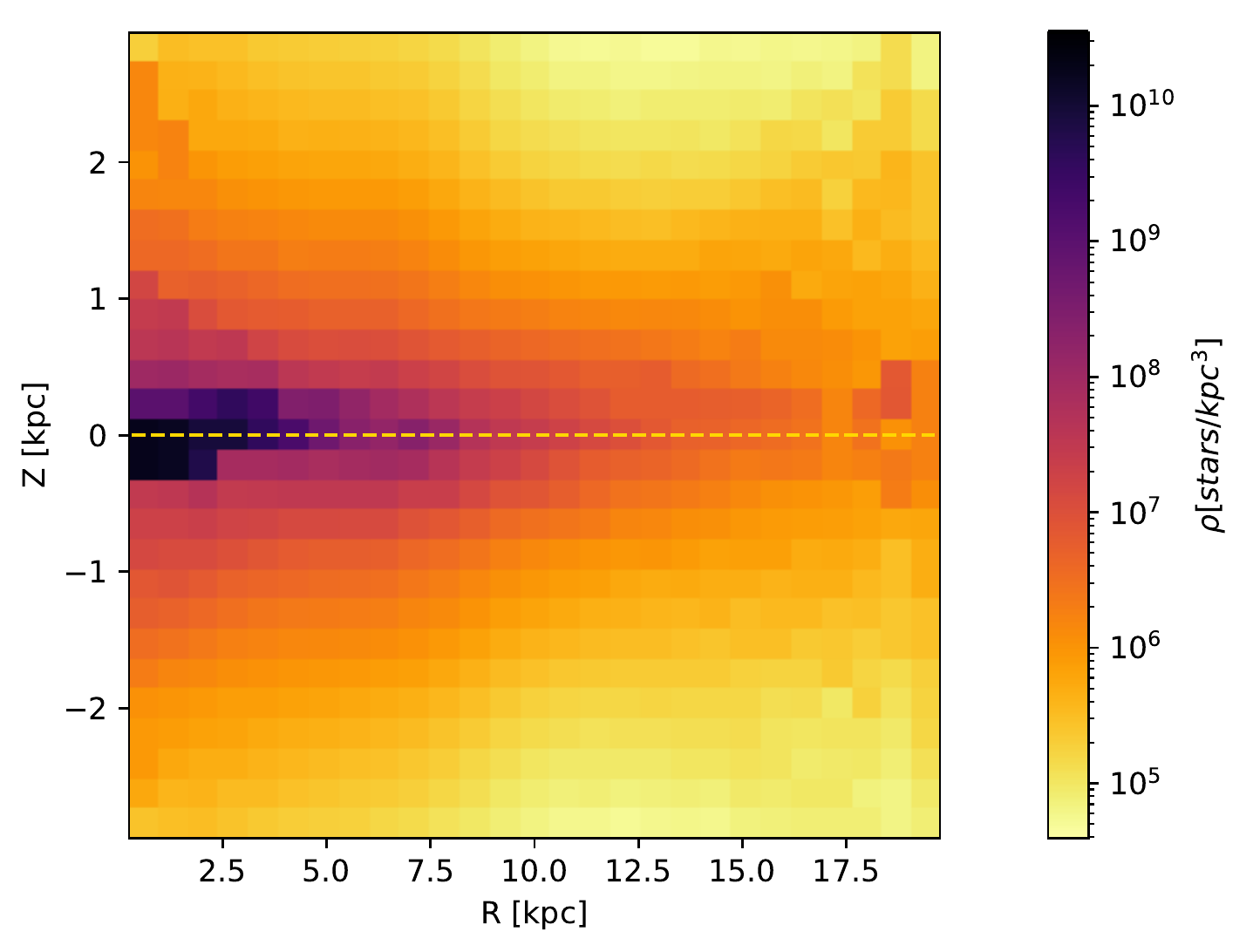}
        }
        \subfloat[$\ang{30}<\phi<\ang{60}$]{
                \includegraphics[width=0.5\textwidth]{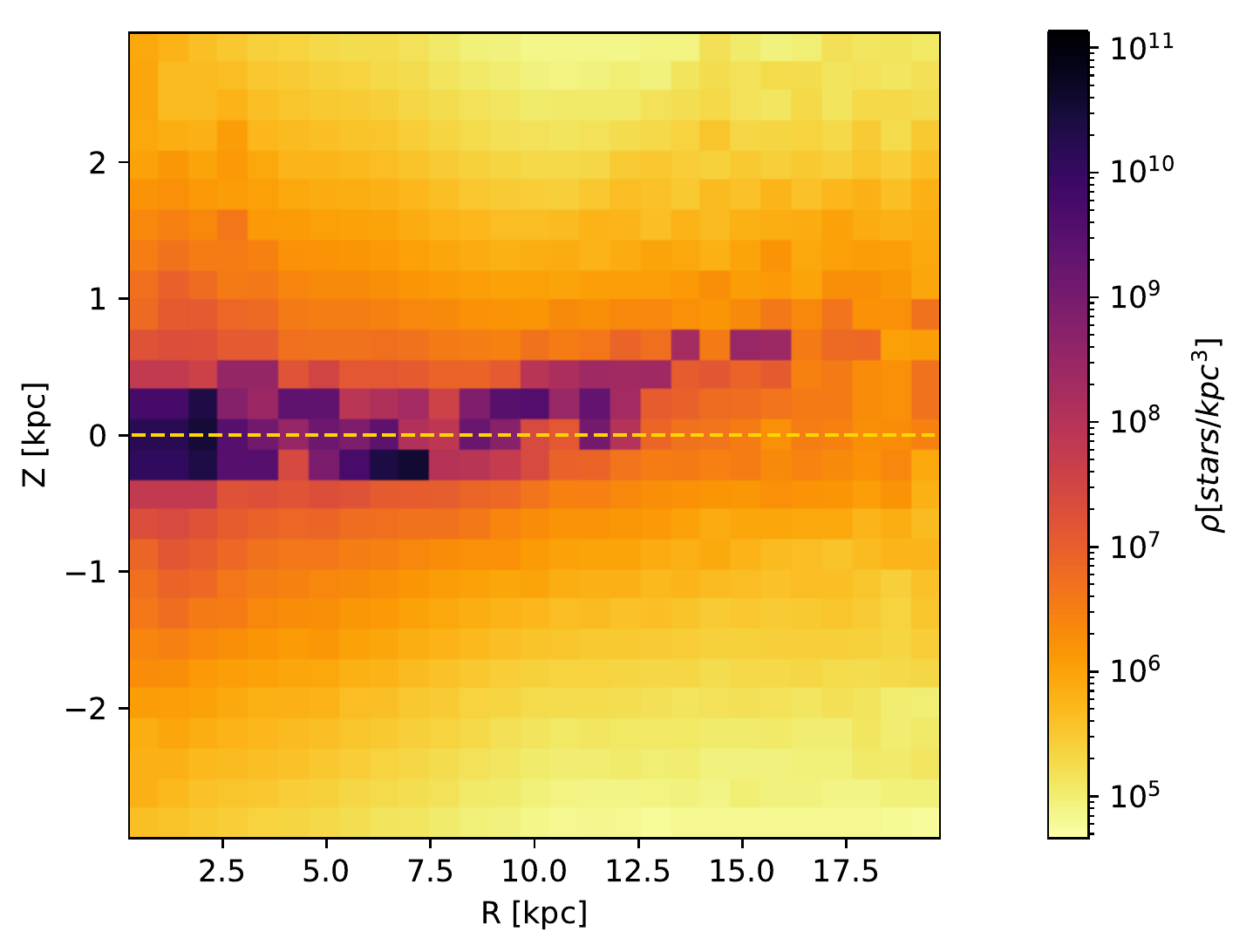}
        }
        \hspace{0mm}
        \subfloat[$\ang{60}<\phi<\ang{90}$]{
                \includegraphics[width=0.5\textwidth]{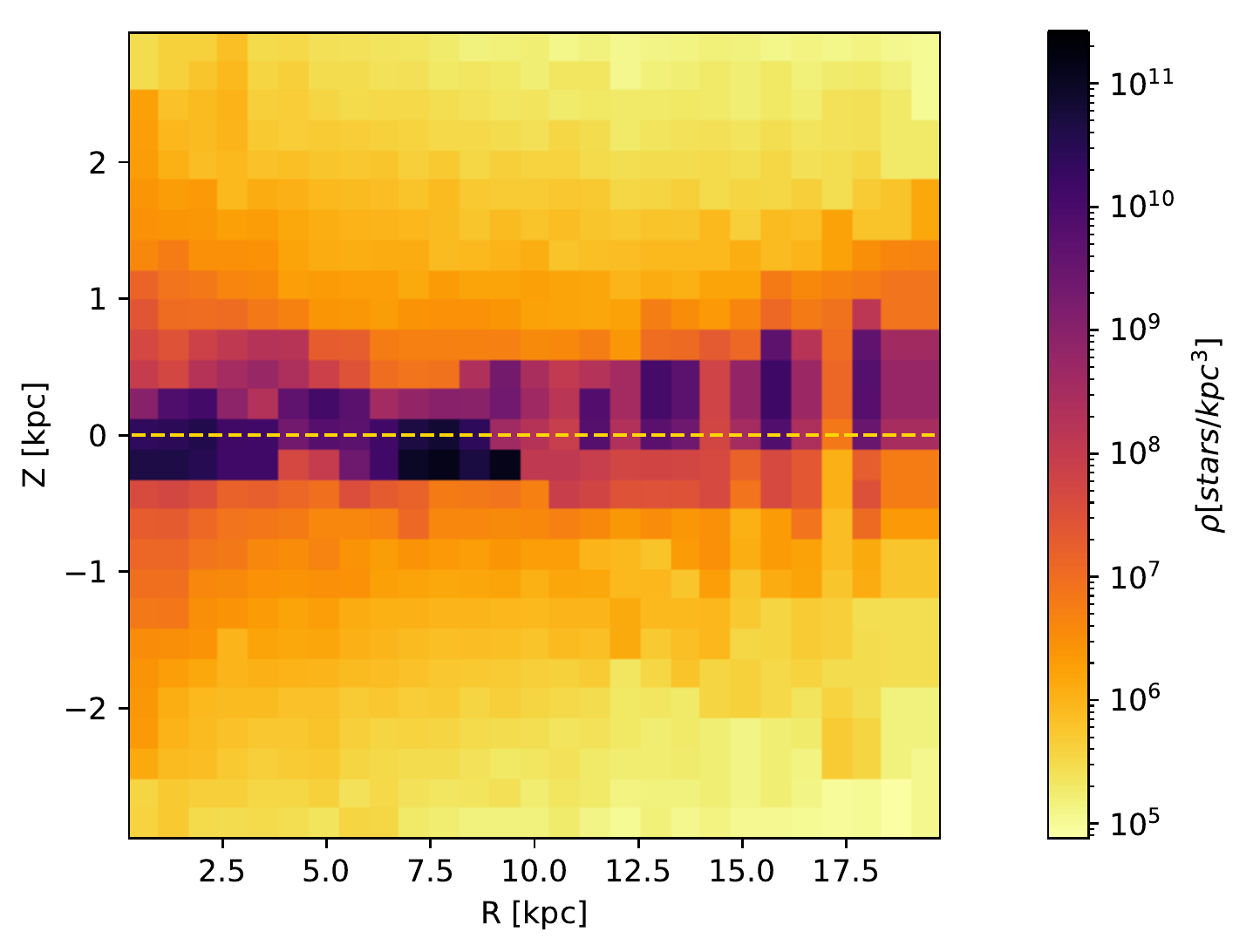}
        }
        \subfloat[$\ang{270}<\phi<\ang{300}$]{
                \includegraphics[width=0.5\textwidth]{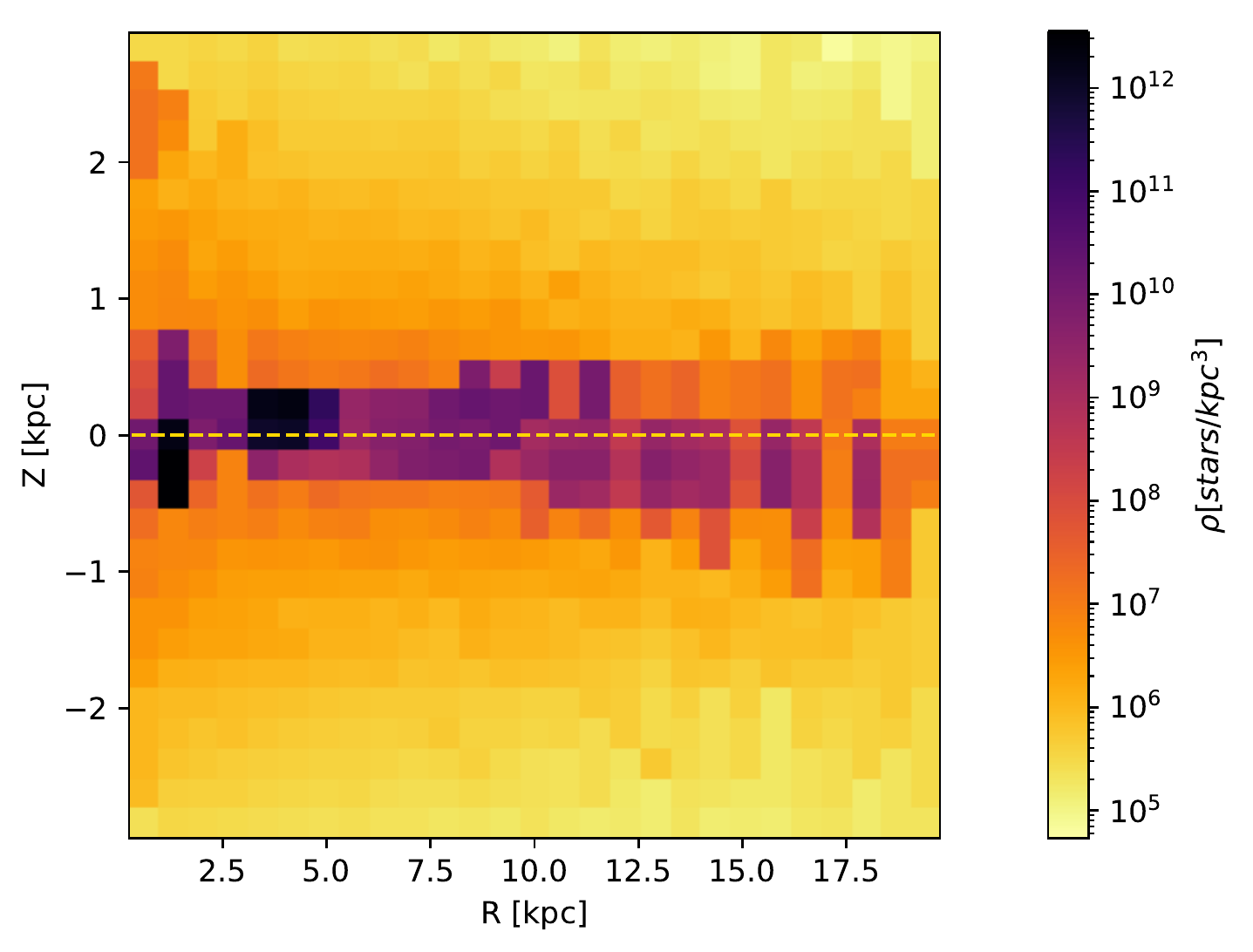}
        }
        \hspace{0mm}
        \subfloat[$\ang{300}<\phi<\ang{330}$]{
                \includegraphics[width=0.5\textwidth]{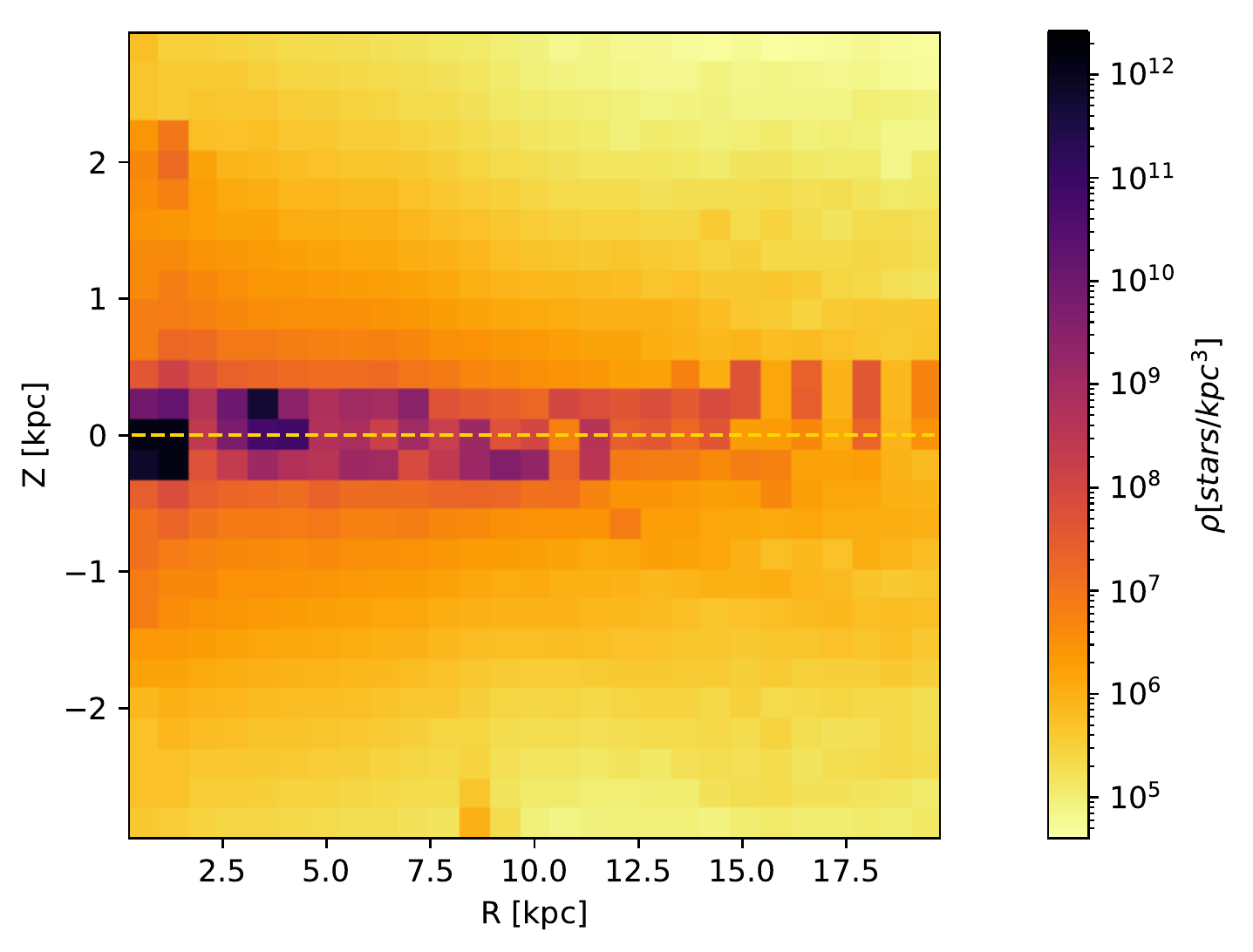}
        }
        \subfloat[$\ang{330}<\phi<\ang{360}$]{
                \includegraphics[width=0.5\textwidth]{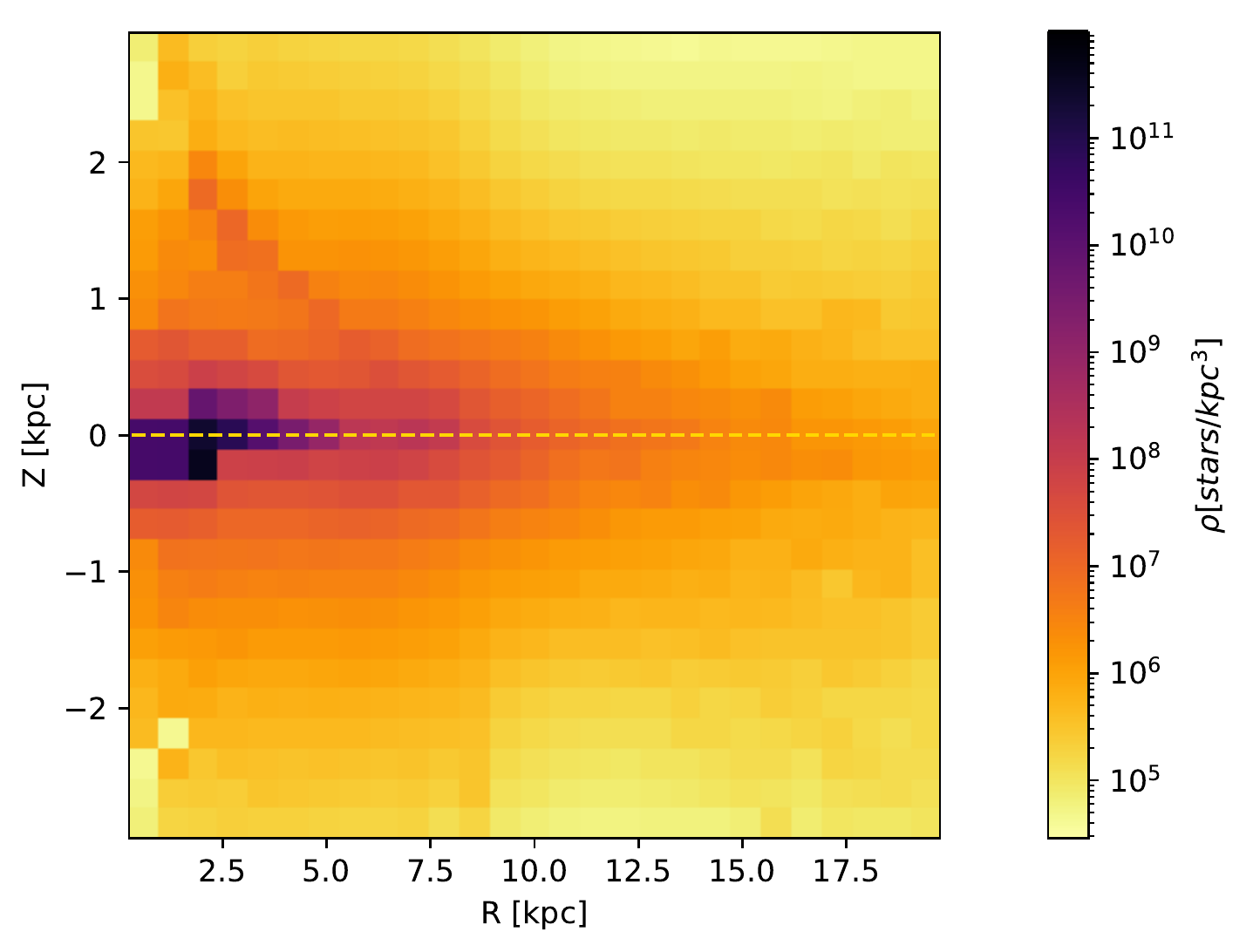}
        }
        \caption{Density maps for various azimuths between $\ang{0}$ an $\ang{360}$.}\label{o7}
\end{figure*}

\begin{figure*}
        \centering
        \subfloat[]{
                \includegraphics[width=0.5\textwidth]{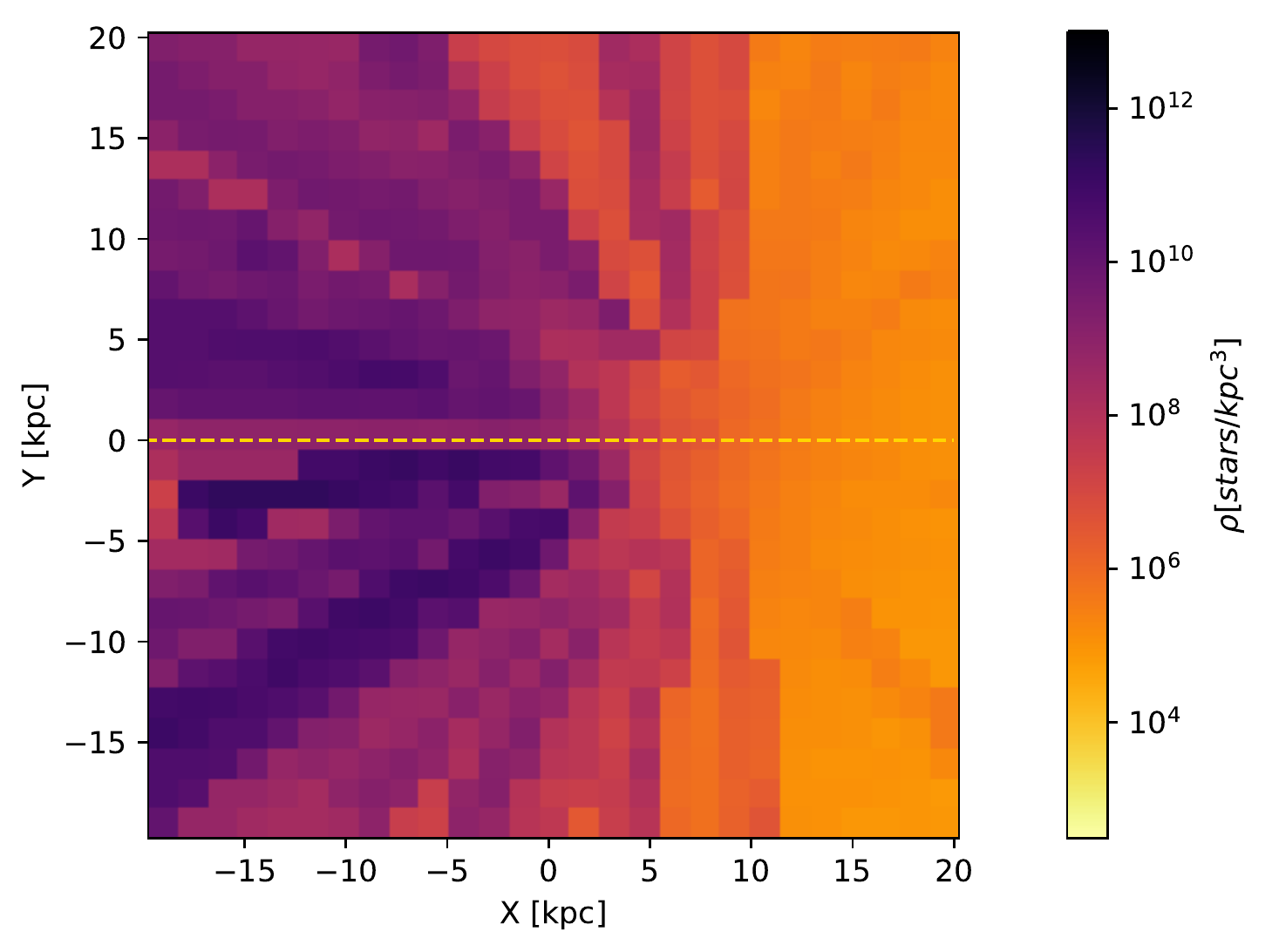}
        }
        \subfloat[]{
                \includegraphics[width=0.5\textwidth]{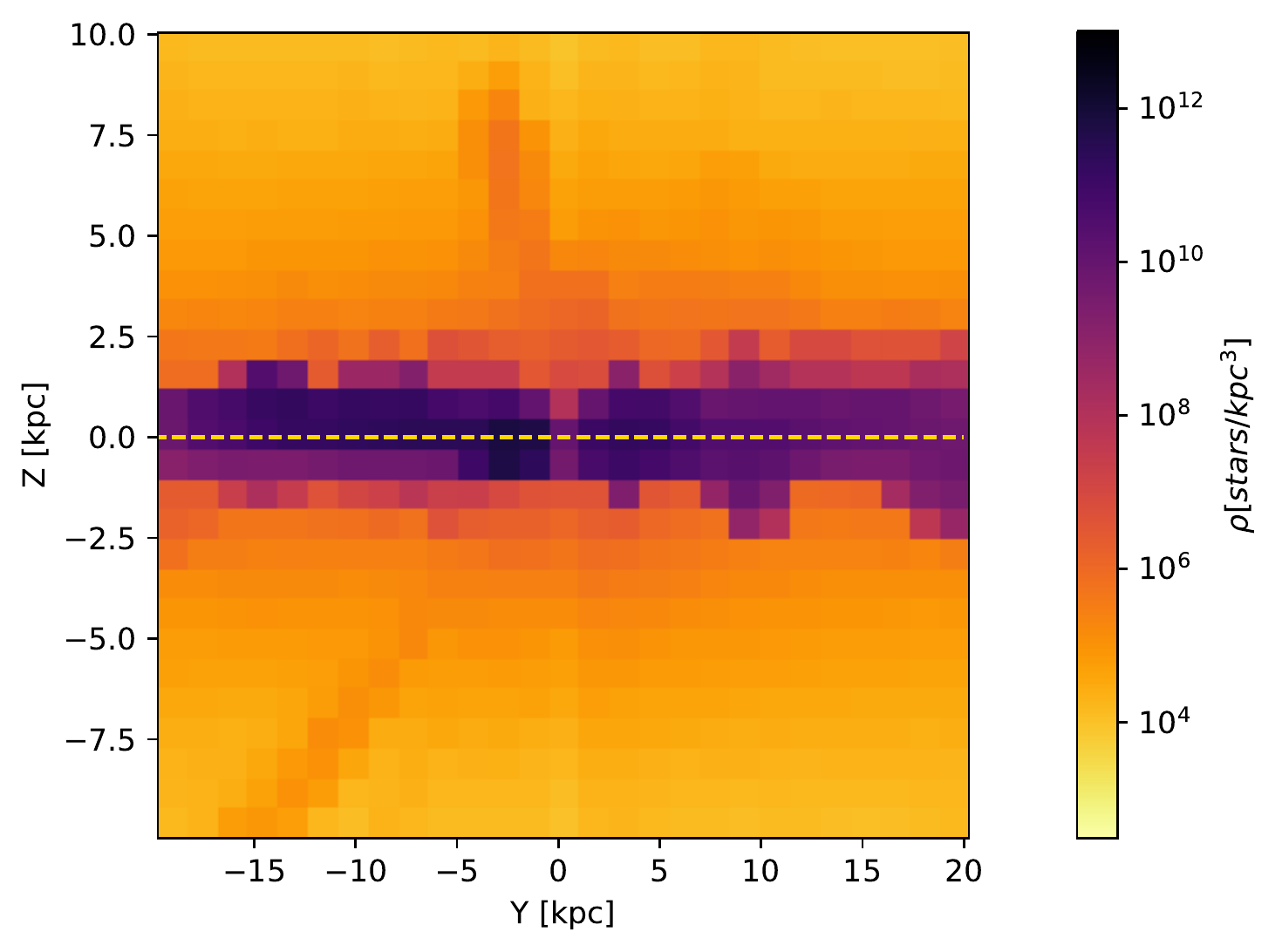}
        }
        \hspace{0mm}
        \subfloat[]{
                \includegraphics[width=0.5\textwidth]{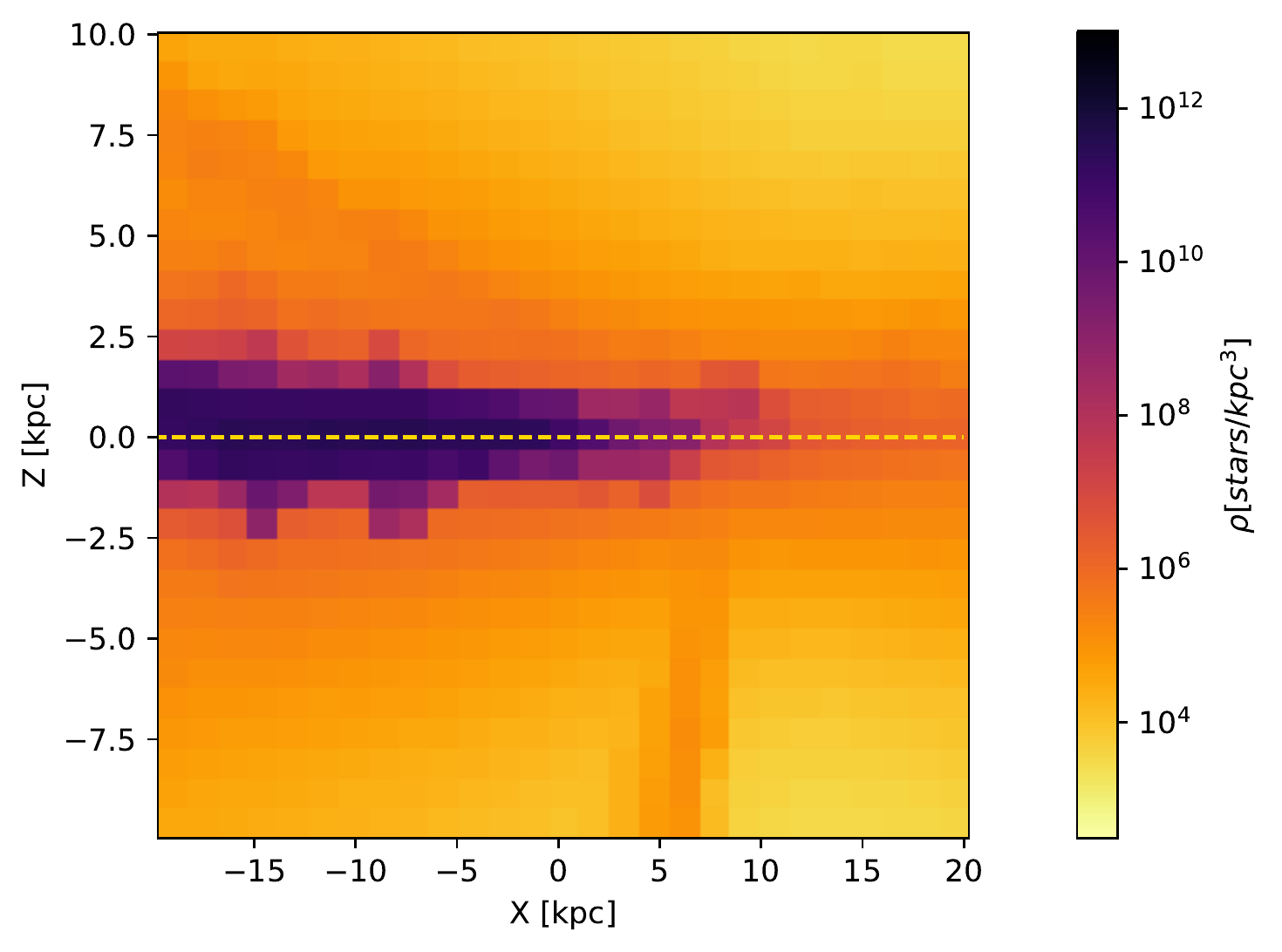}
        }
        \subfloat[]{
                \includegraphics[width=0.5\textwidth]{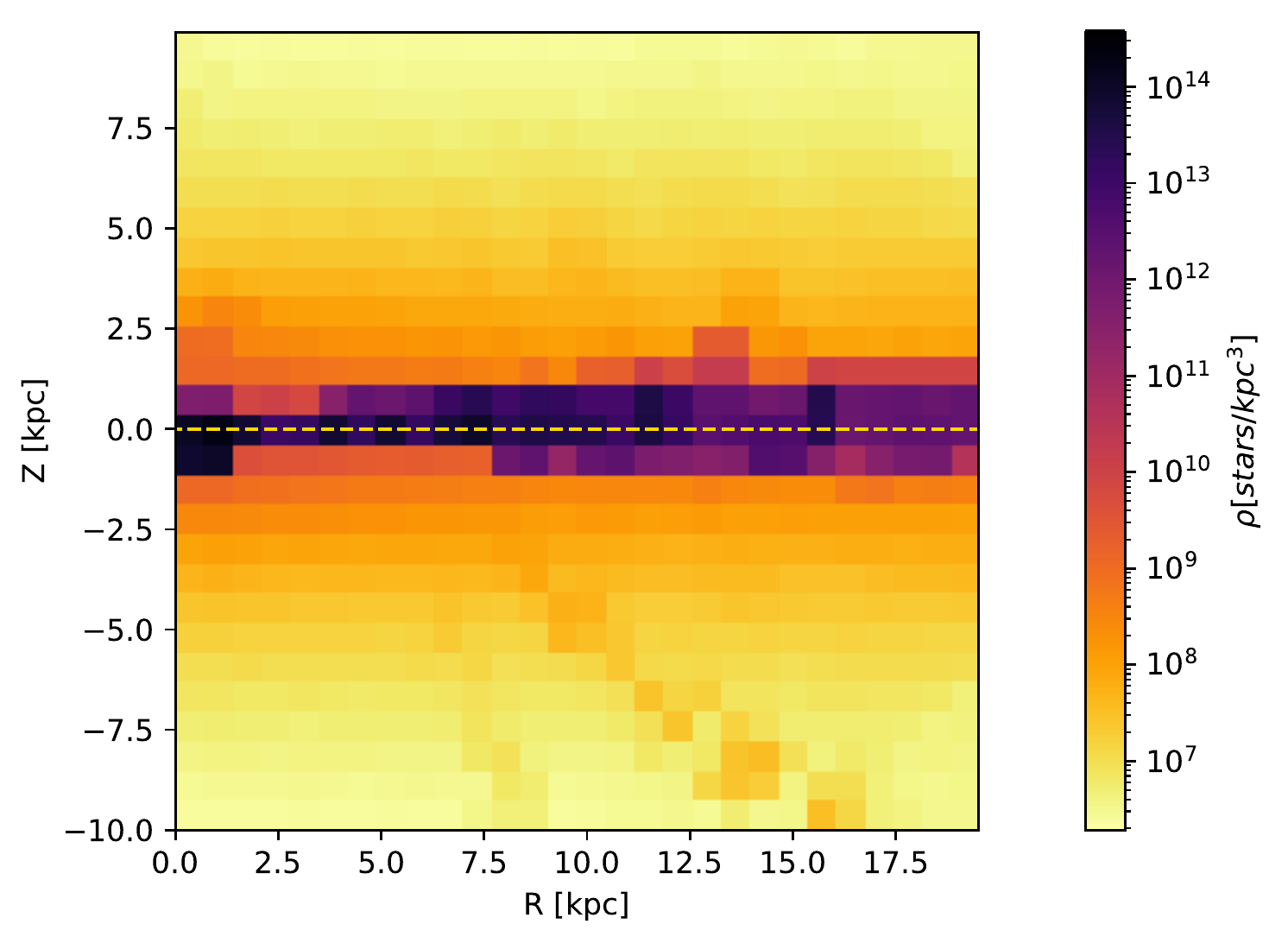}
        }
        \caption{Density maps.}\label{o10}
\end{figure*}

After this, we calculated the weighted mean density for all seven ranges of magnitude in each line of sight. Then we transformed this into cylindrical coordinates and made bins of Galactocentric radius $R$ of length 0.5 kpc, in Galactic height $z$ of 0.1 kpc and in azimuth of $\ang{30}$. We define the azimuthal angle $\phi$ to be measured from the centre-Sun-anticentre direction towards the Galactic rotation, going from \ang{0} to \ang{360}. We interpolated the missing bins with \textit{NearestNDInterpolator} from the python \textit{SciPy} package, which uses nearest-neighbour interpolation in N dimensions.
We plot the resulting density maps in Fig. \ref{o7}-\ref{o10}. In Fig. \ref{o7} we plot the density in cylindrical coordinates as a function of Galactic radius $R$ for different azimuths. We do not plot the results for azimuths $\ang{90}<\phi<\ang{270}$ because in this area the extinction is significant and we observe stars farther than the Galactic centre for which the errors are too large, therefore we cannot see any structure in density. However, we can see even by eye that a northern warp is present in the azimuths $\ang{60}<\phi<\ang{90}$ and a southern warp in the azimuths $\ang{270}<\phi<\ang{300}$. Another structure that can be seen from the plots is the flaring of the disc. We analyse these structures below.
In Fig. \ref{o10}. (a)-(c) we plot the density map in Cartesian coordinates, and in Fig \ref{o10}. (d) we plot the density in cylindrical coordinates, integrated through all ranges of azimuths, except for the areas that were excluded from the analysis. The Cartesian coordinates are defined such that $X_\odot=8.4$ kpc. In these plots we note a flat disc with some fluctuations in density, but no apparent features. However, some slights overdensities both above and below the Galactic plane are visible. The features above the plane are present only in Fig. \ref{o10}. (b)-(c), but not in Fig. \ref{o10}. (d), which suggests that it might be a contamination. The feature below the Galactic plane is present in all the three plots. As the direction of these overdensities is towards the Magellanic Clouds, it might be an effect of the Milky Way pulling stars out of Magellanic Clouds, as suggested by \cite{anders}. Another possible explanation for these overdensities is the finger of God artefact, which is caused by the foreground dust clouds and causes elongated overdensities that point to the Sun. This artefact has previously been seen in Gaia data, as shown in the Gaia DR2 documentation\footnote{\url{https://gea.esac.esa.int/archive/documentation/GDR2/Data_analysis/chap_cu8par/sec_cu8par_validation/ssec_cu8par_validation_additional-validation.html}}.  \\

\subsection{Zero-point correction in parallaxes}\label{ch9}
\begin{figure*}
        \vspace{0.2cm}
        \centering
        \subfloat[]{
                \includegraphics[width=0.5\textwidth]{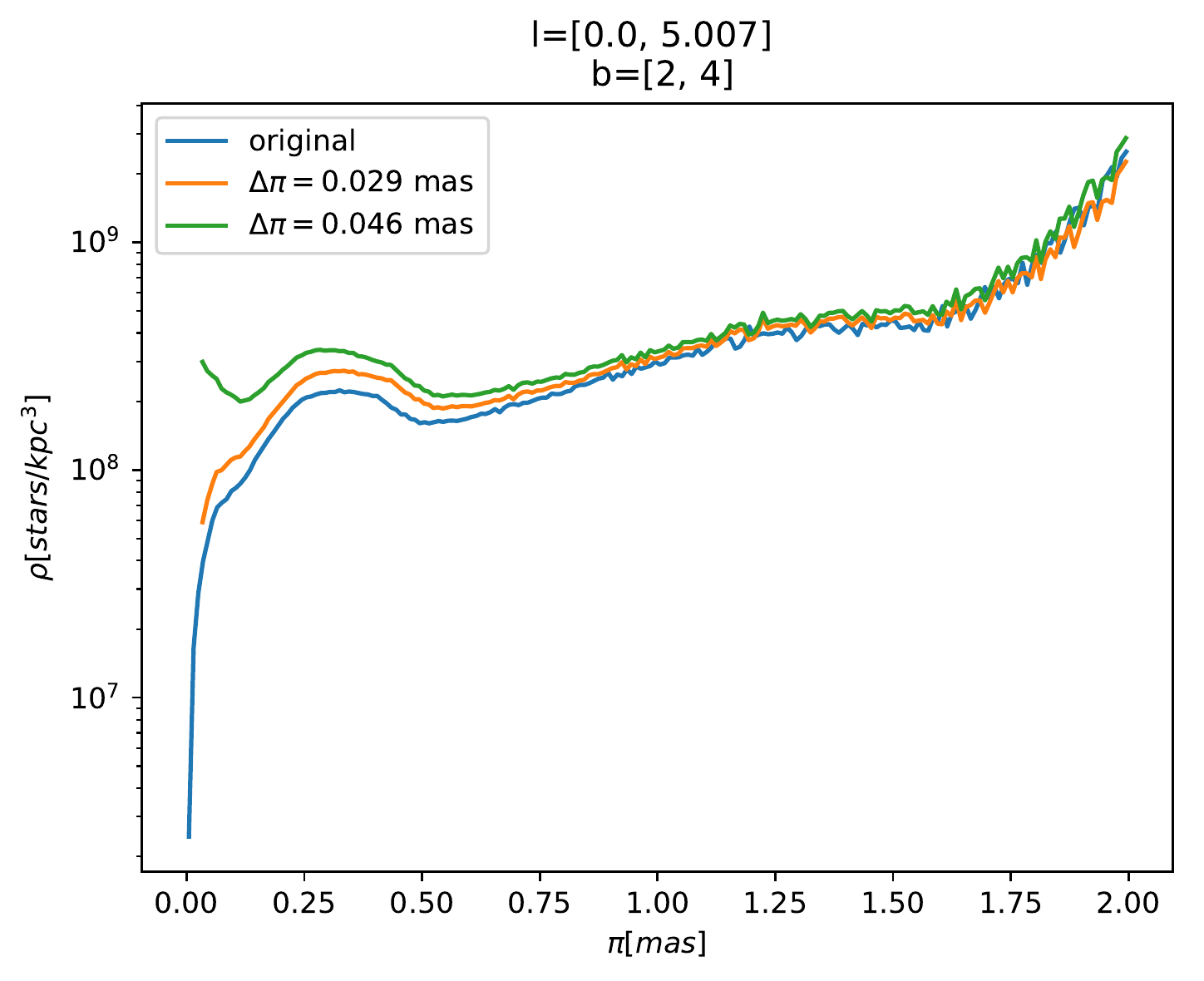}
        }
        \subfloat[]{
                \includegraphics[width=0.5\textwidth]{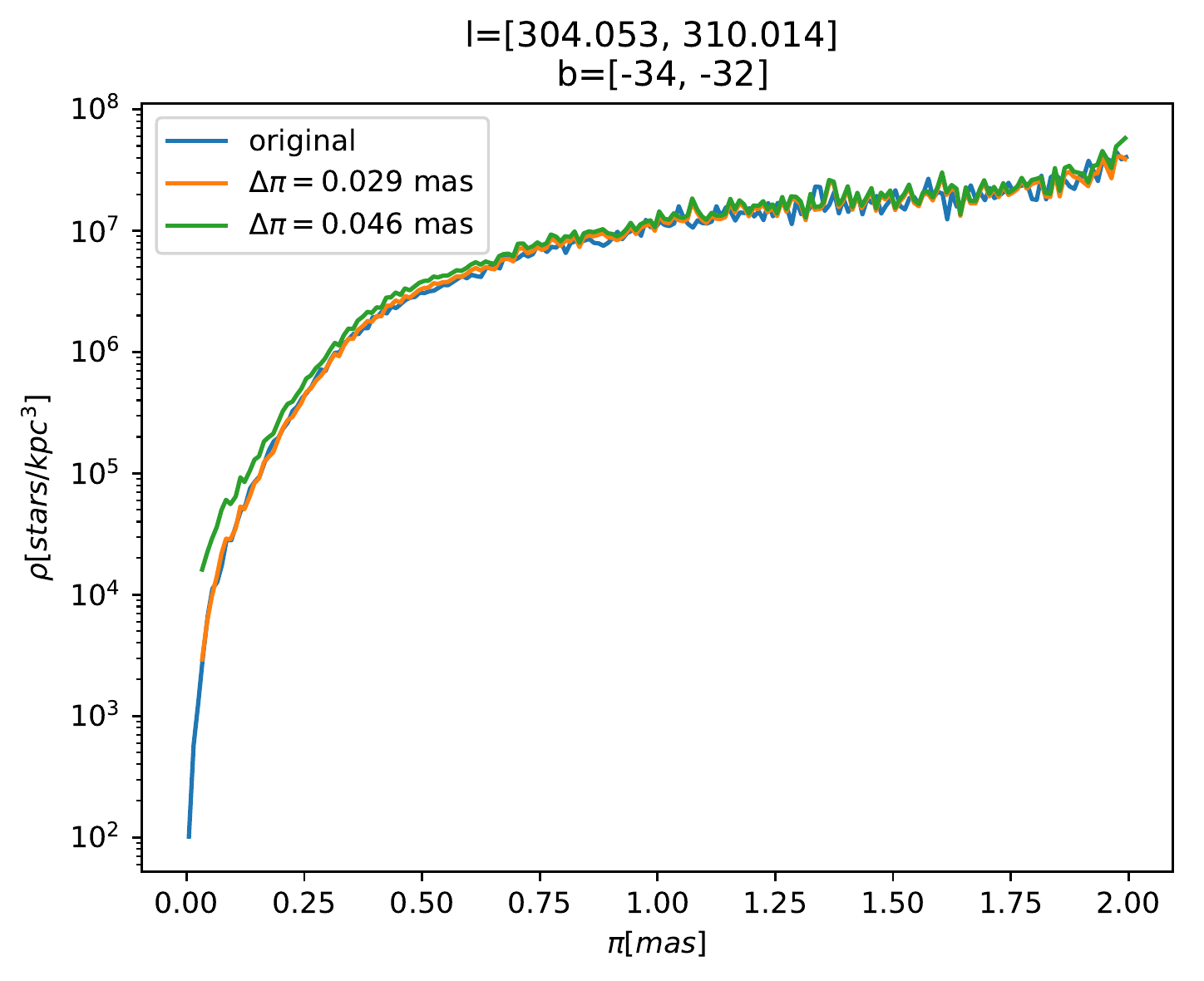}
        }
        \hspace{0mm}
        \subfloat[]{
                \includegraphics[width=0.5\textwidth]{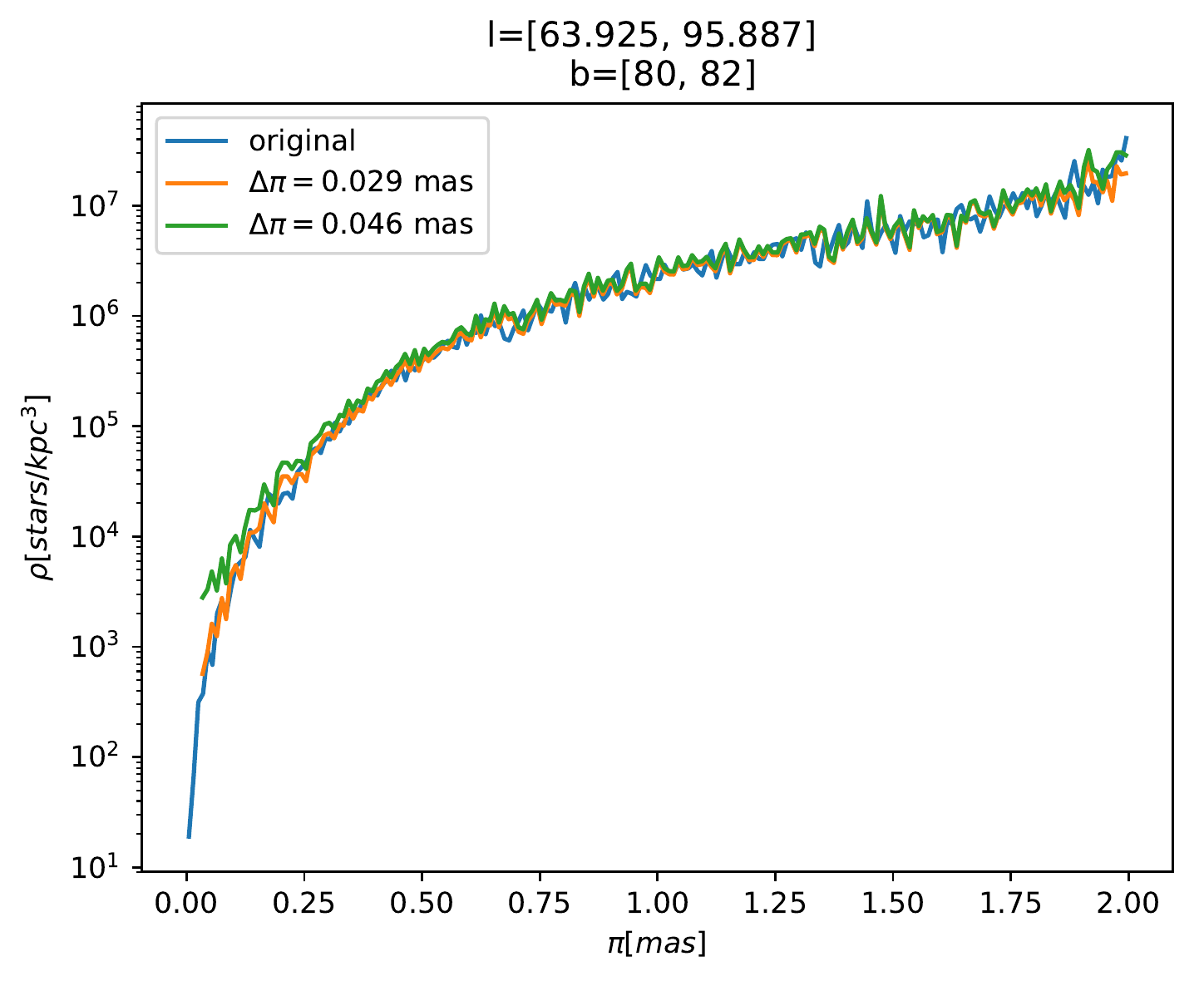}
        }
        \subfloat[]{
                \includegraphics[width=0.5\textwidth]{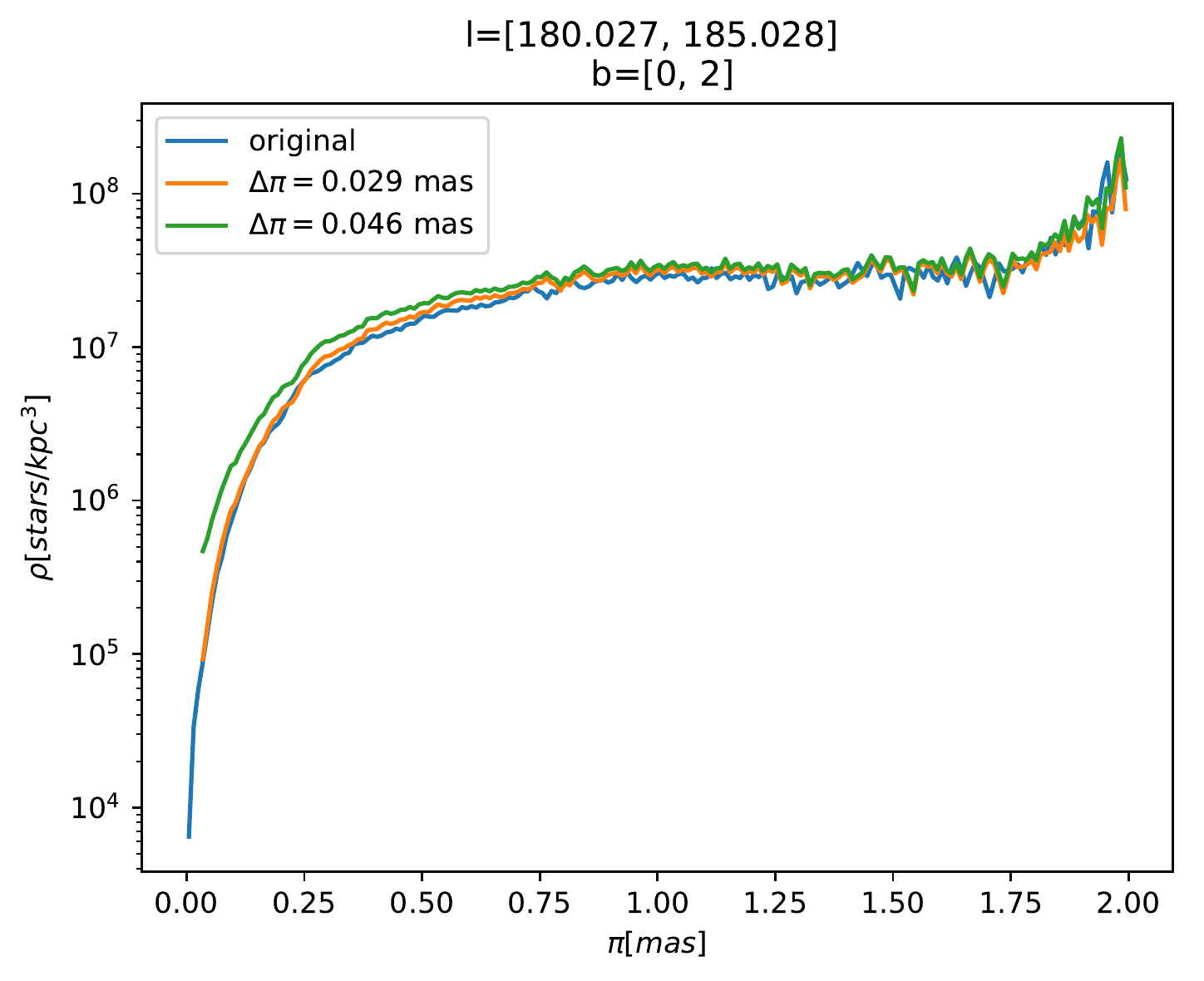}
        }
        \hspace{0mm}
        \subfloat[]{
                \includegraphics[width=0.5\textwidth]{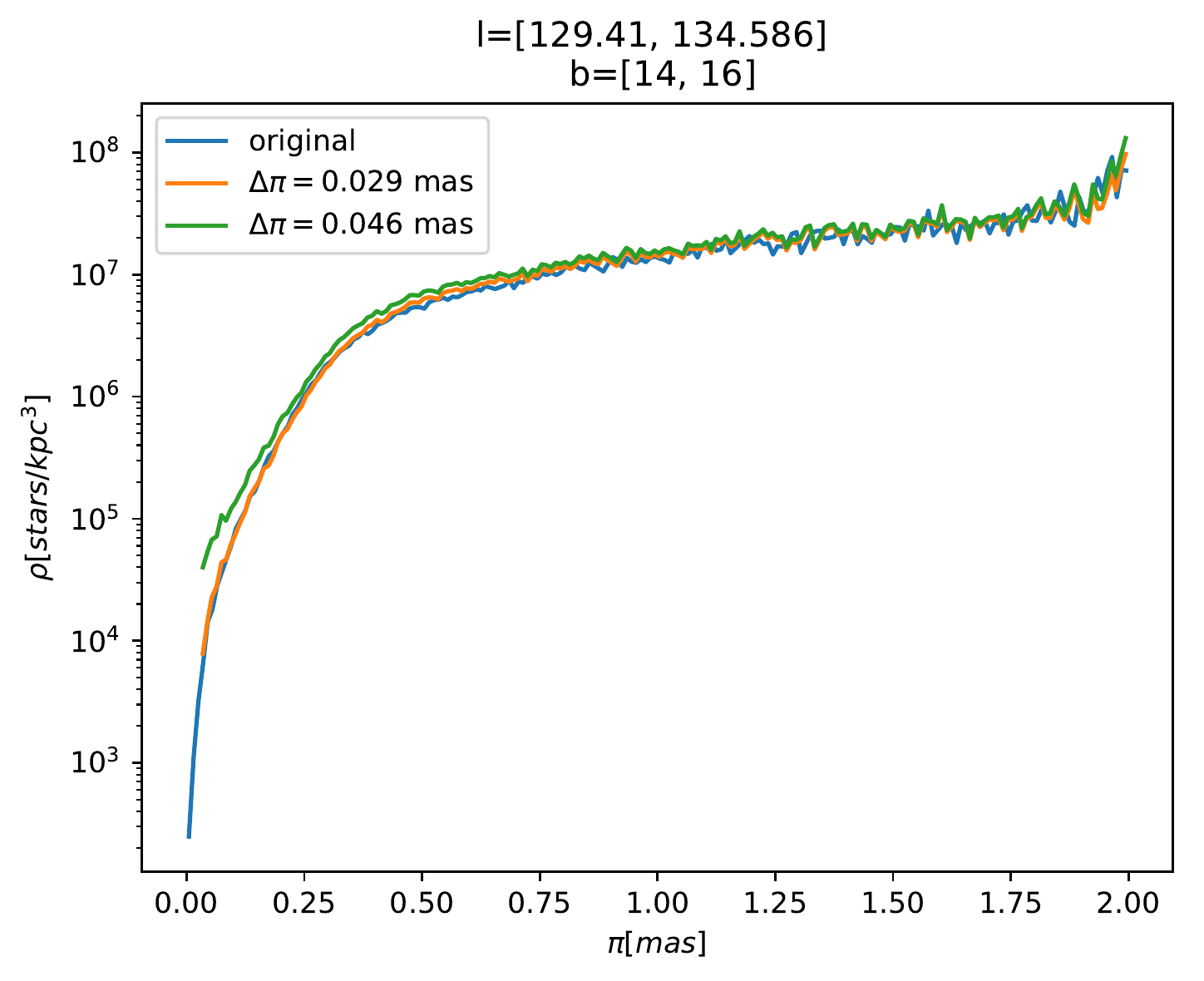}
        }
        \subfloat[]{
                \includegraphics[width=0.5\textwidth]{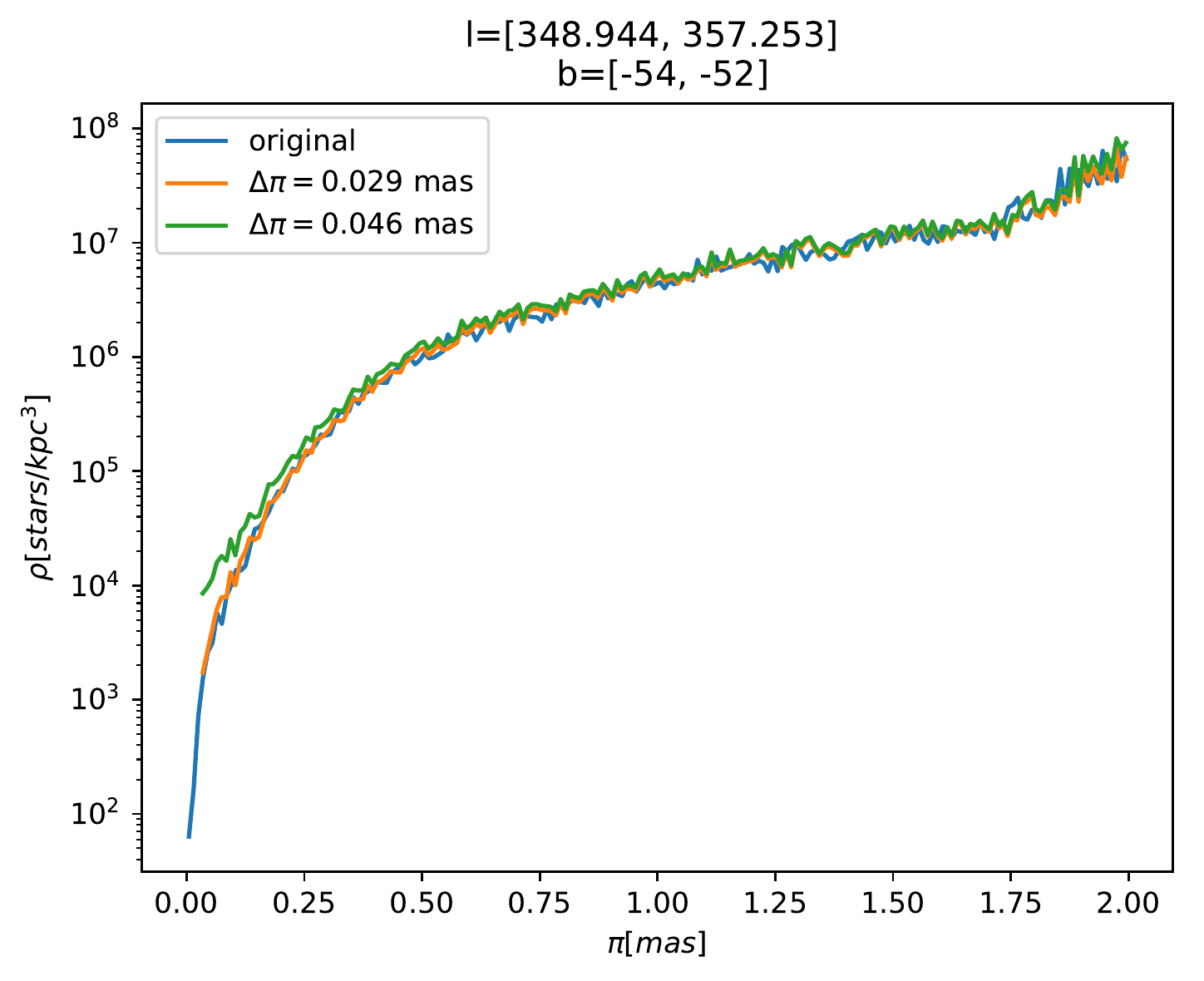}
        }
        \caption{Comparison of densities for various lines of sight. Orange and green curves represent the density including the zero-point correction of the parallax, and the blue curve shows the density without this correction.}\label{o9}
\end{figure*}

So far, we did not consider any zero-point bias in parallaxes. \cite{lindegren} found a global mean offset of $-0.029$ mas, meaning that Gaia DR2 parallaxes are lower than the true value. We repeated our calculations with this correction and present the results in Fig. \ref{o9}, where we chose some of the lines of sight to show the comparison. We find that these results are very similar to our original results, and this correction brings a negligible effect. We also tried a value of $-0.046$ mas, found by \cite{zerop2}. In Fig. \ref{o9} we show that the difference between the different zero-point values is very small, therefore we only use the value of -0.029 mas in the further calculations. \\
For the analysis of the warp in Sections 5.5 and 5.6., we repeated the analysis of Section 4 with the value of parallax corrected for the zero-point. We find that this brings a small correction to the warp parameters, which we state as the systematic error in the results.

\subsection{Error of the extinction}
To test how accurate the extinction map is, we analysed the map of \cite{green} using the function {\tt query}, which returns the standard deviation $\sigma_G$ for a given line of sight. We calculate a new extinction as

\begin{eqnarray}\label{ext_corr}
A_G^*(r)=A_G(r)+f*\sigma_G(r)~,
\end{eqnarray}
where $A_G$ is the extinction given by the map, $r$ is the distance, and $f$ is a factor chosen randomly from a Gaussian distribution with $\mu=0$ and $\sigma=1$. \\

\begin{figure*}
        \vspace{-0.5cm}
        \centering
        \subfloat[]{
                \includegraphics[width=0.5\textwidth]{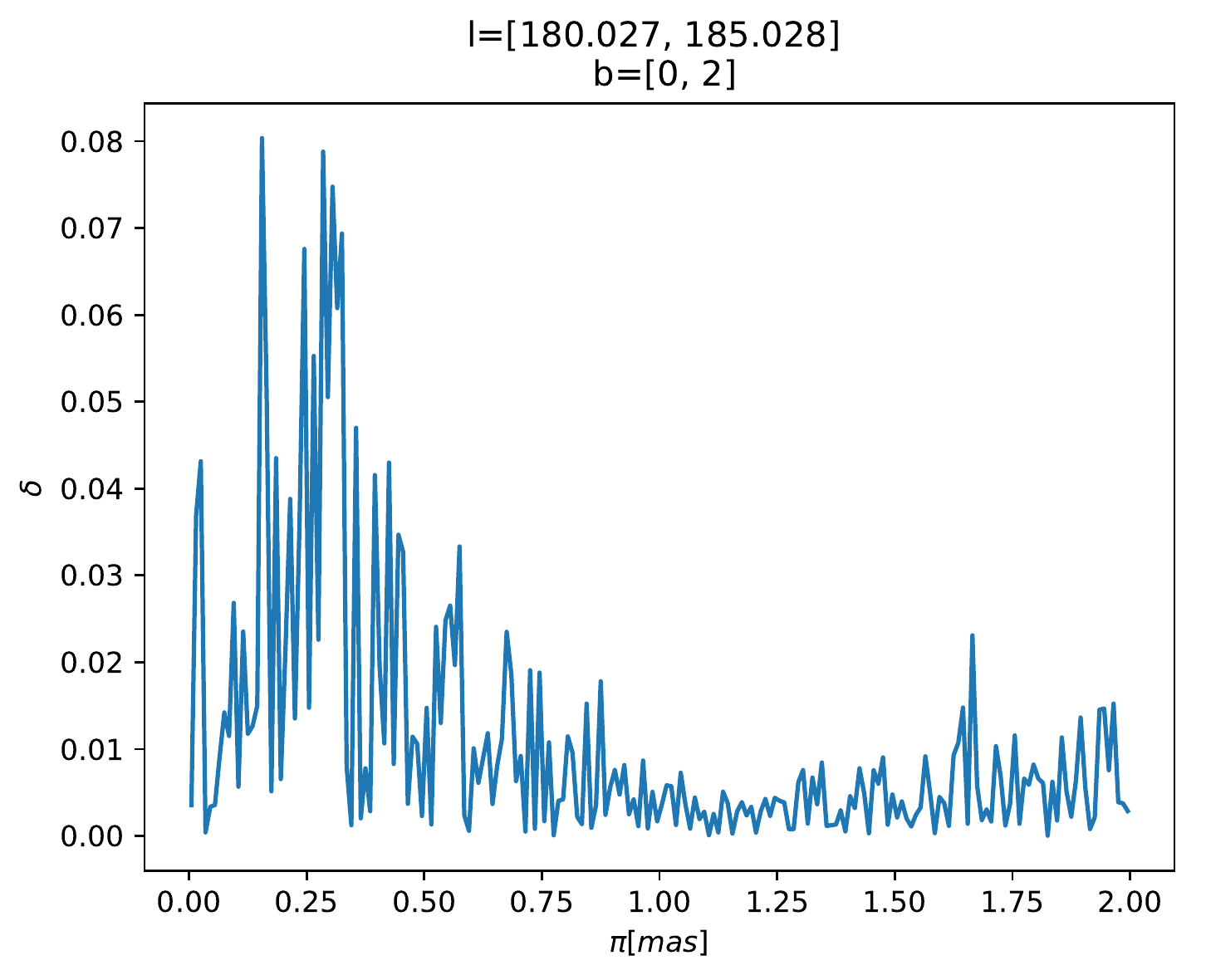}
        }
        \subfloat[]{
                \includegraphics[width=0.5\textwidth]{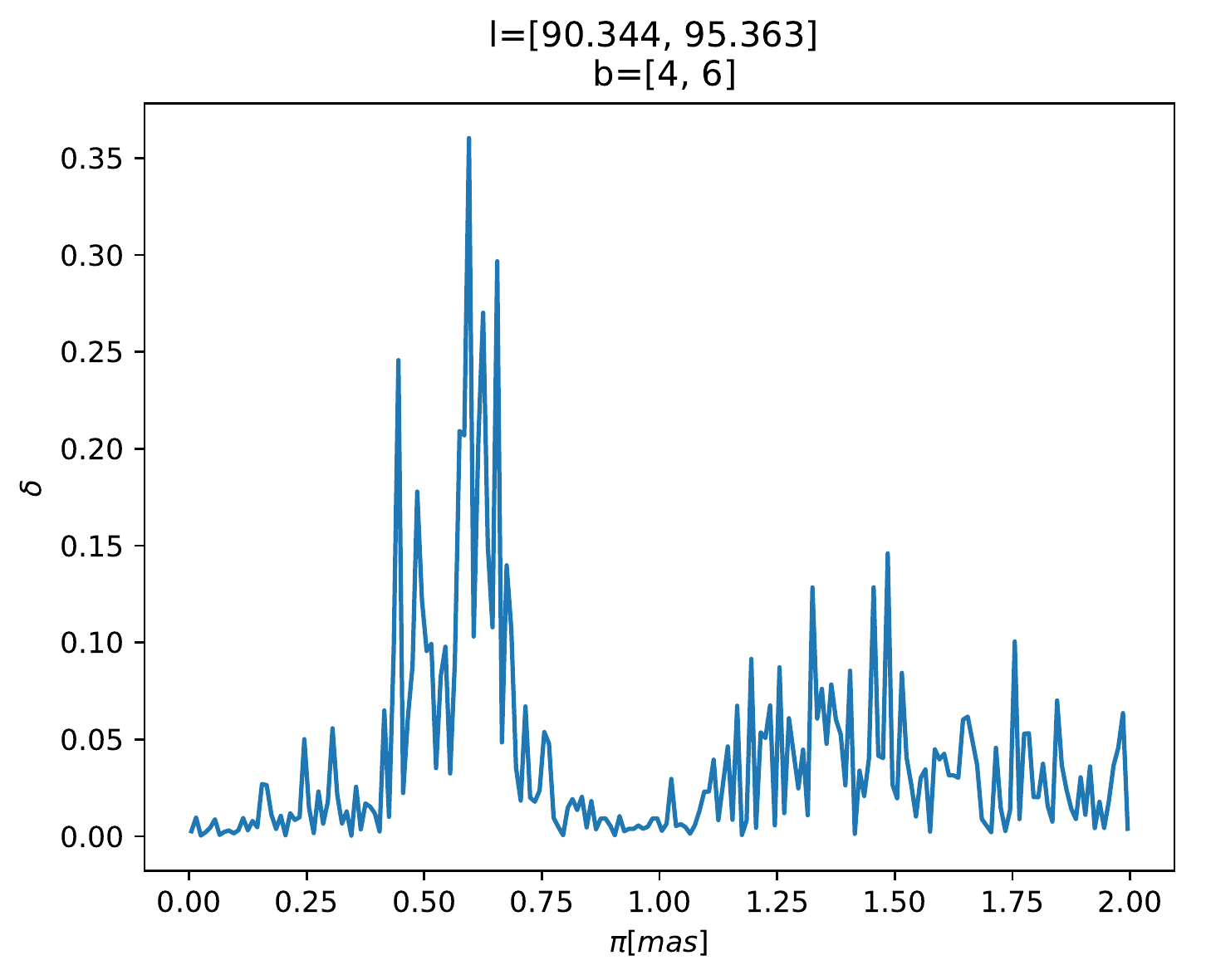}
        }
        \hspace{0mm}
        \subfloat[]{
                \includegraphics[width=0.5\textwidth]{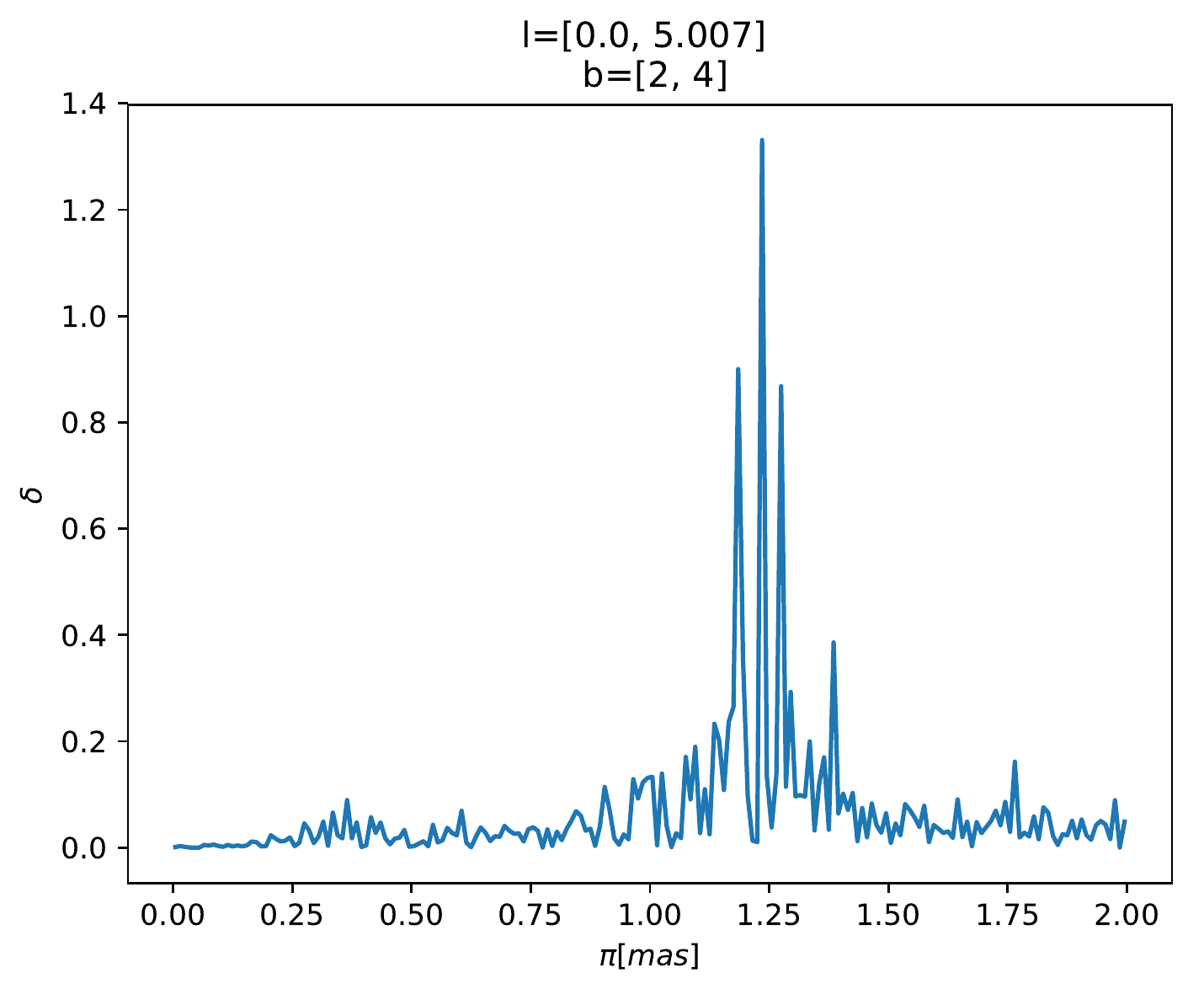}
        }
        \subfloat[]{
                \includegraphics[width=0.5\textwidth]{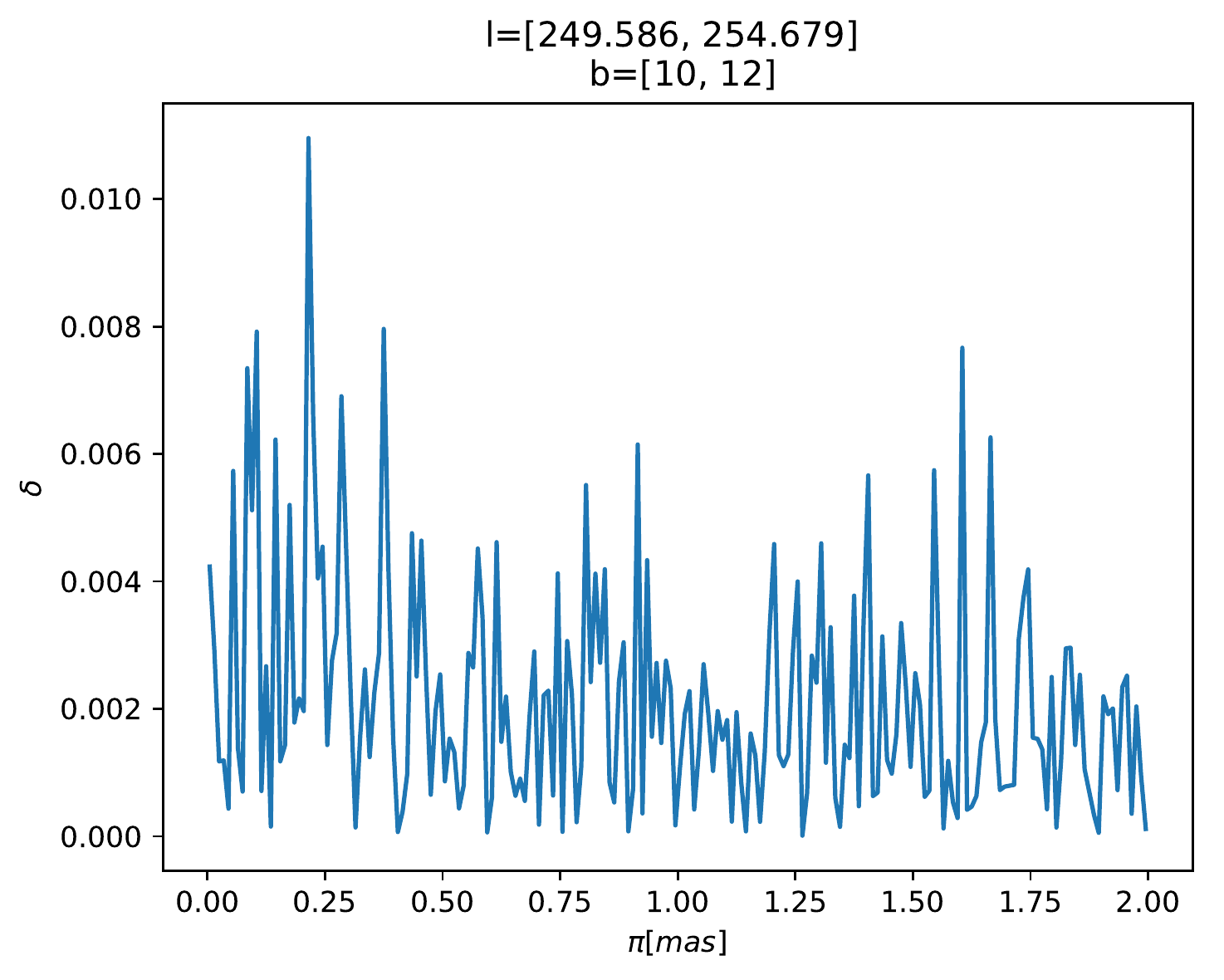}
        }
        \hspace{0mm}
        \subfloat[]{
                \includegraphics[width=0.5\textwidth]{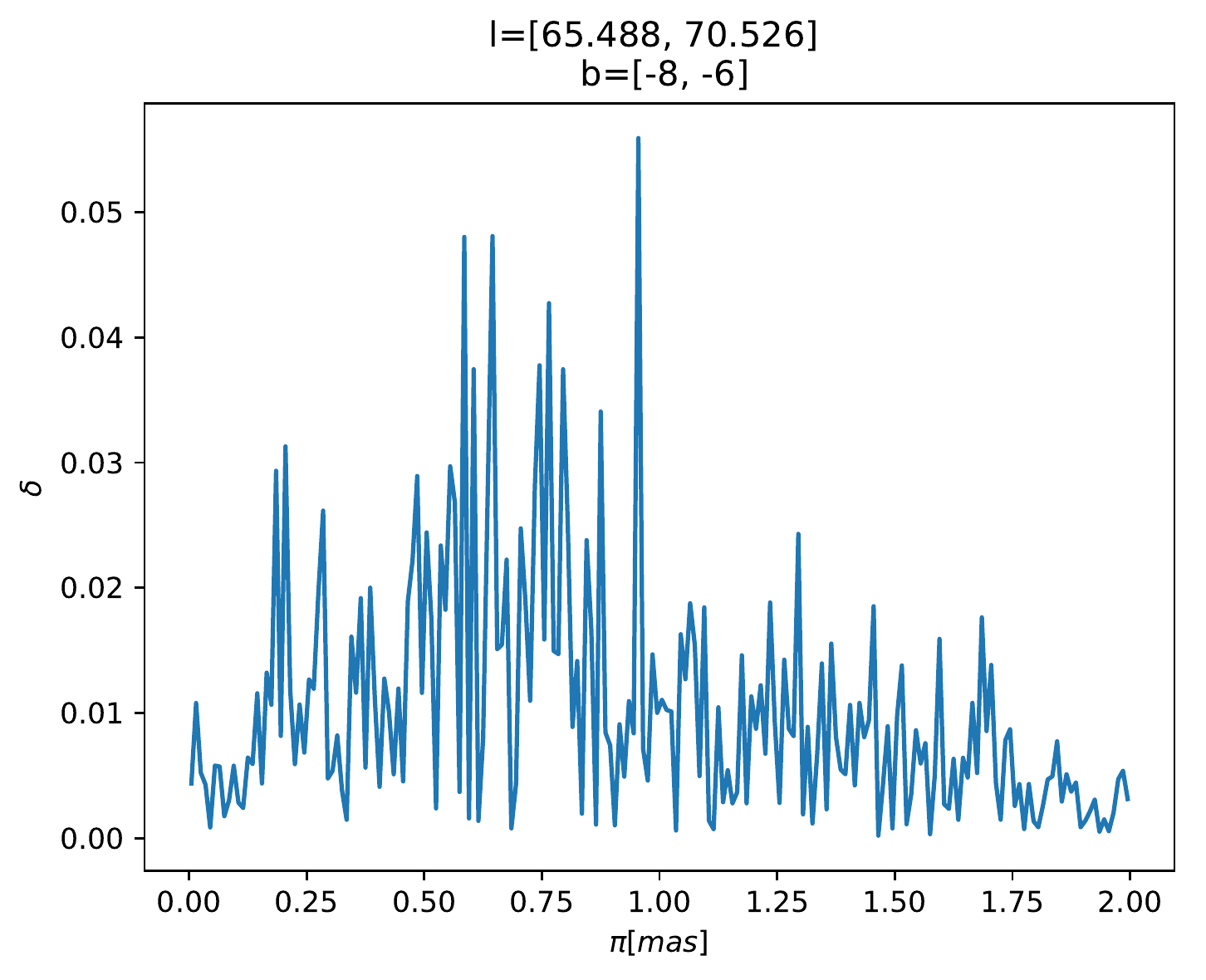}
        }
        \subfloat[]{
                \includegraphics[width=0.5\textwidth]{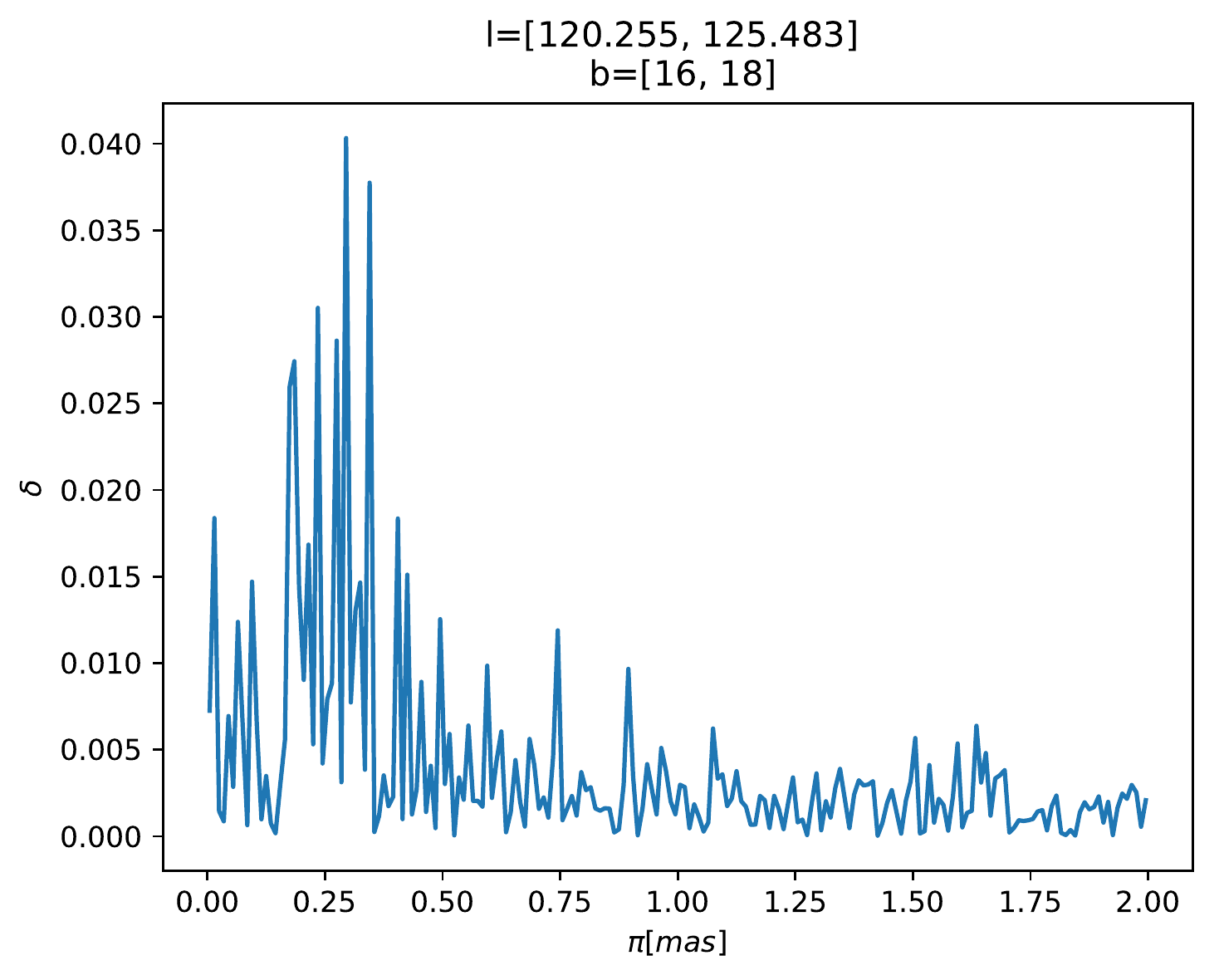}
        }
        \caption{Relative error $\delta=(\rho(A_G)-\rho(A_G^*))/\rho(A_G)$ of the density calculated including the standard deviation of the extinction.}\label{ext}
\end{figure*}

In Fig. \ref{ext} we show the relative error of the density $\delta=(\rho(A_G)-\rho(A_G^*))/\rho(A_G)$. For all lines of sight that we tested, the difference is negligible, except for the area in the centre of the Galaxy, which we know is problematic. However, in the outer disc, where we carried out our analysis, the extinction is determined quite accurately. We must of course take into account that we used the map of \cite{bovy_extinkcia}, which combines different maps and is less accurate and therefore can give different results than \cite{green} in some areas. Moreover, we estimated only the statistical error of the extinction, but we recall that we do not have information about the systematic error of the extinction map. However, for our purposes, the extinction map gives satisfying results in the area we analysed. The stellar warp has been studied using star counts by many other authors \citep[][and others]{martin_warp,reyle,amores}, therefore this method is most likely not especially flawed.

\subsection{Thick-disc areas}\label{ch10}
\begin{figure}
        \includegraphics[width=0.5\textwidth]{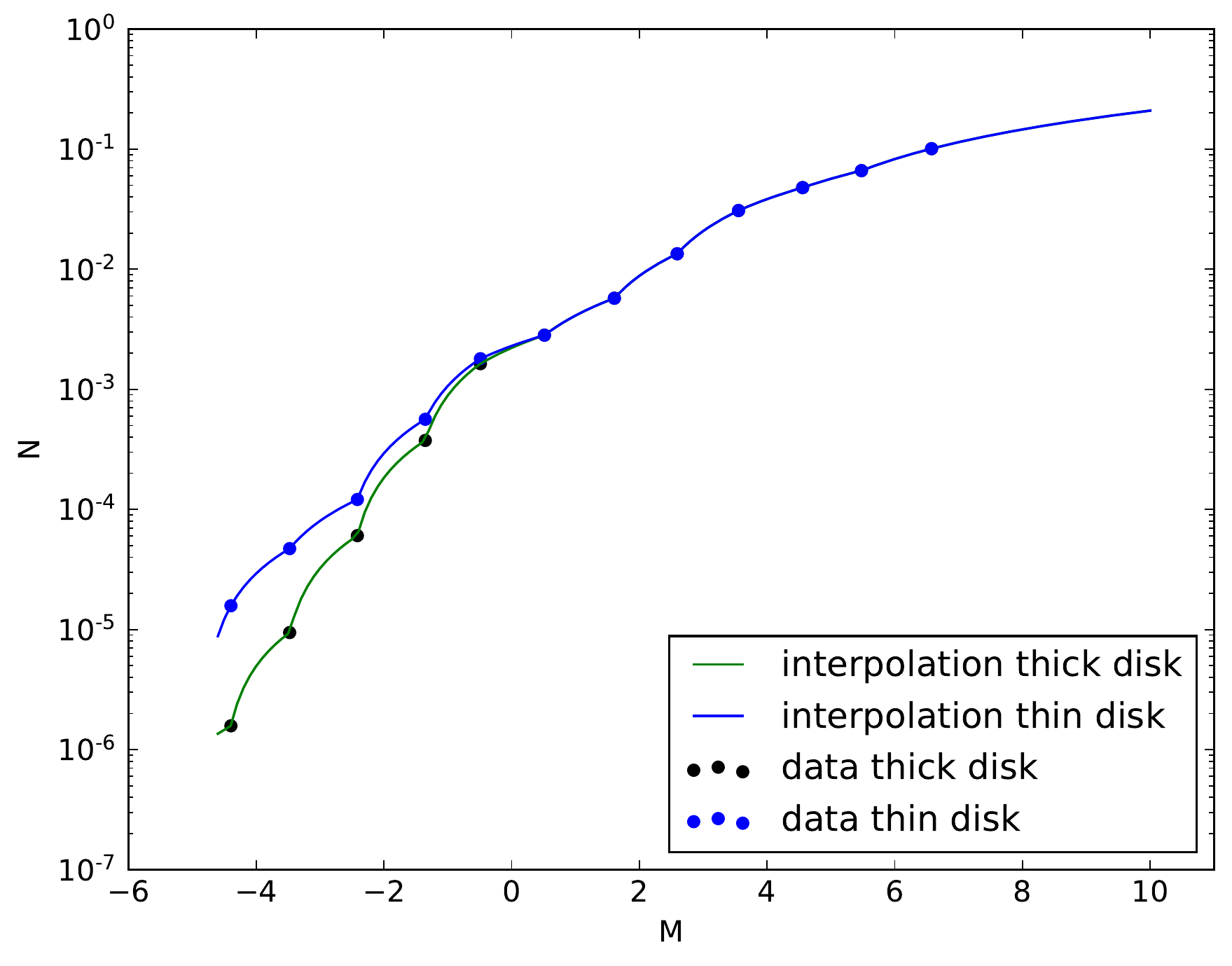}
        \caption{Comparison of luminosity functions for the thin and thick disc.}\label{o21}
\end{figure}

\begin{figure*}
        \vspace{0.3cm}
        \centering
        \subfloat[]{
                \includegraphics[width=0.5\textwidth]{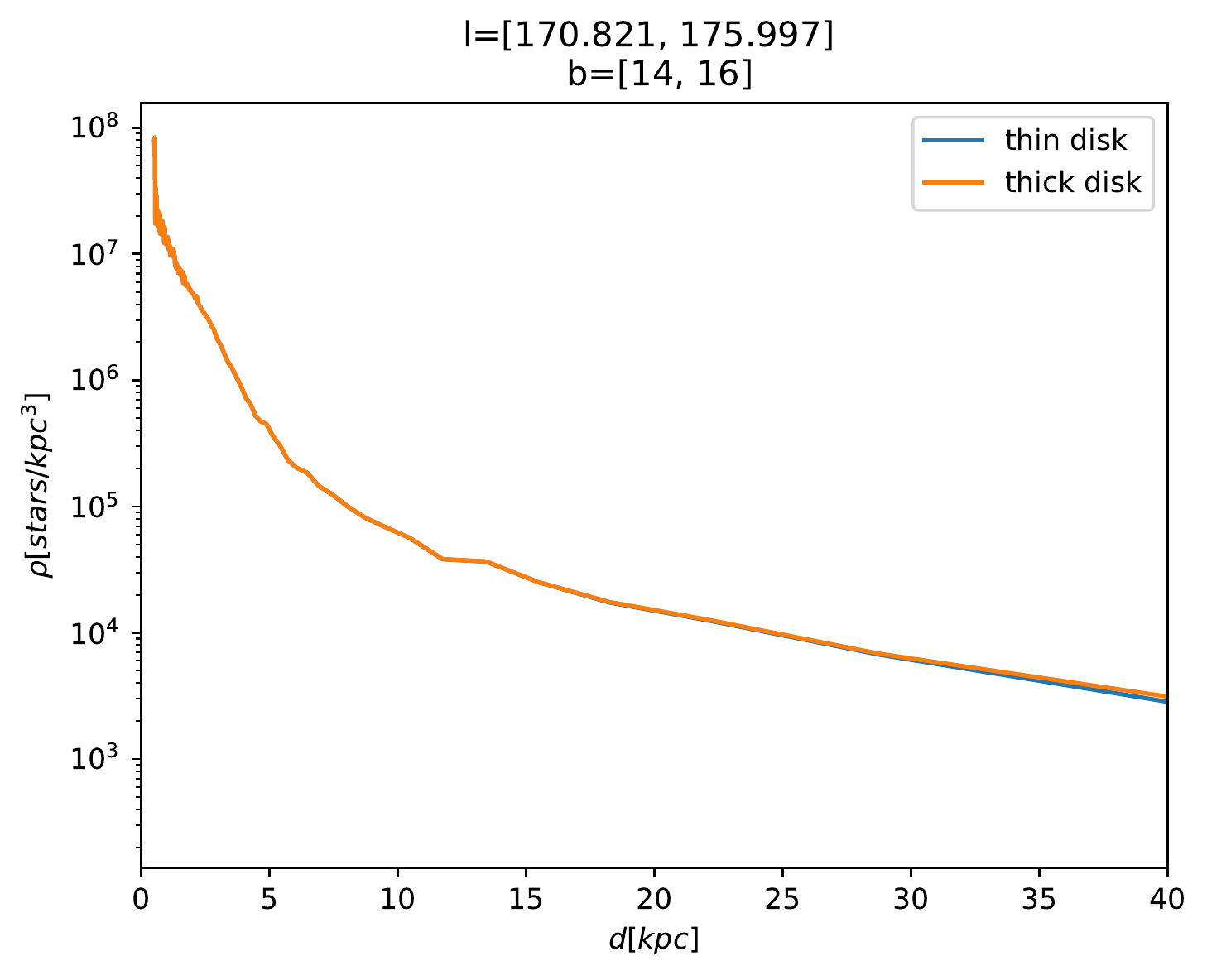}
        }
        \subfloat[]{
                \includegraphics[width=0.5\textwidth]{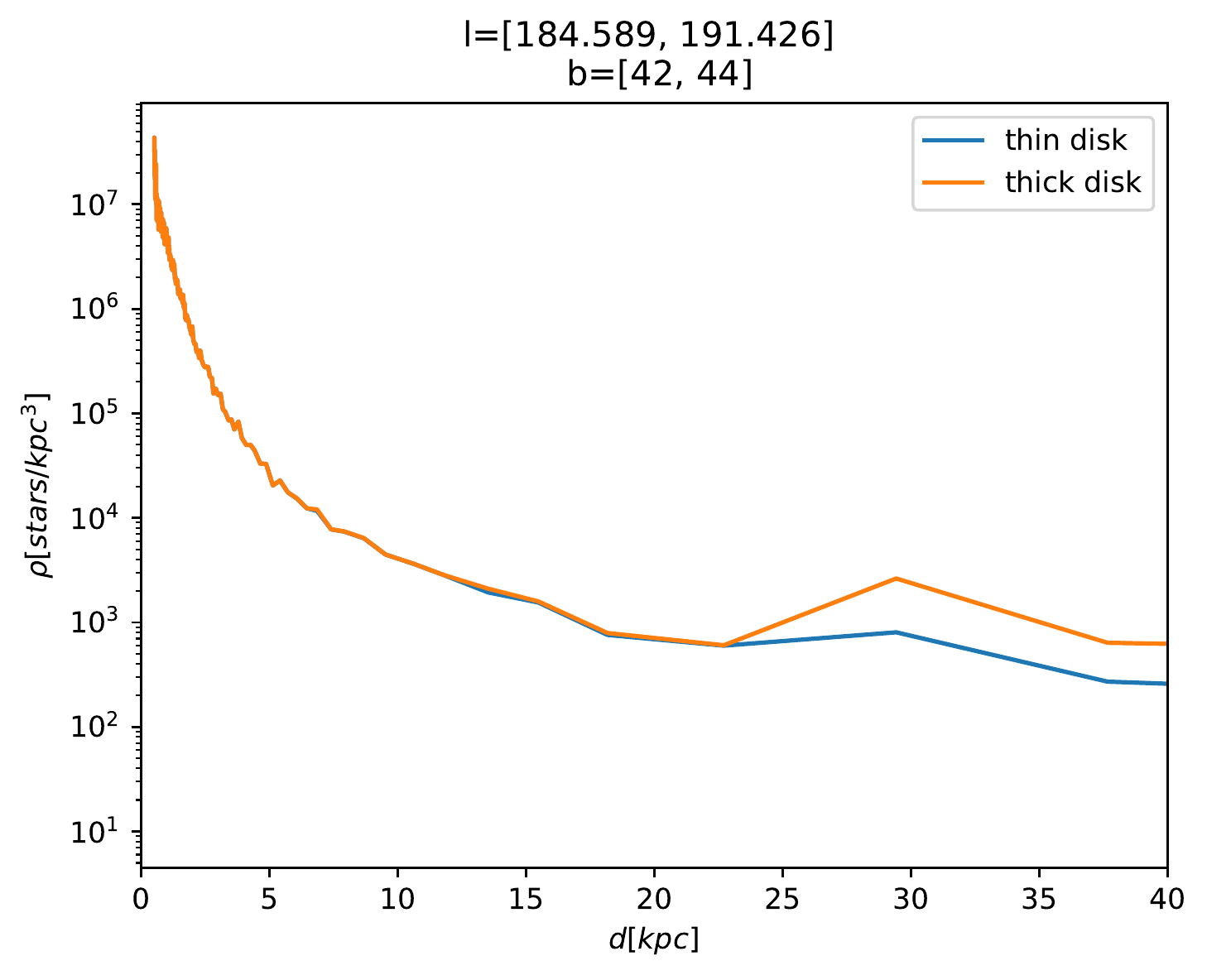}
        }
        \hspace{0mm}
        \subfloat[]{
                \includegraphics[width=0.5\textwidth]{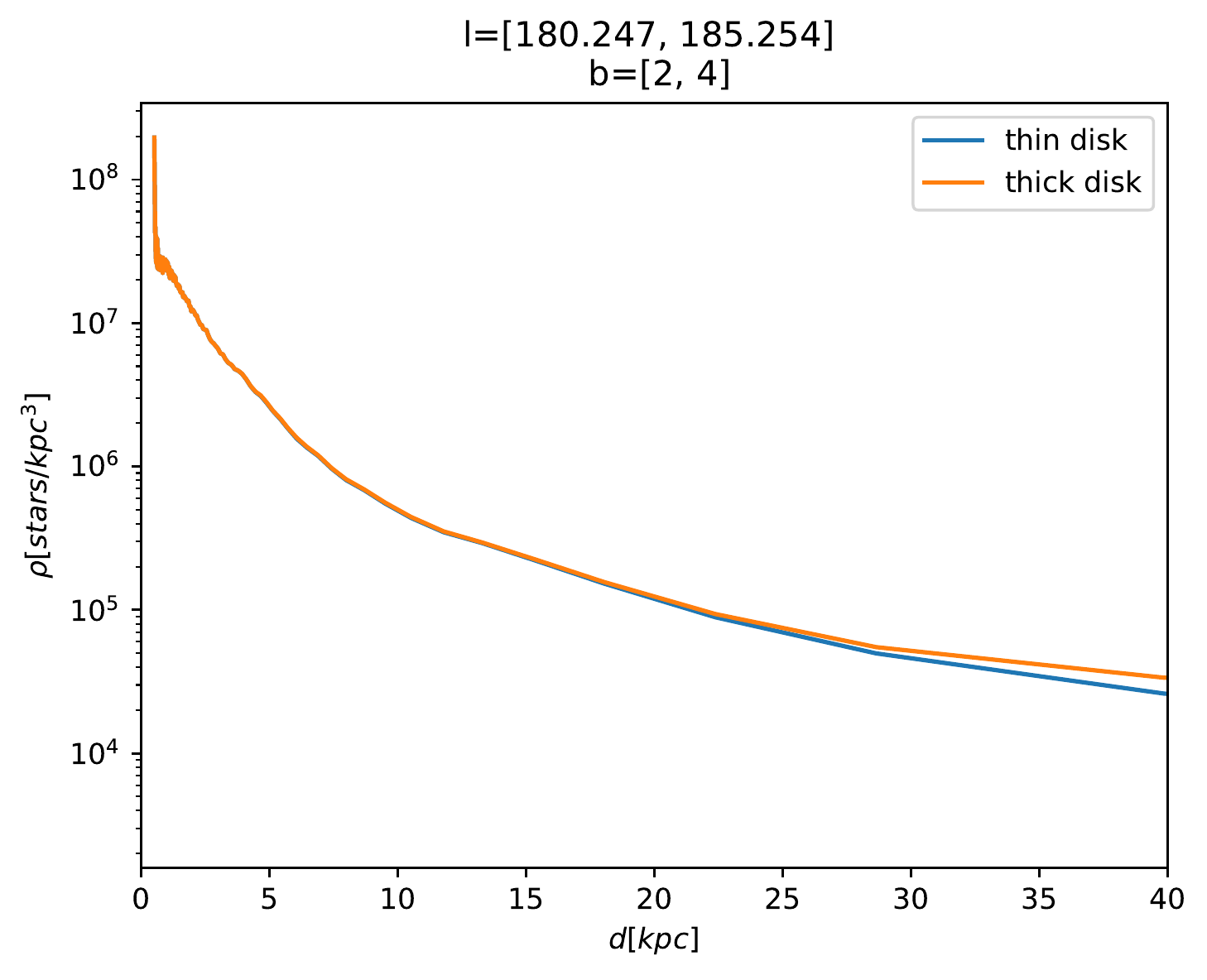}
        }
        \subfloat[]{
                \includegraphics[width=0.5\textwidth]{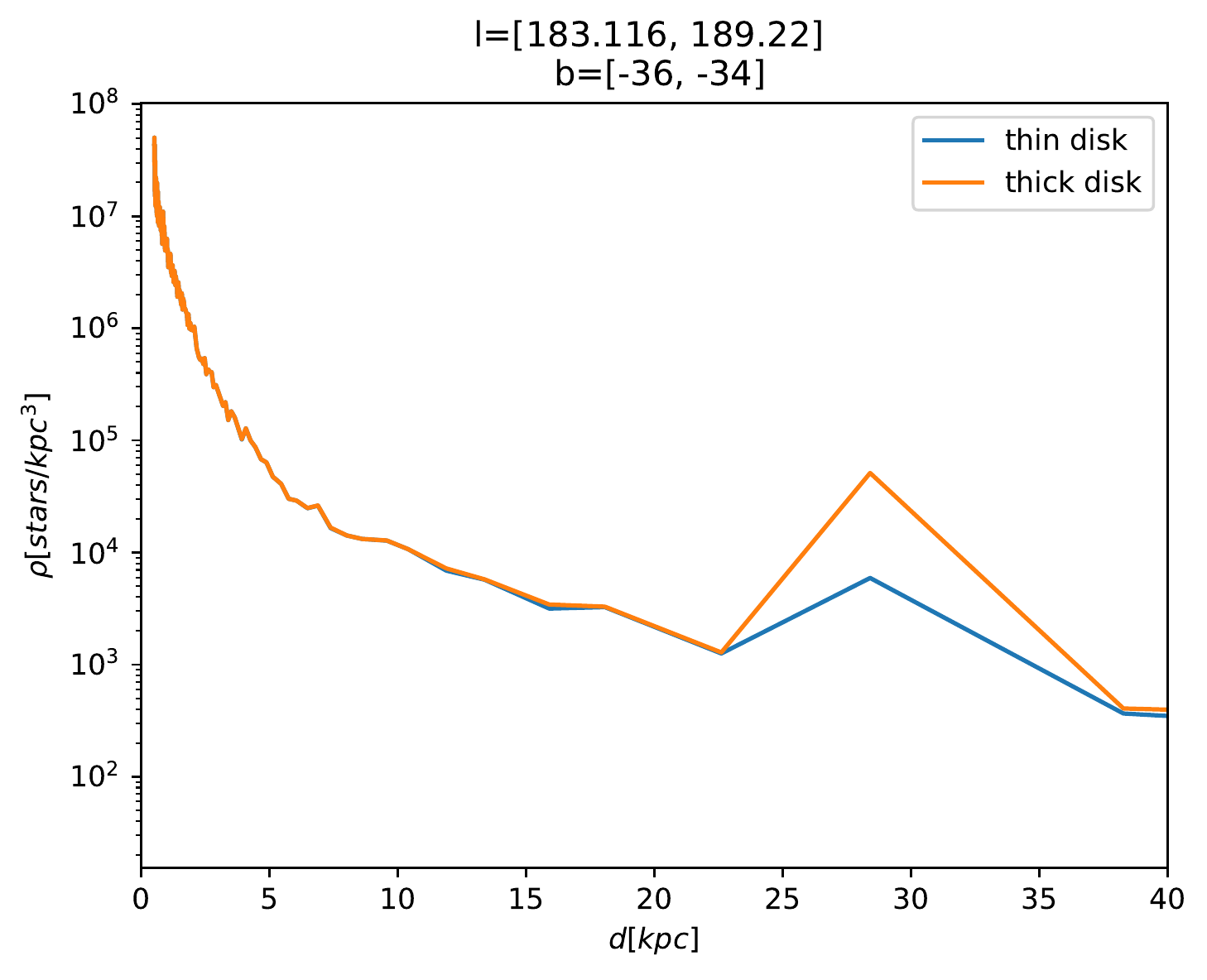}
        }
        \hspace{0mm}
        \subfloat[]{
                \includegraphics[width=0.5\textwidth]{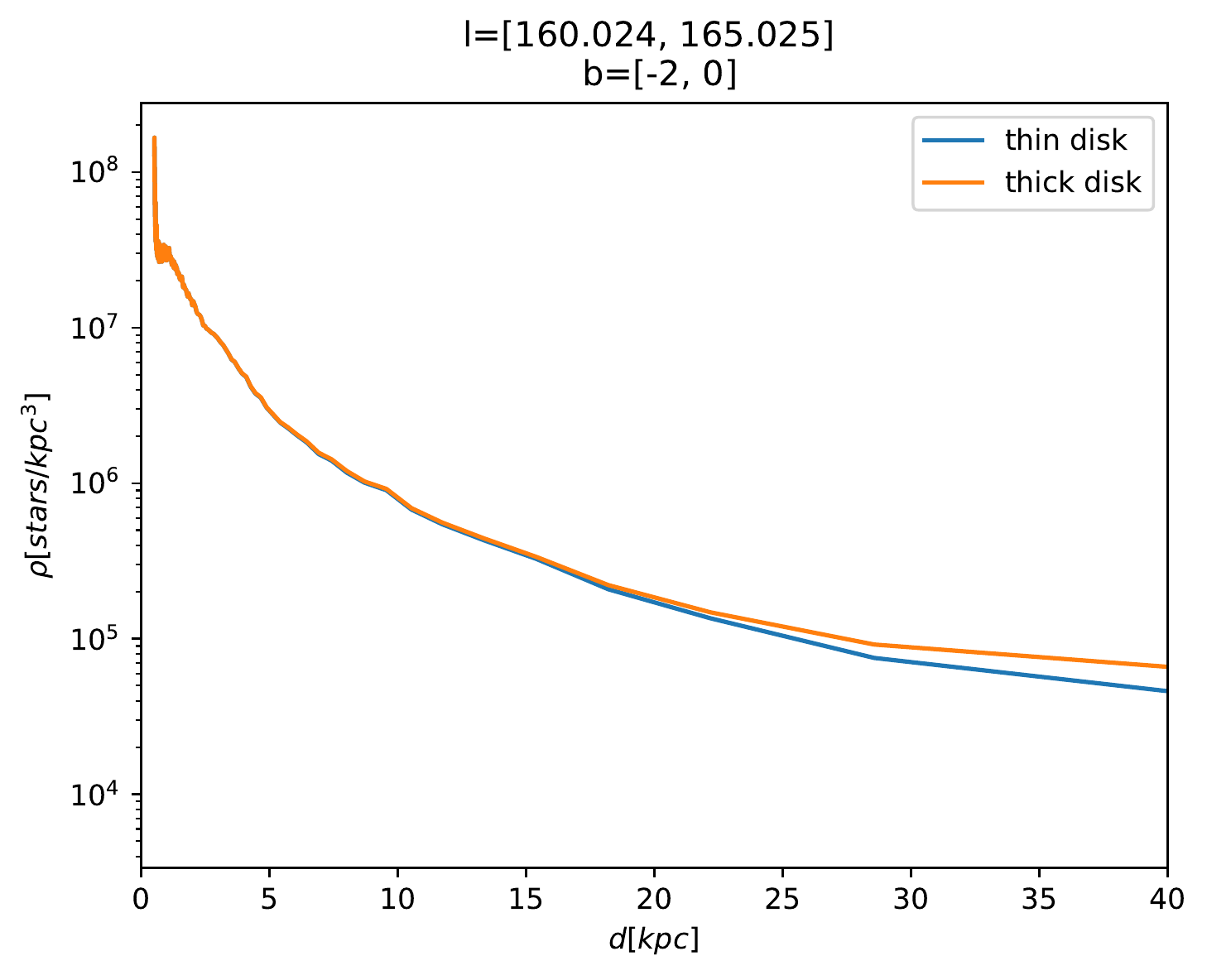}
        }
        \subfloat[]{
                \includegraphics[width=0.5\textwidth]{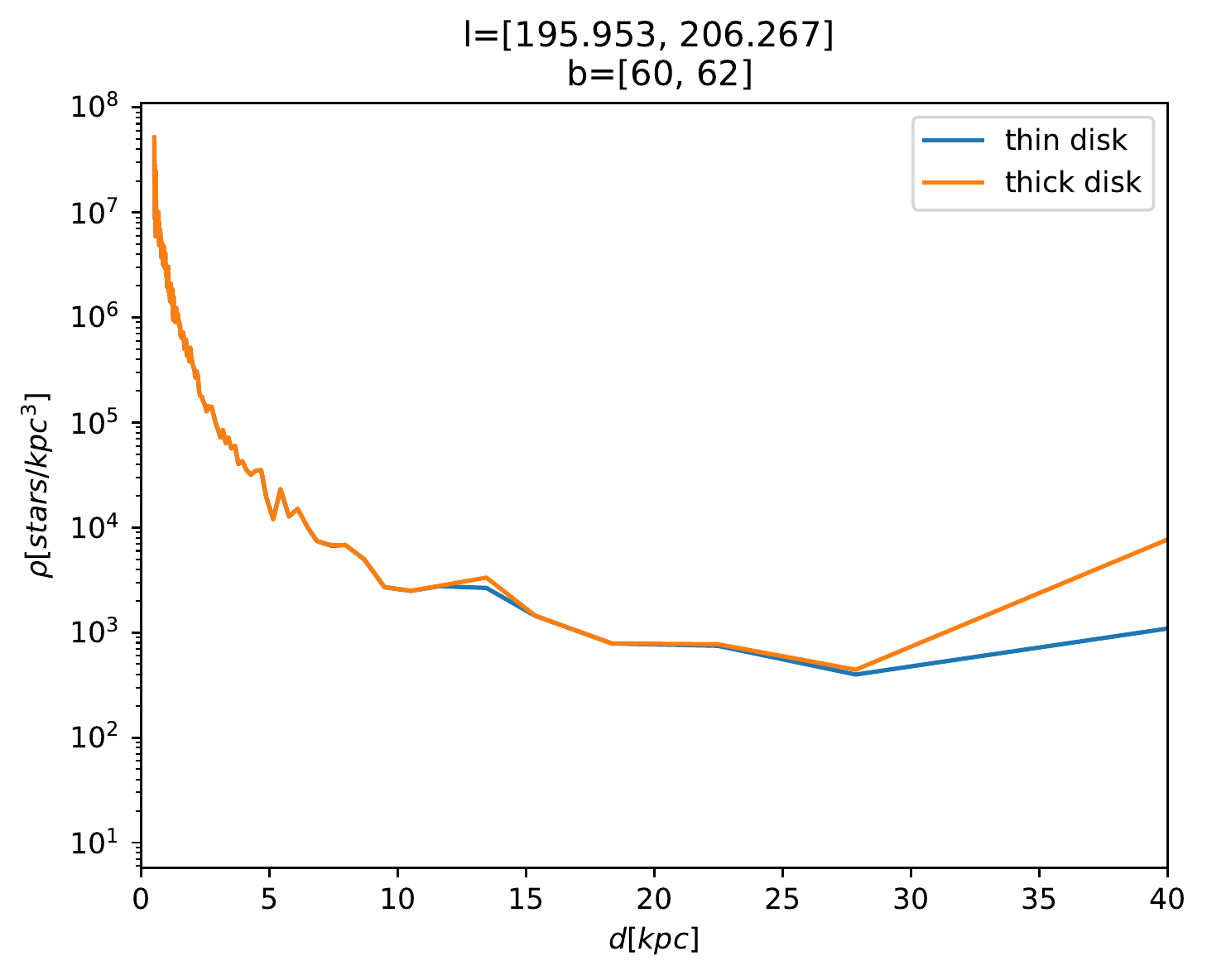}
        }
        \caption{Densities for different lines of sight. Blue curves are densities calculated with the thin-disc luminosity function. Orange curves are densities calculated with the thick-disc luminosity function.}\label{o18}
\end{figure*}

In the previous analysis we considered only the thin-disc population because the luminosity function presented in Section \ref{ch4} is calculated in thin-disc regions. However, we can also analyse high Galactic heights, where the influence of the thick disc is significant. To test the importance of the change in luminosity function, we tested the density calculations with a tentative thick-disc luminosity function that reduces the number of bright stars. We used the source table of \cite{wainscoat}, who give the ratio of all the components of the Galaxy for all stellar classes. Based on this comparison, we altered our luminosity function to construct a theoretical thick-disc luminosity function, as depicted in Fig. \ref{o21}. Then we repeated our calculation with this new luminosity function. In Fig. \ref{o18} we show the result for some lines of sight. In the area where we carried out the analysis, the difference between the two approaches is clearly visible starting at $\sim20$ kpc. Our density analysis is made in the area below 20 kpc, where the difference between the two densities is negligible. We note that this difference changes with line of sight, which is caused by the extinction. In the areas where the extinction is significant, the difference between densities derived from thin- and thick-disc luminosity functions is more important, but these areas are removed from our analysis. Therefore our maps are also valid for thick-disc areas.

\section{Analysis of the density maps}\label{results}

\subsection{Comparison with the maps of \cite{anders}}
Recently, similar maps were created by \cite{anders}. In their analysis, they used the code {\tt StarHorse}, originally developed to determine stellar parameters and distances for spectroscopic surveys \citep{queiroz}. This code compares observed quantities to a number of stellar evolutionary models. It finds the posterior probability over a grid of stellar models, distances, and extinctions. To do this, it needs many priors, including stellar initial mass function, density laws for main Milky Way components, and the broad metallicity and age of those components. Afterwards, the authors applied various criteria on their sample to choose only accurate results.  \\
When we compare our results, we can observe similar structures, except in the area of the Galactic bulge, where our data are not reliable and the data of \cite{anders} are much more accurate. However, because data with high errors in parallax were removed, \cite{anders} were unable to reach such high distances, which are necessary to study features of the outer disc such as the flare or the warp. Another advantage of our method is that we did not assume any priors about the Milky Way. Furthermore, our density maps are a representation of the complete number of stars per unit volume up to some given absolute magnitude, taking into account the luminosity function, whereas \cite{anders} gave the stars observed by Gaia, a much larger number in the solar neighbourhood, thus not useful to quantify absolute trends in the density distribution. Nevertheless, we consider the results of \cite{anders} very useful because they improve the accuracy of the data significantly and can be used to study parts of Milky Way where our data fail.

\subsection{Cut-off in the Milky Way}
There has been some discussion about the cut-off in the Milky Way. Some authors have reported to find a cut-off starting at about 14 kpc from the Galactic centre \citep{robin1,robin2,minniti}. However, \cite{carraro} argued that these finding are erroneous either because the dataset is biased or because the warp and flare is confused with the cut-off. The absence of the cut-off has been confirmed by several studies \citep{martin_cutoff,sale_cut_off,brand_cut_off}. Our results show that there is no cut-off in the Galactic disc, at least up to 20 kpc. \\

\subsection{Stellar density in the solar neighborhood}
We define the solar neighbourhood as the area where 7.5 kpc < R < 8.5 kpc and $\lvert z \rvert < 0.05 $ kpc and calculate the average density in this area. We find  $\rho_\odot=0.064~stars/pc^3$, which is close to other values in the literature, for example $0.03~stars/pc^3$ obtained by \cite{chang_solar_neighb}, who used a three-component model to fit data from 2MASS. \cite{eaton_solar_neigh} found $\rho_\odot=0.056~stars/pc^3$, which is lower than our result, but this value is influenced by the range of the luminosity function, which is where the difference between the values stems from. In our case, we measured stars with $M_G<10$.
        
\subsection{Exponential fits of the density}\label{chdensity}

To describe the radial volume mass density distribution in the Galactic equatorial plane, we used a modified exponential disc with a deficit of stars in the inner in-plane region adopted from \cite{corr1} in the following form:
\begin{eqnarray}\label{dens1}
\rho(R) = \rho_0 \times\exp\left(\frac{R_{\odot}}{h_r}+\frac{h_{r,hole}}{R_{\odot}}\right)\times \exp\left(-\frac{R}{h_r}-\frac{h_{r,hole}}{R}\right)~,
\end{eqnarray}
where $h_r$ is the scale length, $h_{r,hole}=3.74~kpc$ is the scale of the hole, $R_{\odot}$ is the Galactocentric distance of the Sun, and $R$ is the Galactocentric distance. We neglected the contribution of the thick disc and analysed only the thin disc.
We divided the Galactic equatorial plane into three regions according to the Galactic azimuth $\left[\ang{-45},\ang{-15}\right],\left[\ang{-15},\ang{15}\right],\left[\ang{15},\ang{45}\right]$. We focused on the Galactic equatorial plane, therefore we considered stars in the close vicinity of the plane with a vertical distance $|z|<0.2$ kpc and R>6 kpc. We fitted the density for various azimuths with the corresponding exponential fits based on Eq. (\ref{dens1}). The scale length slightly depends on the Galactic azimuth; it reaches the highest value for the Sun-anticentre direction and $\phi = +30^\circ$, $h_r=2.78$$\pm$$0.13$ kpc, and $h_r = 2.29$$\pm$$0.21$ kpc. On the other hand, the lowest value of the scale length is $h_r = 1.88$$\pm$$0.12$ kpc for $\phi = -30^\circ$ . This results in an average of $h_r=2.29\pm0.08$ kpc, with small dependence on the azimuth. We can compare the results with published papers. \cite{martin_cutoff} used SDSS-SEGUE (Sloan Digital Sky Survey - Sloan Extension for Galactic Understanding and Exploration) data to investigate the density distribution in the Galactic disc. They obtained the scale length for the thick and for the thin disc, $h_{r,thin} = 2.1$ kpc and  $h_{r,thick} = 2.5$ kpc for the azimuth $\phi \le 30^{\circ}$, which is consistent with our results. \cite{li} studied OB stars using Gaia DR2 data and the derived scale length of the Galactic disc, and found $h_{r} = 2.10 \pm 0.1$ kpc, which is in accordance with our results.

\begin{figure}  
        \includegraphics[width=9cm]{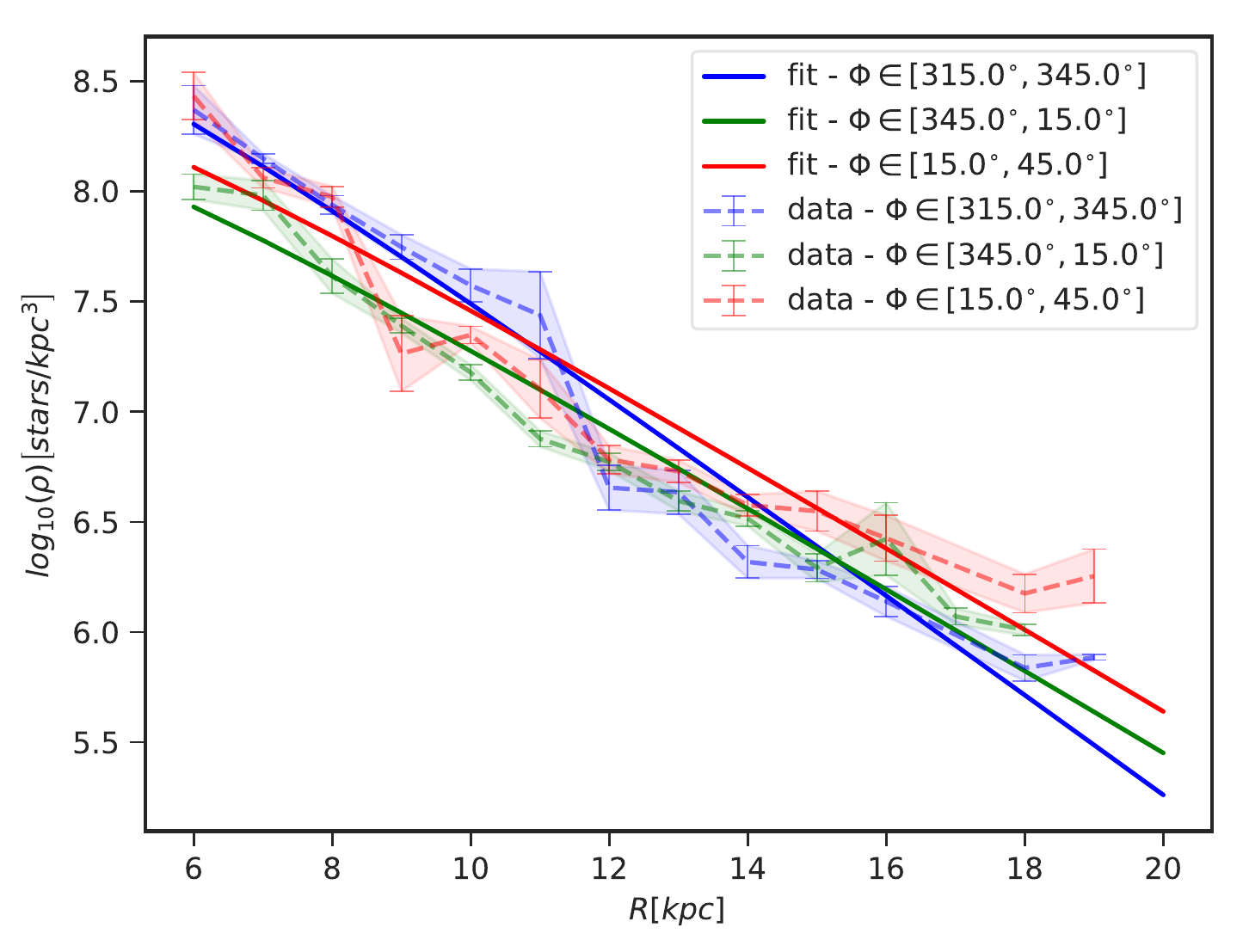}
        \caption{Dependence of the density on azimuth near the centre-Sun-anticentre direction for various values of Galactocentric distance. The data points are obtained as weighted mean in bins of size 1 kpc in R and 0.4 kpc in $\lvert z \rvert$. Only bins with a number of points $N \geq 50$ points are plotted.}\label{d2}
\end{figure}

We also plot the dependence of the density in the Galactic equatorial plane on azimuth for various values of Galactocentric distance in Fig. \ref{d2}. The density is slightly dependent on the Galactic azimuth for all radii, but this dependence is very small. An analysis of the scale height and its corresponding flare will be given in a forthcoming paper (Nagy et al. 2020, in preparation).

\subsection{Warp}\label{ch11}
The density maps (Fig. \ref{o7}) directly show a northern warp in azimuth $\ang{90}$ and southern warp in azimuth $\ang{270}$. Here, we analyse these structures in greater detail. We removed the azimuths $\ang{150}<\phi<\ang{240}$ and radii R<6 kpc from our analysis because these data have low quality and influence the results negatively. \\
We calculated the average elevation above the plane $z_w$ as

\begin{eqnarray}\label{zw}
z_w=\frac{\int_{z_{min}}^{z_{max}} \rho z \mathrm{d}z }{\int_{z_{min}}^{z_{max}} \rho \mathrm{d}z }
\end{eqnarray}
and fit this quantity with models of the warp. In our first approach, we used the model by López-Corredoira et al. (2002b, Eq. 20),

\begin{eqnarray}\label{9}
z_w=[C_wR(pc)^{\epsilon_w}sin(\phi-\phi_w)+17]~pc~.
\end{eqnarray}

\begin{figure}
        \includegraphics[width=0.5\textwidth]{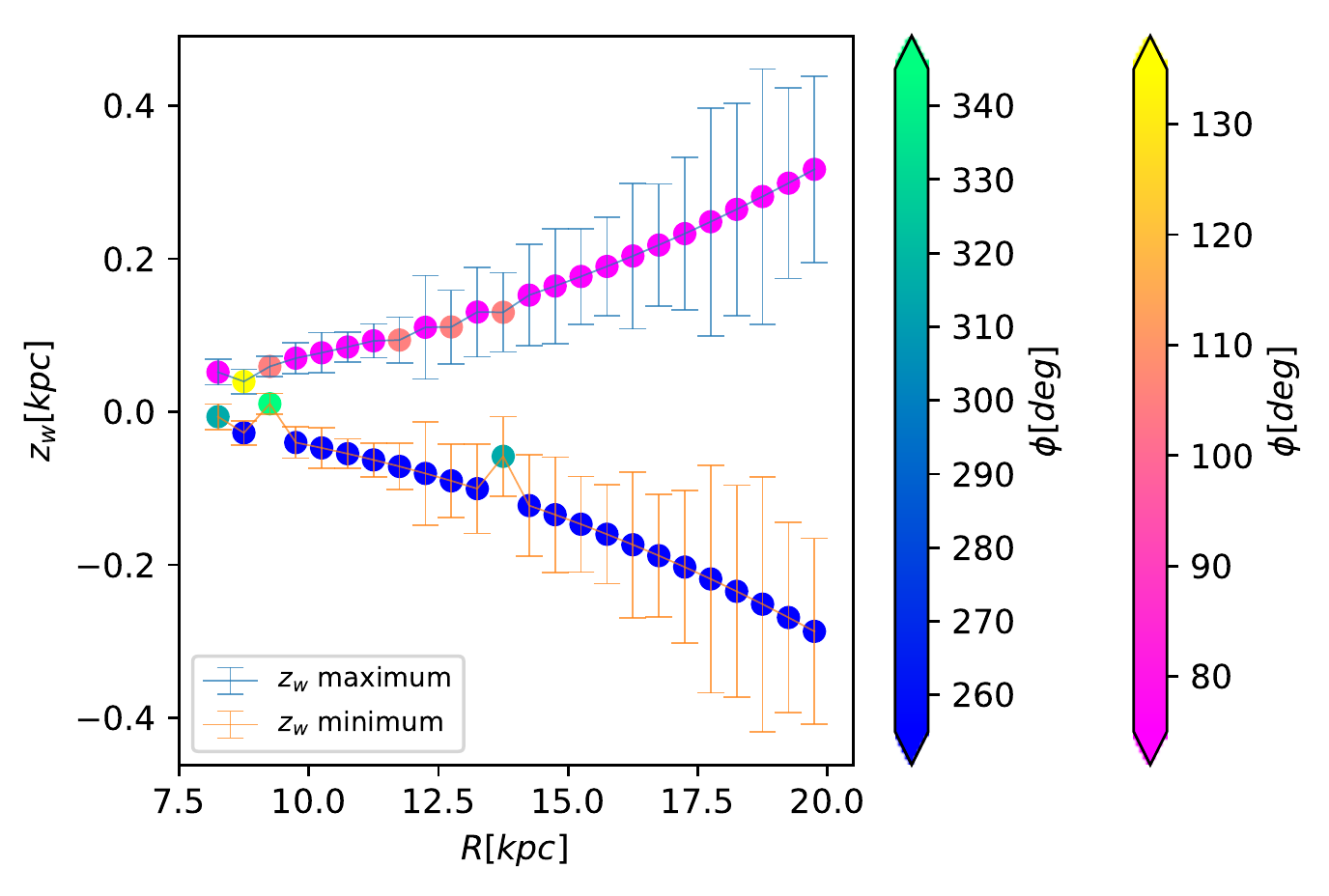}
        \caption{Minimum and maximum average elevation of the plane as a function of radius. The warp fit is based on Eq. (\ref{9}), and the error bars represent the uncertainty in the distance in the Lucy method.}\label{o11}
\end{figure}

The 17 pc term compensates for the elevation of the Sun above the plane \citep{z_slnko}. $C_w,\epsilon_w$ , and $\phi_w$ are free parameters of the model, which were fitted to our data. An asymmetry is observed between the northern and southern warp for the gas \citep{voskes} and for the young population \citep{amores}, therefore we also explore the northern and southern warp separately here.
The fit of our data yields maximum amplitudes $z_w=0.317$ kpc for the northern and $z_w=-0.287$ kpc for the southern warp, both at a distance R=[19.5,20] kpc, revealing a small asymmetry between the north and south. For the fit, we used the function \textit{curve fit} from the python \textit{SciPy} package, which uses non-linear least squares to fit a function to data.
The parameters of the best fit for this model for the whole dataset are 

\begin{eqnarray}\label{parametre}
C_w&=&1.17\cdot10^{-8} \mathrm{pc} \pm 1.34\cdot10^{-9} \mathrm{pc} (stat.) \nonumber \\
&-& 2.9\cdot10^{-10} \mathrm{pc} (syst.)~, \nonumber \\
\epsilon_w&=&2.42\pm 0.76(stat.) + 0.129 (syst.)~, \\
\phi_w&=&\ang{-9.31}\pm \ang{7.37} (stat.) +\ang{4.48} (syst.)~. \nonumber
\end{eqnarray}
Here, the error of $c_w$ stands for the error of the amplitude alone, without the variations of $\epsilon_w$ and $\phi_w$. The plot of the results is shown in Fig. \ref{o11}, where we show the comparison of minimum and maximum value of $z_w(R)$. The average elevation of the plane is highest for azimuths $[\ang{60},\ang{90}]$ and $[\ang{90},\ang{120}]$ in most of the cases, whereas the minimum is reached for azimuths $[\ang{240},\ang{270}]$ in most of the cases. A slight asymmetry between the northern and southern warp is also clearly visible.


\begin{figure*}
        \begin{center}
                \subfloat[$R$=$13.25kpc$]{\includegraphics[width=6cm]{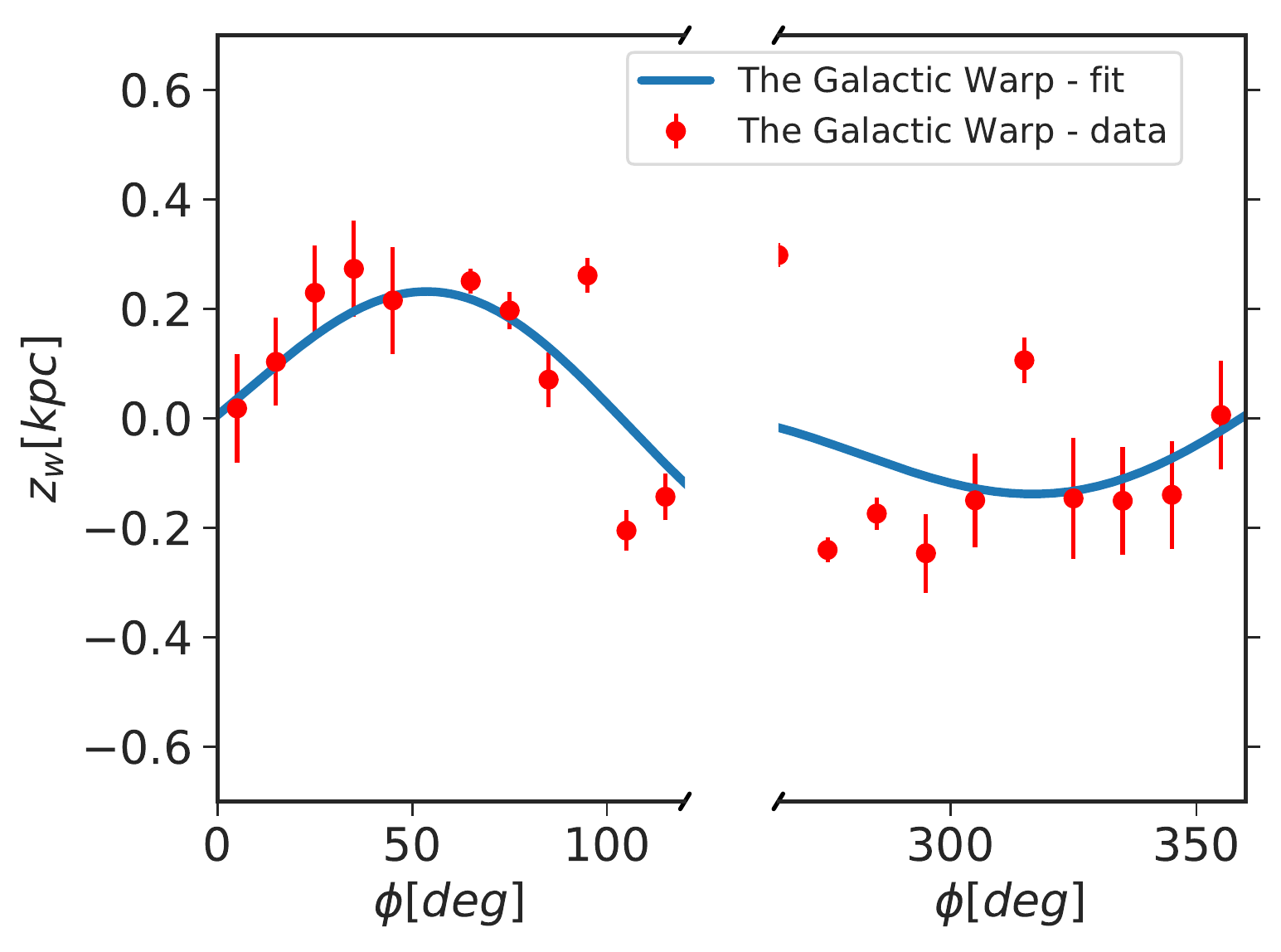}}
                \subfloat[$R$=$16.25kpc$]{\includegraphics[width=6cm]{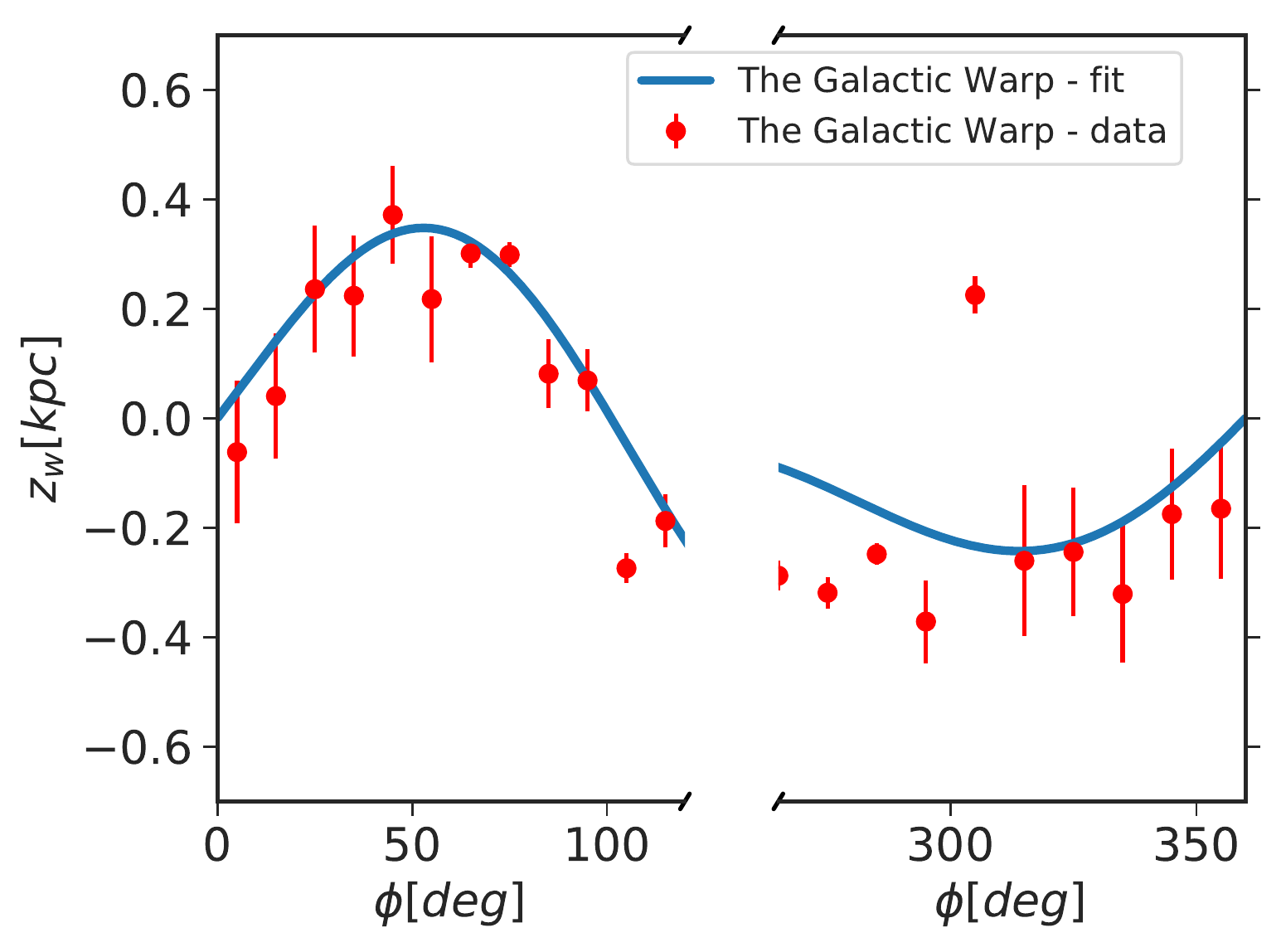}}
                \subfloat[$R$=$19.25kpc$]{\includegraphics[width=6cm]{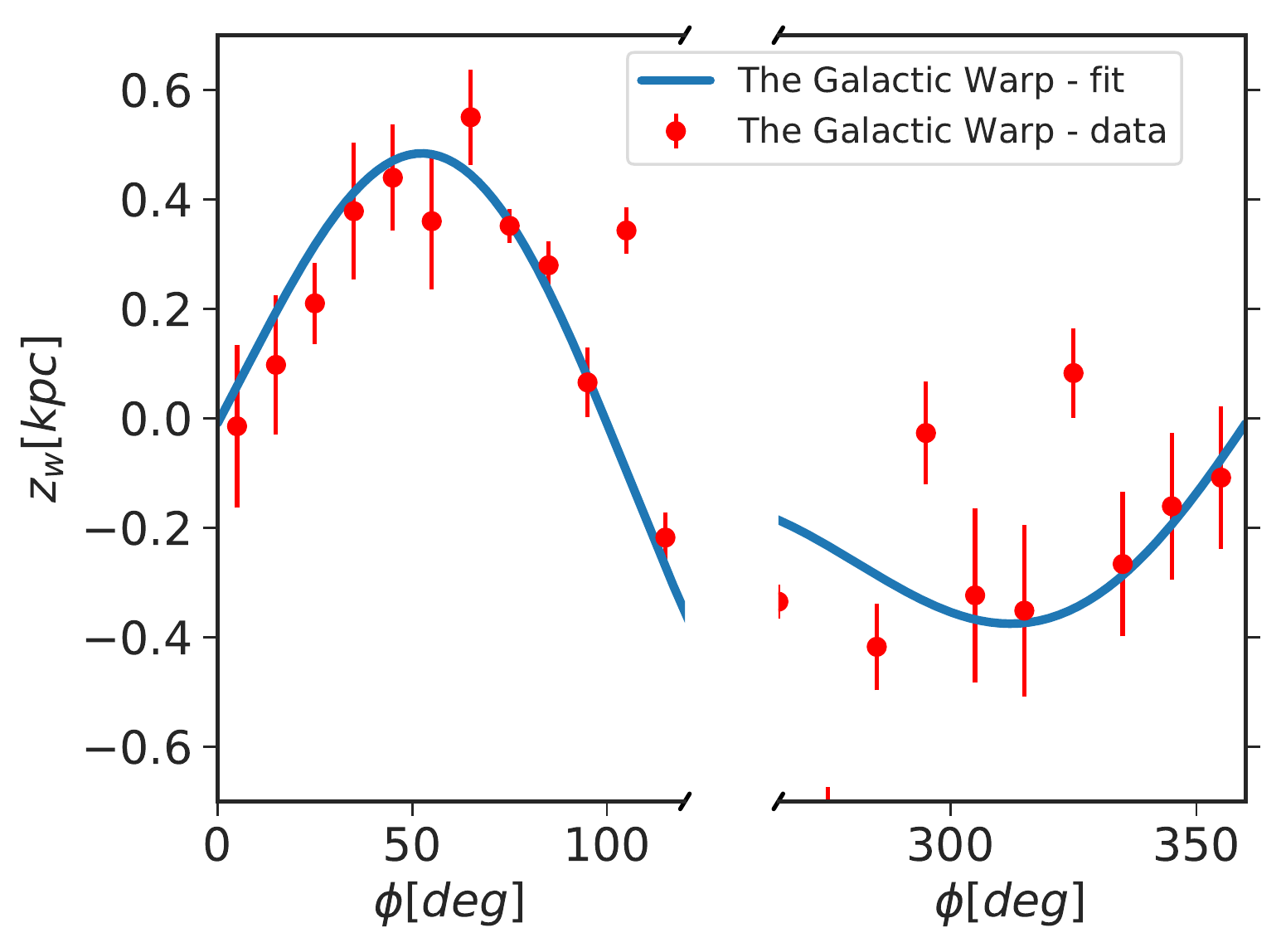}}
                \caption{Average elevation of the plane as a function of the Galactic azimuth for various values of the Galactocentric distance. Red markers represent values of binned data, and the blue line represents a fit to the data.}\label{w2}
        \end{center}
\end{figure*}

\begin{figure}  
        \includegraphics[width=9cm]{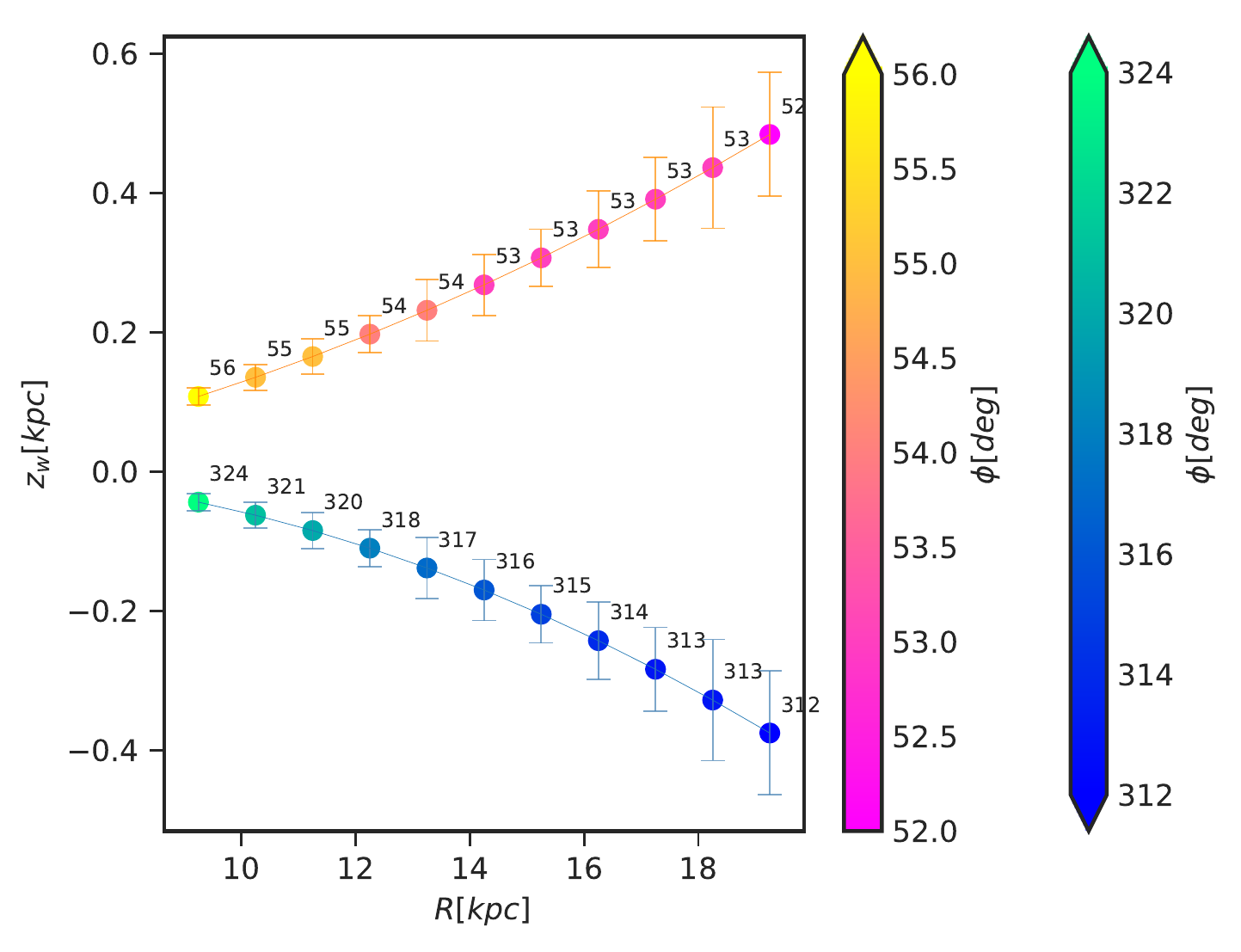}
        \caption{Minimum and maximum of the average elevation of the plane as a function of Galactocentric distance. The warp fit is based on Eq. \ref{warp2}. The colours code the azimuth of the minimum and  maximum of the warp fit, and the error bars represent the uncertainty on the distance in the Lucy method.}\label{w3}
\end{figure}

Another approach that we used is based on the work of \cite{levine}, who studied the vertical structure of the outer disc of the Milky Way by tracing neutral hydrogen gas. They analysed the Galactic warp using a Lomb periodogram analysis. They concluded that the first two Fourier modes are the strongest modes. We use the expression derived by \cite{levine} in the following form:
\begin{eqnarray}\label{warp2}
z_w=z_0 + z_1\cdot\sin{\left(\phi-\phi_1\right)} + z_2\cdot\sin{\left(2\phi-\phi_2\right)}~,
\end{eqnarray}
where $z_w$ is the average elevation above the plane, $z_i$ for $i\in\left(0,1,2\right)$ are the amplitudes of the warp, $\phi_i$ for $i\in\left(1,2\right)$ are the phases. The dependence of the amplitudes of the warp on the Galactocentric distances is
\begin{eqnarray}\label{warp2b}
z_i=k_0 + k_1\cdot\left(R-R_k\right) + k_2\cdot\left(R-R_k\right)^2~\text{for}~i=0,1,2~,
\end{eqnarray}
where  $k_i$ and $R_k$ are free parameters of the fit. We fitted our data with Eqs. (\ref{warp2}) and (\ref{warp2b}) for various values of Galactocentric distances $R<20$ kpc. We plot the data and the fits for $R \in \left(13.25, 16.25,19.25\right)$ kpc in Fig. \ref{w2}. Fig. \ref{w3} shows the azimuth of the maximum and minimum of the Galactic warp as a function of the Galactocentric distance. In our analysis, we excluded data for the Galactic azimuths  $\phi\in\left(120^{\circ},240^{\circ}\right)$ because of the high error values in our data. We used a $10^\circ$ binning in azimuth. Fig. \ref{w2} shows that the data for $\ang{250} < \phi < \ang{270}$ are somewhat noisy, which can be caused by problems with extinction or with the Lucy method in a particular line of sight. Therefore we tested a fit without these points, which turned out to produce an insignificant difference. For instance, the minimum amplitude obtained without these points changed by  ~10\% in the worse case, and the maximum amplitude changed by ~2\%.

Figs. \ref{w2} and \ref{w3} clearly show that the warp is present in our analysis. The azimuth of the maximum of the warp (the northern warp) is an increasing function of the Galactocentric distance ($52^\circ<\phi<56^\circ$). On the other hand, the azimuth of the minimum of the warp is in $312^\circ<\phi<324^\circ$ and corresponds to the southern warp. The strongest deviation of the average elevation of the Galactic plane from the Galactic equatorial plane rises with  Galactocentric distance. The highest amplitude of the northern and southern warp is $z_w=0.48$ kpc and $z_w=-0.38$ kpc, respectively. An asymmetrical warp is clearly present.

\begin{figure}  
        \includegraphics[width=9cm]{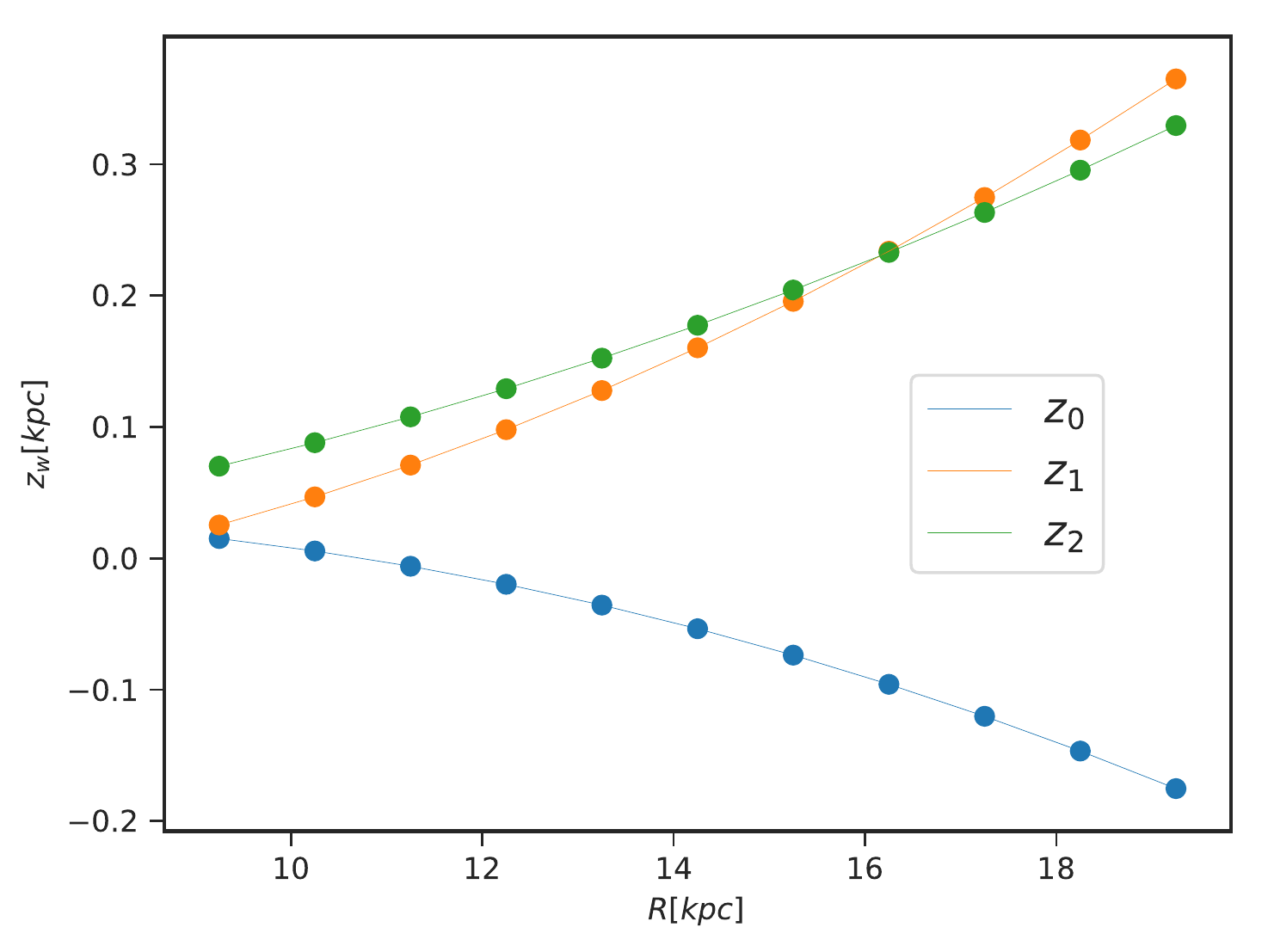}
        \caption{Changes of the amplitudes of the Galactic warp fit according to Eqs. \ref{warp2} and \ref{warp2b}.}\label{w5}
\end{figure}

The value of the line of nodes from the fit is $\phi_0=\ang{-1.18}$.
We plot the changes in amplitude of the Galactic warp fit [Eq.(\ref{warp2b})] with Galactocentric distance in Fig. \ref{w5}. 

\begin{figure}  
        \includegraphics[width=9cm]{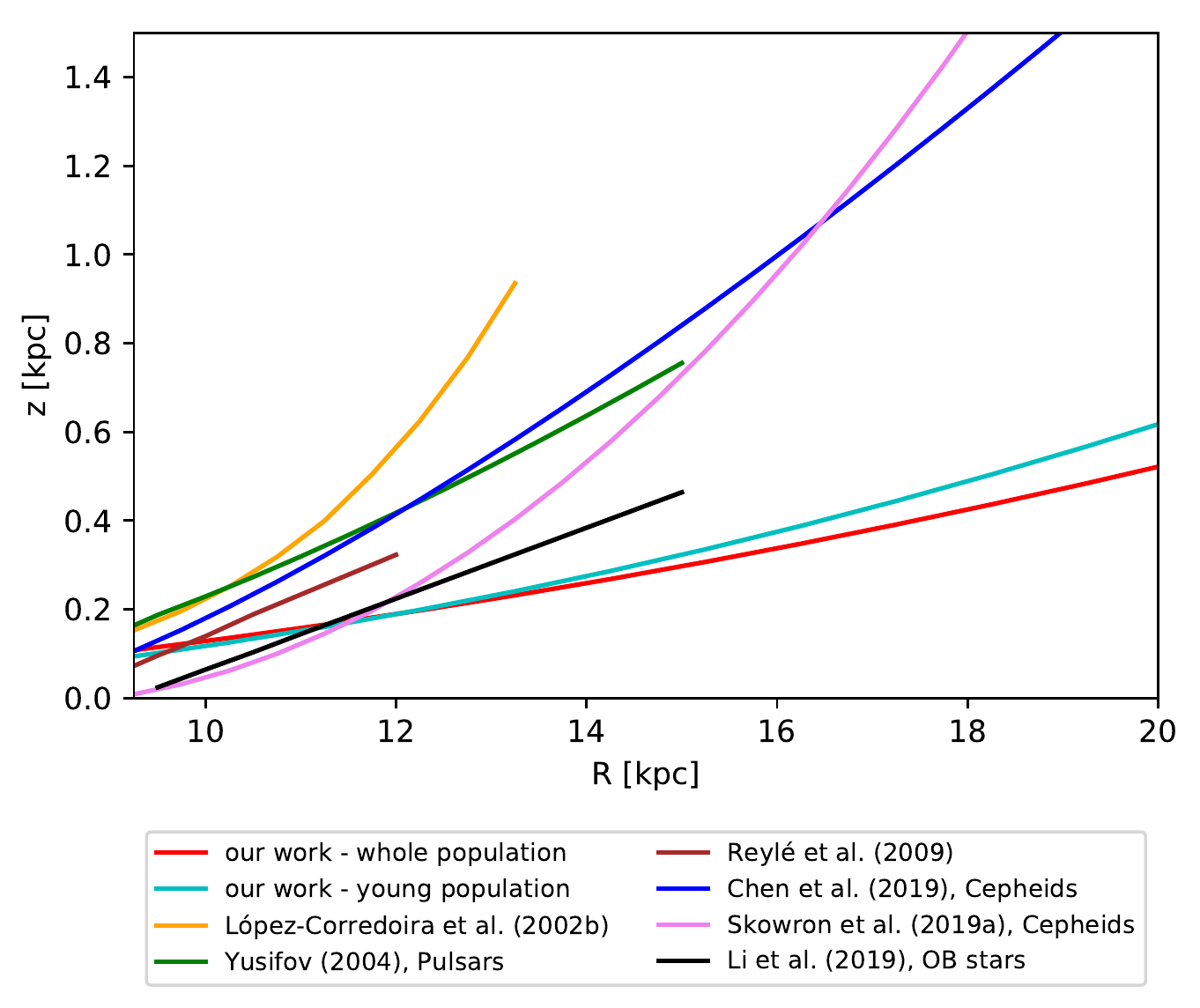}
        \caption{Comparison of maximum amplitudes of our model (based on Eq. (\ref{warp2})) with other works.}\label{amplit}
\end{figure}

Similar results were obtained by \cite{li}, who used OB stars of Gaia DR2 to measure the warp. They fit their data with a sinusoidal function similar to ours and obtained a warp with a mean magnitude up to $z=0.5$ kpc. However, they did not account for the asymmetry of the warp, therefore they found the same result for the north and south. \cite{chen_warp} used Cepheids from the WISE (Wide-field Infrared Survey Explorer) catalogue and traced the warp up to R=20 kpc. Their results show a warp extended up to $\lvert z \rvert=1.5$ kpc, which we cannot confirm using the whole population. \cite{poggio} studied the kinematics of the Milky Way using Gaia DR2 and found the warp up to 7 kpc from the Sun. This agrees with our results, but we show that the warp extends to a higher radius, at least up to 20 kpc. In Fig. \ref{amplit}, we compare the maximum amplitudes of our model with other works. We obtain a very low amplitude, especially in comparison with Cepheids. On the other hand, the closest result is that of \cite{chen_warp}, who used OB stars from Gaia DR2. This significant difference between the amplitude of various populations is in favour of the formation of the warp through accretion onto the disc \citep{martin_accretion}, which causes the gas and young stars to warp more strongly than the remaining population. \\
\cite{momany} studied the stellar warp using 2MASS red clump and red giant stars, selected at fixed heliocentric distances of 3, 7, and 17 kpc. They found a rather symmetric warp and argued that a symmetric warp can be observed as asymmetric for two reasons. First, the Sun is not located at the line of nodes, and second, the northern warp is located behind the Norma-Cygnus arm, which can cause variation in extinction that can produce an apparent asymmetric warp. As for the first point, the position of Sun on the line of nodes is a problem when we observe the warp at a fixed distance. However, we have a 3D distribution, which ensures that the position from which we look does not influence how we perceive the warp. As for the second remark, as we showed in Section \ref{ch3} that the extinction is determined quite accurately by the extinction map of \cite{green}. However, some variations might influence the final shape of the warp and may not have been taken into account, therefore we need to keep that in mind when we interpret our results.
        
\subsection{Young population}\label{ch12}
\begin{figure}
        \includegraphics[width=0.5\textwidth]{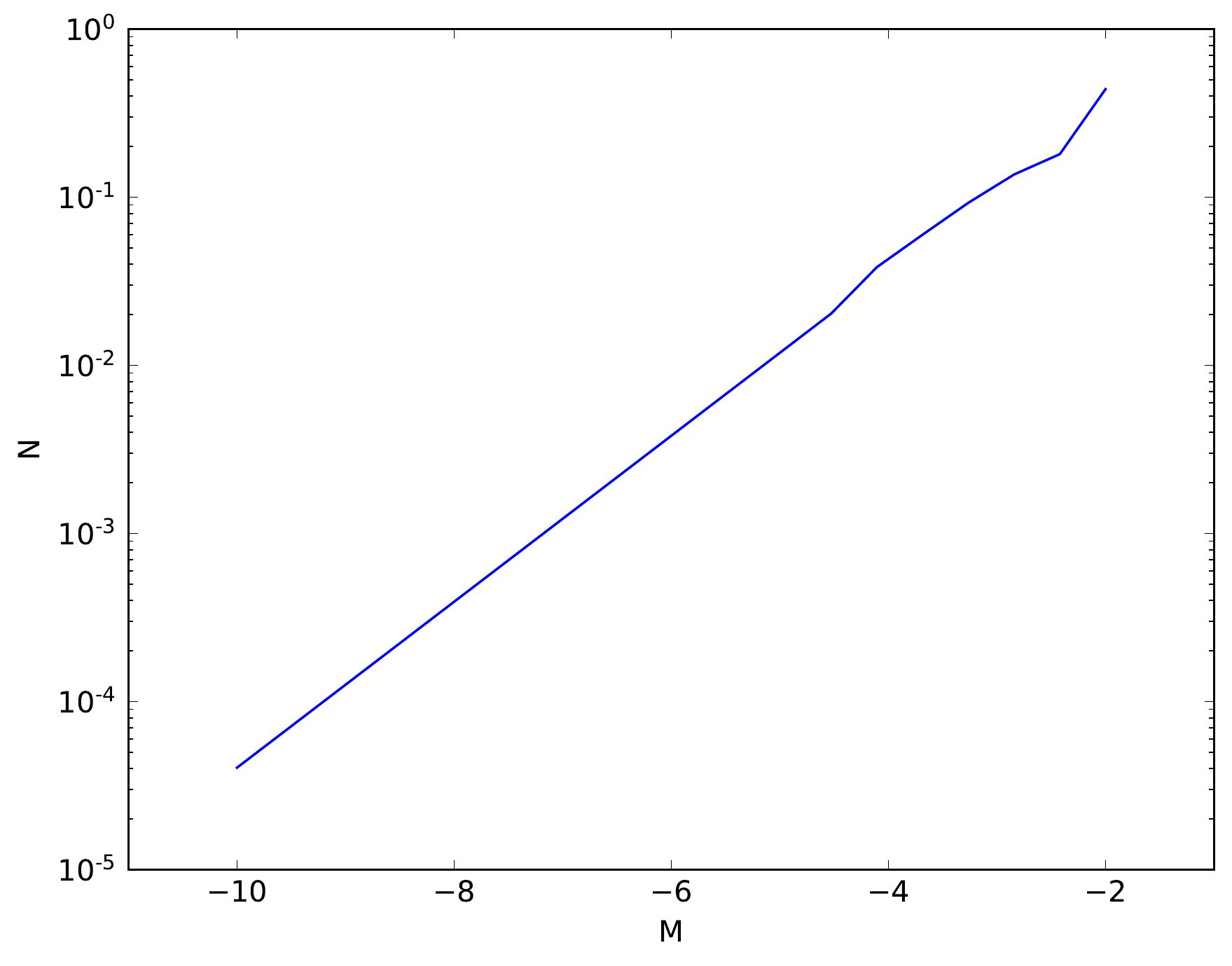}
        \caption{Luminosity function used in the Eq. \ref{4} for the analysis of the young population.}\label{o19}
\end{figure}

In this section, we apply the previous analysis to the young population. To do so, we only chose stars brighter than an absolute magnitude $M_G=-2$ (see the luminosity function in Fig. \ref{o19}) and repeated all the steps as described in Section \ref{ch8}. Then we produced density maps and analysed the scale length and the warp of this population using methods from Sections \ref{chdensity} and \ref{ch11}. \\

\begin{figure}
        \includegraphics[width=0.5\textwidth]{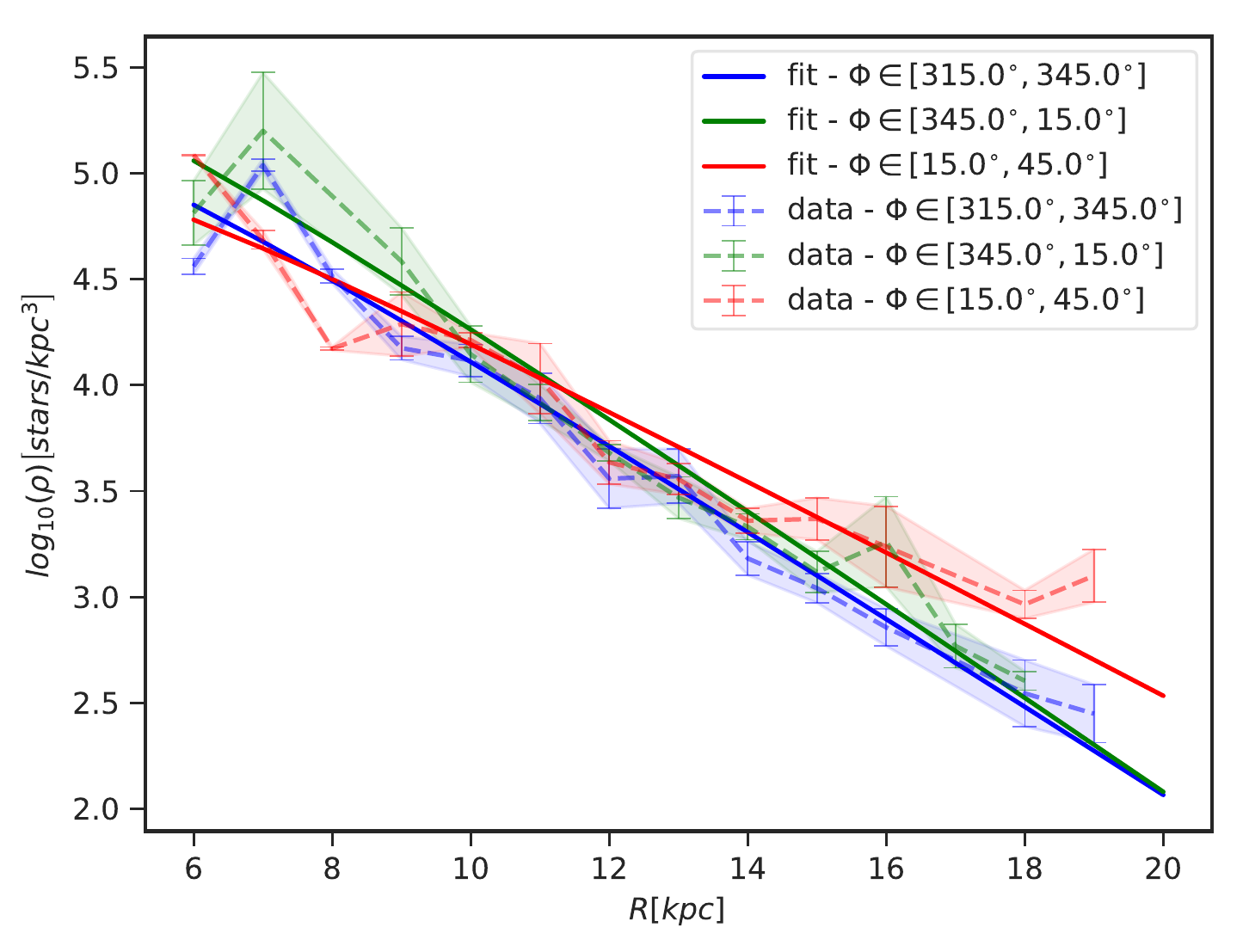}
        \caption{Same as Fig. \ref{d2}, but the young population alone is considered.}\label{hr_young}
\end{figure}

The exponential fits of the density for the young population yield $h_r=2.5\pm0.22$ kpc for $\phi=\ang{30}$, $h_r=1.92\pm0.15$ kpc for $\phi=\ang{0}$ , and $h_r=2.04\pm0.15$ kpc for $\phi=\ang{30}$, which is similar to the whole population. This results in $h_r=2.09\pm0.09$ kpc on average. The variation with azimuth is still insignificant, as in the case of the entire population. Fig. \ref{hr_young} shows that the variation of density with azimuth is also negligible in the case of the young population. \\  
For the warp, as previously, we removed the azimuths $\ang{150}<\phi<\ang{240}$ from the analysis. The fit of Eq. (\ref{9}) to the young population yields 

\begin{eqnarray}
C_w&=&4.85\cdot10^{-14} \mathrm{pc} \pm 6.33\cdot10^{-15} \mathrm{pc} (stat.) \nonumber \\
&+&5.4\cdot 10^{-15} \mathrm{pc} (syst.)~, \nonumber \\
\epsilon_w&=&3.69\pm 1.19 (stat.) -0.373(syst.)~, \\
\phi_w&=&\ang{-1.64}\pm \ang{8.85} (stat.) - \ang{2.803} (syst.)~. \nonumber
\end{eqnarray}  

\begin{figure*}
        \begin{center}
                \subfloat[$R$=$13.25kpc$]{\includegraphics[width=6cm]{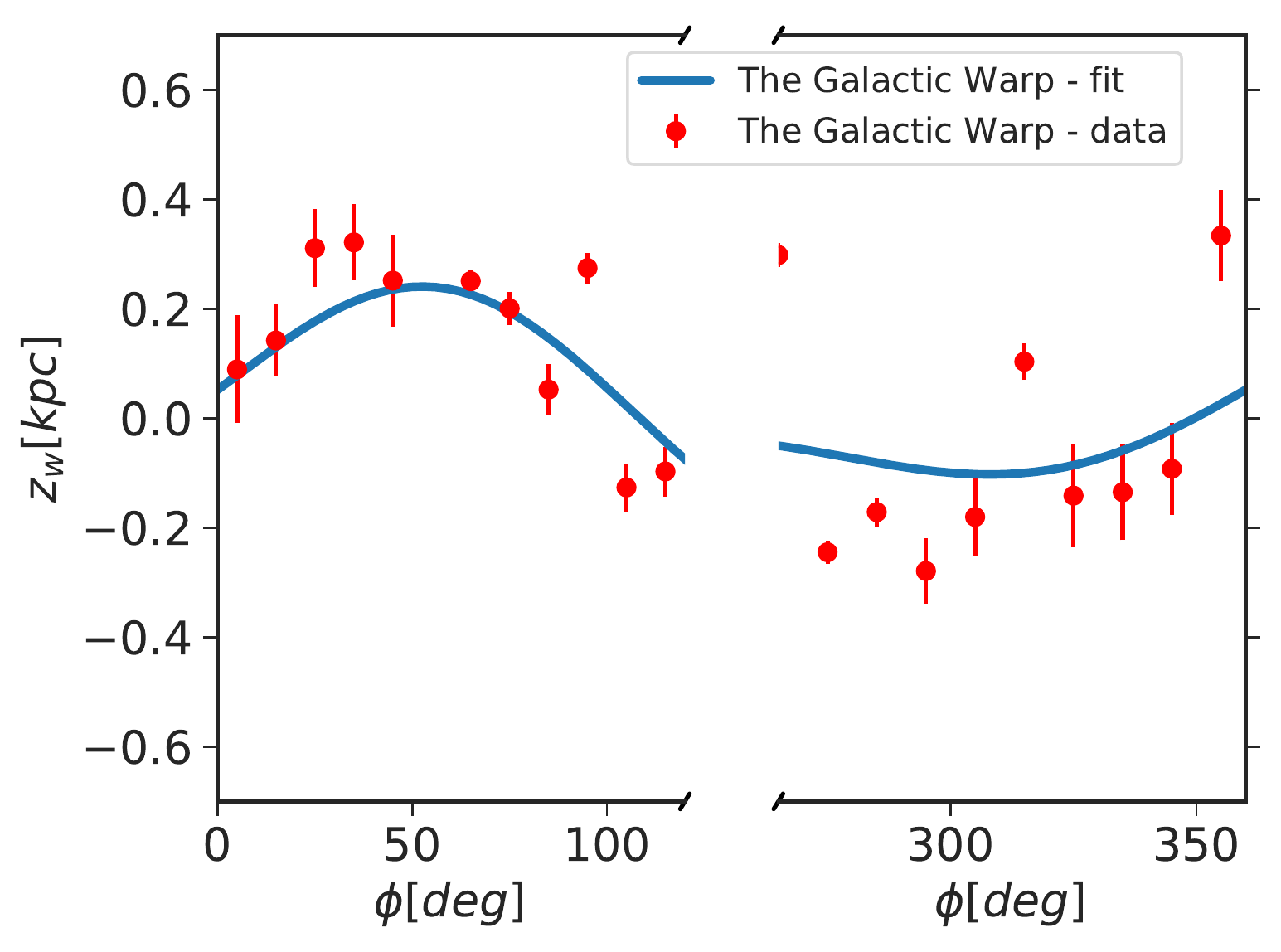}}
                \subfloat[$R$=$16.25kpc$]{\includegraphics[width=6cm]{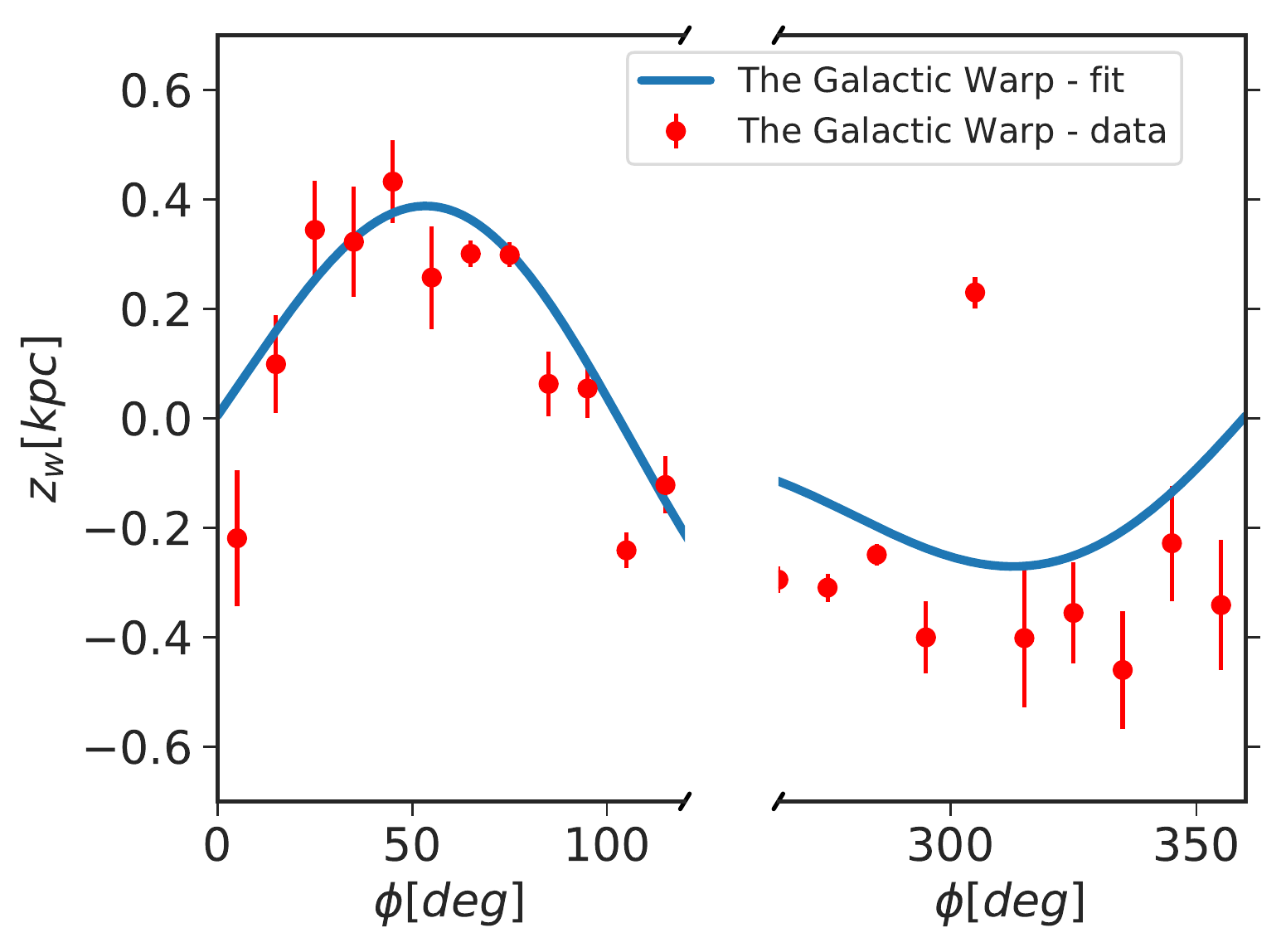}}
                \subfloat[$R$=$19.25kpc$]{\includegraphics[width=6cm]{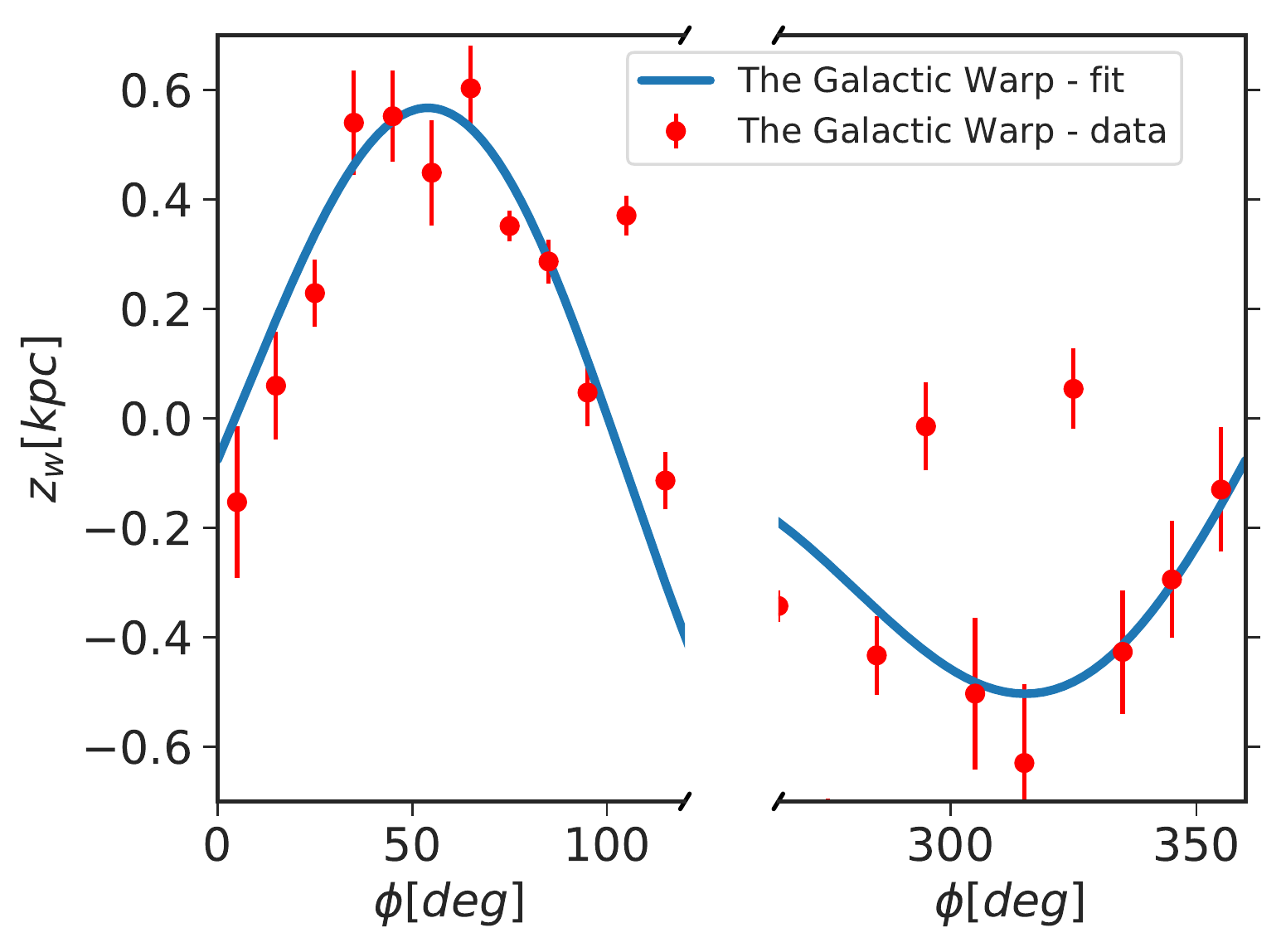}}
                \caption{Dataset containing a young population of stars. The average elevation of the plane as a function of the Galactic azimuth for various values of the Galactocentric distance. Red markers represent values of binned data, and the blue line represents a fit to the data.}\label{w2y}
        \end{center}
\end{figure*}

\begin{figure}  
        \includegraphics[width=9cm]{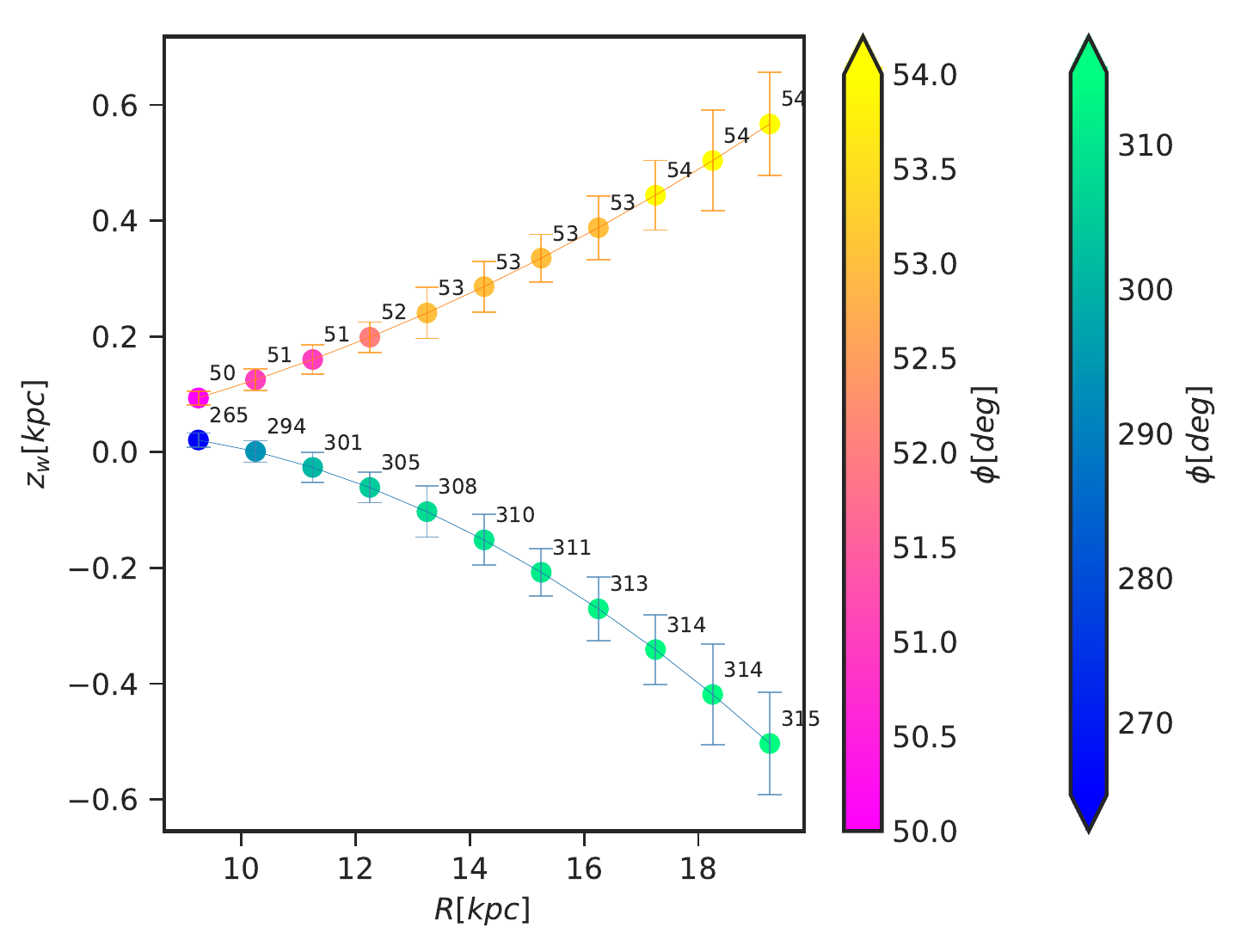}
        \caption{Minimum and maximum of the average elevation of the plane as a function of the Galactocentric distance. The warp fit is based on Eq. (\ref{warp2}). The colours code the azimuth of the minimum and the maximum of the warp fit. The dataset containing a young population of stars is considered, and the error bars represent the uncertainty on the distance in the Lucy method.}\label{w3y}
\end{figure}

We also repeated the analysis  with the approach using Eq. (\ref{warp2}). Fig. \ref{w2y} presents the Galactic warp of the young stellar population for various Galactocentric distances, and Fig. \ref{w3y} shows the amplitudes of the fits of the Galactic warp and the azimuth of the maximum and minimum. In this case, the warp of the young stellar population is stronger than the case considering all stars in our dataset. The azimuth of the maximum of the northern warp is an increasing function of the Galactocentric distance ($50^\circ<\phi<54^\circ$), and the azimuth of the minimum of the warp is in $265^\circ<\phi<315^\circ$. The highest amplitude of the northern and the southern warp is $z_w=0.57$ kpc and $z_w=-0.5$ kpc, respectively. For the line of nodes, we find $\phi_0=\ang{-6.56}$, which agrees with the whole population. 

\begin{figure}  
        \includegraphics[width=9cm]{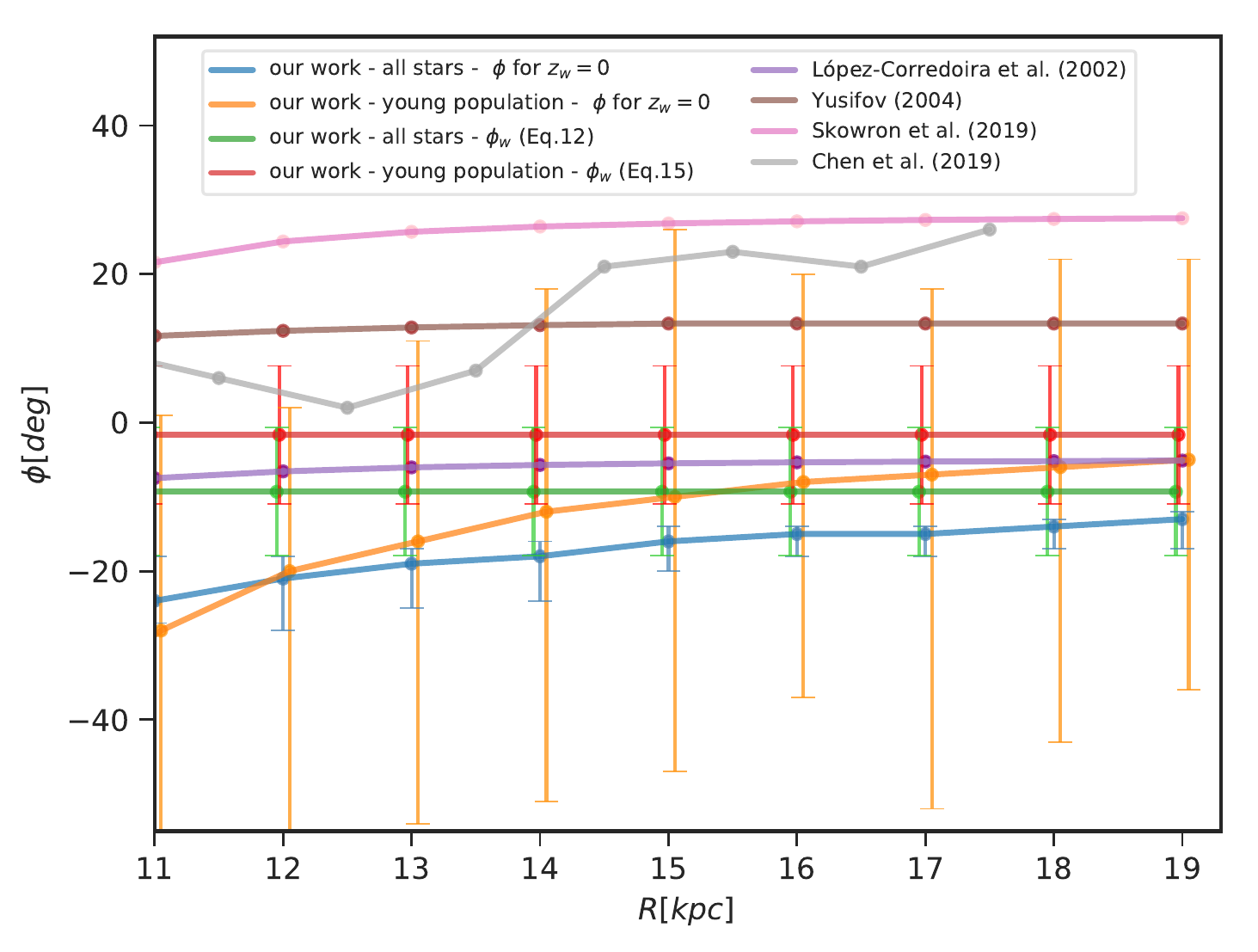}
        \caption{Comparison of line of nodes for our model (based on Eq. \ref{9}) with other works. We use two different methods to plot the line of nodes for our work. First, we plot the angle $\phi_w$ for Eq. (\ref{parametre}). Another method is to use the Eq. (\ref{9}) to find the value of the angle $\phi$ when $z_w=0$.}\label{lon}
\end{figure}

\cite{chen_warp} used Cepheids from the WISE survey and a number of optical surveys to measure the warp, and \cite{skowron} used Cepheids from the OGLE catalogue supplemented by other surveys. \cite{chen_warp} obtained a rather symmetric warp with an amplitude of about 1.5 kpc in R=20 kpc. \cite{skowron} obtained a similar result with an amplitude 0.74 kpc in R=15 kpc. These values are much higher than our findings, which is probably due to differences in the population: our young population is older than the Cepheids. In Fig. \ref{lon} we plot the variation of the line of nodes with radius for the whole and the young population, compared with other works. We use two different methods to plot the line of nodes for our work. First, we plot the angle $\phi_w$ for Eq. (\ref{parametre}). Another method is to use the Eq. (\ref{9}) to find the value of the angle $\phi$ when $z_w=0$. We would expect that our young population lies between the total population and the young Cepheids. This is true only for R>12 kpc. At shorter distances, the warp is not very strong and is more difficult to detect, therefore the error bars are larger in this area. Moreover, the error bars of the young populations are very large because of the lower number of stars in the sample combined with possible problems in determining extinction. For these reasons, the value of the line of nodes for R<12 kpc is rather unreliable.

\section{Conclusions}\label{concl}
We produced density maps from Gaia DR2 data and analysed them to study the Galactic warp. The density maps directly show a northern warp in the azimuths $\ang{60}<\phi<\ang{90}$ and a southern warp in the azimuths $\ang{270}< \phi <\ang{300}$. Our maps reach a Galactocentric radius of 20 kpc, and we note that up to this distance, the density decreases exponentially and we do not observe a cut-off. Another feature in the density maps is  a Galactic flare, that is, an increase in scale height towards the outer Galaxy. The analysis of the flare will be given in a forthcoming paper (Nagy et al. 2020, in preparation). We used the maps to calculate the scale length, where we find $h_r=2.29\pm0.08$ kpc, with a small dependence of $h_r$ from the Galactic azimuth. The lowest value of $h_r$ that we found is $1.88\pm0.12$ kpc for $\phi\approx\pm\ang{-30}$ and the highest value is $2.78\pm0.13$ kpc for the Sun-anticentre direction and $2.29\pm0.21$ $\phi\approx\pm\ang{30}$.  \\

From our maps, we calculated the average elevation of the plane and fitted it with different warp models. We fitted the northern and southern warp separately with a simple sinusoidal model, and we found a small asymmetry: the northern warp reaches an amplitude $0.317$ kpc for the azimuth $\ang{60}<\phi<\ang{90}$ and the southern warp reaches $-0.287$ kpc for the azimuth $\ang{240}<\phi<\ang{270}$, both at R=[19.5,20.0] kpc. Then we fitted the warp with a model combining two sinusoids to detect the asymmetry without assuming its existence, and we found values of amplitude $\sim0.5$ for the northern and $\sim-0.4$ for the southern warp both at R=[19.5,20.0] kpc, revealing the asymmetry found with the previous approach. The azimuths of the warp maximum and minimum for this model are $\ang{52}<\phi<\ang{56}$ and $\ang{312}<\phi<\ang{324}$ , respectively. In terms of Galactocentric radius, we find that warp starts to manifest itself from about 12 kpc and extends at least up to 20 kpc. We repeated this analysis on the young population, where we find that it follows the result for the whole population, but reaches a higher amplitude of warp and similar values of scale height. The comparison of our amplitude of the warp with other works showed that we obtain a significantly lower amplitude than an analysis carried out with very young stars such as Cepheids. This supports the formation of the warp through accretion onto the disc \citep{martin_accretion}. \\
A future analysis of the next Gaia data release combined with the deconvolution method based on Lucy's method of inversion, as described in Section \ref{ch6}, will allow us to explore distances larger than 20 kpc. The future data release will provide a much deeper magnitude limit and much lower parallax errors, which will allow us to extend the range of Galactocentric distances and study the morphology of the disc and of the stellar halo at very large distances.

\begin{acknowledgements}
We thank the anonymous referee for helpful comments, which improved this paper, and Astrid Peter (language editor of A\&A) for proof-reading of the text. ZC and MLC were supported by the grant PGC-2018-102249-B-100 of the Spanish Ministry of Economy and Competitiveness (MINECO). RN was supported by the Scientific Grant Agency VEGA No. 1/0911/17. This work made use of the IAC Supercomputing facility HTCondor (http://research.cs.wisc.edu/htcondor/), partly financed by the Ministry of Economy and Competitiveness with FEDER funds, code IACA13-3E-2493. This work has made use of data from the European Space Agency (ESA) mission {\it Gaia} (\url{https://www.cosmos.esa.int/gaia}), processed by the {\it Gaia} Data Processing and Analysis Consortium (DPAC, \url{https://www.cosmos.esa.int/web/gaia/dpac/consortium}). Funding for the DPAC has been provided by national institutions, in particular the institutions participating in the {\it Gaia} Multilateral Agreement. The reduced catalogue of Gaia with $m_G<19$ was produced by Pedro Alonso Palicio.
\end{acknowledgements}

\clearpage
\bibliographystyle{aa} 
\bibliography{Refer_notes_gaia_maps}

\appendix

\section{Lucy's method for the inversion of Fredholm integral equations 
        of the first kind}
\label{.Lucy}

The inversion of Fredholm integral equations of the first kind such as Eq. (\ref{2}) 
is ill-conditioned. Typical analytical methods for solving these equations \citep{balazs} cannot achieve a good solution because the kernel is sensitive to the noise of the star counts \citep[chapter 5]{craig}.
Because the functions in these equations have a stochastic rather than analytical interpretation, it is to be expected that statistical inversion algorithms are more robust \citep{turchin,jupp,balazs}.
These statistical methods include the iterative method of Lucy's algorithm 
\citep{lucy,turchin,balazs,martin2}, which is
appropriate here. Its key feature is the interpretation of the kernel as a conditioned probability and the application of Bayes' theorem.

In Eq. (\ref{2}), $N(\pi )$ is the unknown function, and the kernel is $G(x)$, whose
difference $x$ is conditioned to the parallax $\pi '$. The inversion is carried out as
\begin{equation}
N (\pi )=\lim _{n\rightarrow \infty}N _{n}(\pi )
,\end{equation}
\begin{equation}
N_{n+1}(\pi)=N_n(\pi )\frac{\int _0^\infty \frac{\overline{N}(\pi ')}
        {\overline{N_n}(\pi ')}G _{\pi '}(\pi -\pi')d\pi '}
{\int _0^\infty G_{\pi '}(\pi -\pi')d\pi '}
,\end{equation}
\begin{equation}
\overline{N_n}(\pi )=\int _0^\infty N_n(\pi ')G_{\pi '}(\pi -\pi')d\pi '
.\end{equation}
The iteration converges when $=\overline{N_n}(\pi )\approx \overline{N}(\pi )$ 
$\forall \pi$, that is, when $N _{n}(\pi )\approx N (\pi )$ $\forall \pi$. 
The first iterations produce a result that is close to the final answer, with the subsequent iterations giving only small corrections. In our calculation, we set as
initial function of the iteration $N_0(\pi )=\overline{N}(\pi ),$ and we carry out
a number of iterations until the Pearson $\chi ^2$ test
\begin{equation}
\frac{1}{N_p-2}\sum _{j=2}^{N_p-1}\frac{[\overline{N_n}(\pi _j)-\overline{N}(\pi _j)]^2}{\overline{N_n}(\pi _j)}
,\end{equation}
reaches the minimum value.
Further iterations would enter within the noise.

This algorithm has a number of beneficial properties (Lucy 1974, 1994): all the functions are defined as being positive, the likelihood increases with the number of iterations, the method is insensitive to high-frequency noise in $\overline{N}(\pi )$, and so on.
We note, however, that precisely because this method only works when $N$ are positive functions, it does not work with negative ones.

\nocite{romi}
\nocite{martin_warp}
\nocite{yusifov}
\nocite{bahcall_lum}
 
\end{document}